\documentclass[11pt,a4paper]{article}
\usepackage{amssymb}
\usepackage{graphicx}
\usepackage{tikz}
\usepackage{pgflibraryarrows}
\usepackage{pgflibrarysnakes}

\newcommand\qed{\hfill$\sqcap\kern-7.5pt\hbox{$\sqcup$}$}
\newcommand{\CC}{\mathbb{C}}

\newcommand{\RR}{\mathbb{R}}

\newcommand{\ZZ}{\mathbb{Z}}
\newtheorem{theo}{Theorem}
\newtheorem{prop}[theo]{Proposition}
\newtheorem{lem}[theo]{Lemma}
\newtheorem{cor}[theo]{Corollary}
\newtheorem{rem}[theo]{Remark}
\newtheorem{defin}[theo]{Definition}

\newcommand{\re}{R}
\renewcommand{\Re}{\re}
\newcommand{\im}{I}
\renewcommand{\Im}{\im}
\newcommand{\beqn}{\begin{equation}}
\newcommand{\eeqn}{\end{equation}}
\newcommand{\bear}{\begin{eqnarray}}
\newcommand{\eear}{\end{eqnarray}}
\newcommand{\bean}{\begin{eqnarray*}}
\newcommand{\eean}{\end{eqnarray*}}

\begin{document}
\title{On the Fundamental Solution of a Homogeneous Linearized Coagulation Equation.}
\maketitle

\begin{center}
M. Escobedo\footnotemark[1] and J. J. L. Vel\'azquez \footnotemark[2]
\end{center}
\footnotetext[1]{Departamento de Matem\'aticas, Universidad del
Pa{\'\i}s Vasco, Apartado 644, E--48080 Bilbao, Spain. E-mail: {\tt
mtpesmam@lg.ehu.es}} 
\footnotetext[2]{ICMAT (CSIC-UAM-UC3M-UCM) Facultad de Matem\'aticas, Universidad Complutense. E--28040 Madrid, Spain. E-mail~: {\tt
JJ\_Velazquez@mat.ucm.es}} 

\noindent

\section{Introduction}
\label{Introduction}
\setcounter{equation}{0}
\setcounter{theo}{0}

Under rather general conditions on the kernel $K(x, y)$, a symmetric homogeneous function in $x$ and $y$,  the Cauchy problem for the Smoluchowski coagulation equation:
\bear
{\partial f\over \partial t} & = & Q[f] \label{e1.1}\\ \nonumber \\
Q[f]& = & {1\over 2}\int_0^xK(x-y, y)\, f(x-y)\, f(y)\,
dy-\int_0^{\infty} K(x, y)\, f(x)\, f(y)\, dy \label{e1.1bis}\\ \nonumber \\
f(0, x) & = & f_{in}(x)\in L_1^1(\RR),\label{e1.2}
\eear
has a global  solution $f(t, x)$ for all initial data $f_{in}(x)$ such that $\int_0^{\infty}x\, f_{in}(x)dx<\infty$. Moreover, this solution satisfies the same estimate for all $t>0$. 

Equation (\ref{e1.1}), (\ref{e1.1bis}) describes the aggregation process of particles of mass $x$ and $y$ with probability $K(x, y)$, assuming that the distribution of particles are uncorrelated at all times. In this context the quantity:
\bear
\int_0^\infty x\, f(t, x)\, dx
\eear
represents the total mass of particles in the system.

On the other hand, it is known that, when  the kernel is of the form $K(x, y)=x^\alpha y^\beta +x^\beta y^\beta$ with $\alpha\ge 0$, $\beta\ge 0$ and $\alpha+\beta=\lambda >1$, the solutions to the Cauchy problem for the Smoluchowski equation  undergo the so called gelation phenomenon. This means that there exists a
positive time
$T_g<\infty $ such that, for all $t<T_g$,
\beqn
\label{e1.3}
\int_0^{\infty }x\, f(t, x)\, dx=\int_0^{\infty }x\, f_{in}(x)\, dx
\eeqn
and for all $t>T_g$,
\beqn
\label{e1.4}
\int_0^{\infty }x\, f(t, x)\, dx<\int_0^{\infty }x\, f_{in}(x)\, dx.
\eeqn
It is also known that when $\lambda \le 1$ gelation does not occur and mass is conserved for all time (cf. \cite{DS, S2}).
Two interesting open questions are related with this phenomenon. One is to describe the
solution $f$ as $t\to T_g$.  We consider here the second one, which is to understand the behaviour of the solution $f$ after the gelling time $T_g$.

Althought no general result is known, several partial results indicate that before the gelling time, at least for a large familiy of initial data, the solutions to the coagulation equation decay exponentially fast as $x\to +\infty$. The Smoluchowski equation (\ref{e1.1}) has a discrete counterpart:
for which an explicit exact gelling solution was constructed in \cite{LT} for $K(k, j)= k\, j$. Such a solution decays exponentially fast before the gelling time, and as a power law after that time.
The exponential decay, before the gelling time,  was later shown in \cite{EZH} for the continuous equation (\ref{e1.1}), $\lambda =2$ and several initial data. Moreover, it has also been formally shown in \cite{vDE, EZH} that, for several initial data and $\lambda \in (1, 2]$, the solution of (\ref{e1.1}) decays,  after gelling, like $x^{-(3+\lambda )/2}$ as $x\to +\infty $ (see \cite{Le} for more detailed references). On the other hand, it was proved in \cite{EMP} that $x^{-(3+\lambda )/2}$ is the only possible power law decay for the solutions of  (\ref{e1.1}) after gelation. Our main purpose is to prove that for the  coagulation kernel
\bear
\label{K}
K(x, y)=(xy)^{\lambda/2},\quad \lambda \in (1, 2)
\eear 
and any initial data
$f_{in}$, regular near the origin and such that:
\bear
\label{S1Edatin}
f_{in}(x)\sim x^{-(3+\lambda )/2}\quad \hbox{as}\,\,x\to +\infty,
\eear 
the problem  (\ref{e1.1})-(\ref{e1.2}) has a solution $f$ satisfying
\bear
\label{S1ECompinf}
f(t, x)\sim a(t)\, x^{-(3+\lambda)/2}\,\,\,\hbox{as}\,\,\,x\to +\infty. 
\eear
Moreover, this solution satisfies 
\bear
\label{S1Eflux}
\frac{d}{dt}\int_0^{\infty}x\, f(x, t)dx = -2 \pi a^2(t)\quad \hbox{for all}\,\,t>0
\eear
which was formally shown in \cite{vD} for the discrete equation. 
By (\ref{S1Eflux}) the total mass of the solution $f$ is decreasing. This loss of mass is a characteristic feature of the solutions of (1.1),(1.2) after the gelation time. 
The choice of exponents $\lambda <2$ is natural, because $\lambda \le 2$ excludes instantaneous gelation or non existence of solutions (\cite{CdC,vD, S}) and $\lambda=2$ is one of the ``explicit'' cases which has been treated using the Laplace transform (cf. \cite{EZH}).

In order to prove the existence  of classical solutions of  (\ref{e1.1})-(\ref{e1.2}) after gelation we will use the same approach as in \cite{EMV, EMV2}.
The starting point of this approach is to linearize around an initial data $f_{in}$ satisfying $f_{in} \approx x^{-(3+\lambda)/2}$ for $x$ large and to derive detailed estimates on the solutions of the resulting linear equation. 
\bear
\label{S1ELcurva}
\frac{\partial g}{\partial t}={\cal L}(g, f_{in}).
\eear
 To this end we will need some rather delicate estimates on the asymptotics of the solutions as $x$ tends to infinity. Moreover, even to prove solvability of the linearized problem (\ref{S1ELcurva}) is nontrivial. We will obtain it treating this problem as a perturbation of the problem obtained replacing $f_{in}$ by its asymptotics as x tends to infinity: 
\bear
\label{S1ELrectacauchy}
\frac{\partial g}{\partial t}=L(g).
\eear
In order to carry on this program we need to derive detailed estimates about the solutions of (\ref{S1ELrectacauchy}). This will be the main goal of this paper.

The linearised equation around the weak solution $x^{-(3+\lambda )/2}$ may be introduced more directly as follows. Consider a solution $f(t,  x)$ of the coagulation equation with an initial data $f_{in}$ satisfying (\ref{S1Edatin}). If one is interested in the behaviour of $f(t, x)$ for $x$ large it is natural to scale the variables as follows: $x=R\, \overline x, y=R\, \overline y, t=R^{-(\lambda-1)/2}\overline t$  and $f(t, x)=R^{-(3+\lambda)/2}\, F_R(\overline t, \overline x)$. In these new variables, the equation  (\ref{e1.1}) reads $(F_R)_{\overline t}=Q[F_R]$ and the initial data $F_{in}$ satisfies now: $F_R(0, \overline x)=R^{(3+\lambda)/2}\, f_{in}(R\, \overline x)\sim (\overline x) ^{-(3+\lambda)/2}$ as $R\to +\infty$. The limit of the function $F_R$ as $R\to +\infty$, if it exists,  would then solve the same equation (\ref{e1.1}) with initial data $\overline x\, ^{-(3+\lambda)/2}$. Therefore the linear problem 
(\ref{S1ELrectacauchy}) appears naturally as the linearisation of the coagulation equation (\ref{e1.1}) in the region $x>>1$. Notice, however that in the region where $\overline x$ is small the function $\overline f$ is bounded and the approximation by means of the power law $\overline x\,\, ^{-(3+\lambda)/2}$ cannot be valid. The analysis of that region would lead naturally to the study of a boundary layer whose description requires the analysis of the operator ${\cal L}$. This will be made in a forthcoming work.

On the other hand, the linearised equation (\ref{S1ELrectacauchy}) has some interest by itself. It is indeed a simple model to describe a set of particles at equilibrium, whose density distribution is given by $x^{-(3+\lambda)/2}$, and where a small set of particles is introduced, whose distribution $\varphi(x)$ is considered as a small perturbation. The particles so introduced  start to collide both between themselves and with the particles in the background. The equilibrium density distribution $x^{-(3+\lambda)/2}$ is then perturbed. The distribution density function of the resulting set of particles may then be seen at any time $t$ as the equilibrium distribution $x^{-(3+\lambda)/2}$ and a remaining perturbation $\varphi(t, x)$.
The linear equation (\ref{S1ELrectacauchy}) only takes into account the collisions of the ``particles in the perturbation'' with the background and describes how the distribution of these particles evolves in time. It neglects the collisions between particles in the perturbation. This could be a reasonable approximation as long as the
perturbation $\varphi(t, x)$ remains small.
Notice that the number of clusters in the background as well as the number of particles (the total mass)
are infinite (since nor $x^{-(3+\lambda)/2}$ nor $x^{1-(3+\lambda)/2}$ are integrable in $(0,+\infty$)), but the number of clusters and particles in the initial  perturbation are finite. Our results show the following:
\begin{itemize}
\item There is instantaneously an infinite number of ``perturbed clusters'', although their mass  is finite.

\item As $t\to +\infty$, the number of perturbed particles (the mass in the perturbation) tends to zero, but the number of perturbed clusters remains infinite.

\item The total flux of particles is perturbed at $t$ finite  but  tends to the flux corresponding to the original equilibrium distribution as $t\to +\infty$.

\end{itemize}

Our results are obtained  using classical Fourier analysis and the Wiener Hopf method, in a similar way as we did for the linearized Uehling Uhlenbeck operator in \cite{EMV} although with an important difference. This is the regularising effect of the operator $L$, absent in the operator studied in \cite{EMV}, and coming from the fact that $L$ is similar to the half derivative operator. The fundamental solution of  (\ref{S1ELrectacauchy}) has then very different properties than that obtained in \cite{EMV}. 

In Section \ref{The Linearized Equation} we state our main results and transform the integro differential equation (\ref{S1ELrectacauchy}) to a Carleman equation in the complex plane. In Section \ref{The auxiliary function} we state the fundamental  properties of the auxiliary function $\Phi$ appearing in the Carleman equation. This equation is solved in Sections \ref{sole2} and \ref{solecinco} using the classical Cauchy integral, which gives an explicit solution. In sections \ref{decayghat} and \ref{decayG}, the precise asymptotics of the solutions are obtained. The Section \ref{invalpbm} is devoted to a brief mention of the initial value problem. Some properties of the fluxes of particles described by the solutions are considered in Section \ref{flux}. We have finally added  Appendices I, II and III where are collected some necessary technical results.

\section{The Linearized Equation}
\label{The Linearized Equation}

\setcounter{equation}{0}
\setcounter{theo}{0}

We start this Section writing the precise expression of linearized equation 
(\ref{S1ELrectacauchy}).
\begin{prop} 
\label{S2Th1}
The linearised equation of (\ref{e1.1})-(\ref{e1.1bis}) with $K(x, y)=(x y)^{\lambda/2}$  around the solution $G(x)=x^{-(3+\lambda )/2}$ is
\bear
\label{e2.4bis}
\frac{\partial g }{\partial t}& = & L(g),
\\
\label{e2.5}
L(g) 
 & = &  
\int_0^{x/2}\left((x-y)^{\lambda /2}G(x-y)-x^{\lambda
/2}G(x) \right)y^{\lambda /2}g(y)dy  \\ 
&& + \int_0^{x/2}\left((x-y)^{\lambda /2}g(x-y)-x^{\lambda /2}g(x)
\right)y^{-3/2}dy \nonumber\\
&& - x^{-3/2}\int_{x/2}^{\infty }y^{\lambda /2}g(y)dy-2\sqrt 2 x^{(\lambda -1)/2}g(x).\nonumber
\eear
\qed

\end{prop}

\noindent 
\textbf{Proof.}
By the symmetry of the kernel $K$:
\beqn
\label{e2.3}{1\over 2}\int_0^{x/2}K(y, x-y)f(y)f(x-y)dy={1\over
2}\int_{x/2}^xK(y, x-y)f(y)f(x-y)dy
\eeqn
Then we may then write the equation (\ref{e1.1})-(\ref{e1.1bis}) as follows:
\beqn
\label{e2.4}
{\partial f\over \partial t} =\int_0^{x/2}\left[(x-y)^{\lambda
/2}f(x-y)-x^{\lambda /2}f(x)
\right]y^{\lambda /2}f(y)dy-\int_{x/2}^{\infty }K(x, y)f(x)f(y)dy.
\eeqn
If we linearize around the solution $G(x)=x^{-(3+\lambda )/2}$, define
$f=G+g$ and neglect quadratic terms on $g$ we  obtain (\ref{e2.4bis}), (\ref{e2.5}).

\begin{rem}
\label{S2R1bis}
The second term in the right hand side of (\ref{e2.5}) can be seen as some kind of half derivative operator applied to  function $x^{\lambda /2}\, g(x)$. This will appear again in the Fourier analysis that will be done later on the linearised equation.
\end{rem}

\begin{rem}
\label{S2R1}
In order for the first integral in the right hand side of (\ref{e2.5}) to be defined we need $y^{1+\lambda/2}g(y)$ to be integrable at the origin. For the second integral we need some kind of regularity of $g(x)$ with respect to $x$. For example $y^{\lambda/2} g(y)$ $\gamma$-H\"older continuous with $\gamma > 1/2$. Finally, for the last one we need $y^{\lambda/2} g(y)$ to be integrable as $y\to \infty$. Assuming power like behaviours we then need bounds on $g$ of the form:
\bear
&&g(y)\le C y^{-\lambda/2 -r} \quad \hbox{as}\quad y\to +\infty,\label{S2Est}\\
&&g(y)\le C y^{-\lambda/2 -\rho} \quad \hbox{as}\quad y\to 0,\label{S2Est2}
\eear
for some  $r>1$ and $\rho <2$.
\end{rem}

\begin{defin}
\label{S2Def1}
We will denote as ${\cal V}(r, \rho)$ the set of functions 
\bear
{\cal V}(r, \rho)=\left\{
g\in L^{\infty}_{\hbox{loc}}(\RR^+);\,\, \sup_{0\le x \le 1}g(x) x^{\lambda/2+\rho}<\infty, \,\,\,\,\, \sup_{1\le x }g(x) x^{\lambda/2+r}<\infty
\right\}
\eear
\end{defin}
We state now the main results of this paper. The first one is an existence and uniqueness result of fundamental solutions for the equation (\ref{e2.4bis}), (\ref{e2.5}).

\begin{theo}
\label{S2T1-1} For all $x_0>0$, there
exists a unique solution $g(t, \cdot, x_0)$ of (\ref{e2.4bis}), (\ref{e2.5}) with initial data:
\bear
\label{S2ID}
g(0, x, x_0)=\delta(x-x_0)
\eear such that $g(t, \cdot, x_0)\in 
{\cal V}(3/2, (3-\lambda)/2)$ for all $t>0$. Moreover, $g(t, \cdot, x_0)$ has the self similar form
\bear
\label{S2.SFSM}
g(t, x, x_0)=\frac{1}{x_0}g\left(tx_0^{\frac{\lambda-1}{2}}, \frac{x}{x_0}, 1\right).
\eear
The function $g(t, \cdot, 1)$ satisfies the following behaviours in the different regions of the $k, t$ space for some explicitly known constants $a_i, i=1, \cdots, 5$ and arbitrarily small positive constants $\varepsilon$ and $\delta$.\\ \\
We have the following representation formula for $t\ge 1$:
\bear
\label{S2Ttinfty1}
g(t, x, 1)=t^{\frac{2}{\lambda-1}} \varphi_1(\sigma)+\varphi_2(t, \sigma)
\eear
where $\sigma$ is the self similar variable:
\bear
\label{S2Eautosem}
\sigma=t^{\frac{2}{\lambda-1}}\, x, 
\eear
and the functions $\varphi_1$ and $\varphi_2$ satisfy the following estimates:
\beqn
 \label{S2Ttinfty2}
\varphi_1(\sigma)  = \left\{
\begin{array}{l} a_1\, \sigma^{-\frac{3}{2}}+
{\cal O}\left(\sigma^{-\frac{4-\lambda}{2}+\varepsilon} \right)\quad \hbox{for}\,\,\,0\le \sigma<1  \\ \\
a_2\, \sigma^{-\frac{3+\lambda}{2}}+
{\cal O}\left(\sigma^{-(1+\lambda+\varepsilon})\right)\quad \hbox{for}\,\,\,\sigma > 1 
\end{array}
\right.
\eeqn
where $a_1$ and $a_2$ are two explicit constants and for some positive constant $\varepsilon$ arbitrarily small,
\beqn
 \label{S2Ttinfty3}
\varphi_2(t, \sigma)  = \left\{
\begin{array}{l} 
b_1(t)\, \sigma^{-\frac{3}{2}}+{\cal O}\left(t^{\frac{2}{\lambda-1}-\delta_0}\, \sigma^{-\frac{3}{2}+\varepsilon} \right)\quad \hbox{for}\,\,\,0\le \sigma<1  \\ \\
b_2(t)\, \sigma^{-\frac{3+\lambda}{2}}+{\cal O}\left(t^{\frac{2}{\lambda-1}-\delta_0}\, \sigma^{-\frac{3+\lambda}{2}-\varepsilon} \right)\quad \hbox{for}\,\,\,\sigma>1   
\end{array}
\right.
\eeqn
where  $b_1$ and $b_2$ are two continuous functions such that $|b_1(t)|+|b_2(t)|\le C\, t^{\frac{2}{\lambda-1}-\delta_0}$, $\varepsilon>0$ and $\delta_0>0$.\\ 
On the other hand, for all $0<t<1$ fixed:
\beqn
 \label{S2Ttuno}
g(t, x, 1)=
\left\{
\begin{array}{l}
a_3\, t\, x^{-\frac{3}{2}}+b_3(t)\, x^{-\frac{3}{2}}+
{\cal O}\left(t\, x^{-\frac{3}{2}+ \delta_1} \right)\quad \hbox{for}\,\,\,0\le x\le \frac{1}{2}\\ \\
a_4\, t\, x^{-\frac{3+\lambda}{2}}+b_4(t)\, x^{-\frac{3+\lambda}{2}}+
{\cal O}\left(
t\, x^{-\frac{3+\lambda}{2}-\delta_1} \right)\quad \hbox{for}\,\,\,x\ge \frac{3}{2},
\end{array}
\right.
\eeqn
where  $b_3$ and $b_4$ are continuous functions such that $\left| b_{3}\left( t\right) \right| +\left| b_{4}\left( t\right)
\right| \leq Ct^{1+\delta},$ with $\delta>0,\;\delta_{1}>0$ and $
a_{3},\;a_{4}$ explicit numerical constants. \\
Finally, if $t\to 0$:
\beqn
 \label{S2Ttcero}
g(t, x, 1)=
\left\{
\begin{array}{l}
t^{-2}\Psi \left(\frac{x-1}{t^2} \right)+{\cal O}\left(t^{-2} \right)
\quad \hbox{for}\,\,\,x=1+{\cal O}(t^2)\\ \\
{\cal O}\left(\frac{t^{1-2\delta}}{|x-1|^{\frac{3}{2}-\delta}} \right) \quad \hbox{for}\,\,\,t^2<|x-1|<\frac{1}{2}
\end{array}
\right.
\end{equation}
 where the function $\Psi$ is given by:
 \beqn
  \label{S2Functipsi}
\Psi(\chi)= \left\{
\begin{array}{l}
\frac{2}{\pi}e^{-\frac{\pi}{\chi^{3/2}}},\quad \hbox{for all}\,\,\,\chi\ge 0,\\ \\
0\quad \hbox{for all}\,\,\,\chi< 0.
 \end{array}
\right.
\eeqn
\end{theo}
\begin{rem}
For $t\ge 1$ large the behaviour if $g$ in terms of the variable $x$ is :
\beqn
 \label{S2Remtinfty}
g(t, x, 1)  = \left\{
\begin{array}{l} a_1\, t^{\frac{2}{\lambda-1}}\, \sigma^{-\frac{3}{2}}+b_1(t)\sigma ^{-\frac{3}{2}}+
{\cal O}\left(t^{\frac{2}{\lambda-1}}\, \sigma^{-\frac{4-\lambda}{2}+\varepsilon}+t^{\frac{2}{\lambda-1}-\delta_0}\, \sigma^{-\frac{3}{2}} \right)\quad \hbox{for}\,\,\,\,0\le \sigma<1  \label{S2Ttinfty}\\ \\
a_2\, t^{\frac{2}{\lambda-1}}\, \sigma^{-\frac{3+\lambda}{2}}+b_2(t)\sigma ^{-\frac{3+\lambda}{2}}+
{\cal O}\left(t^{\frac{2}{\lambda-1}}\, \sigma^{-(1+\lambda+\varepsilon)}+t^{\frac{2}{\lambda-1}-\delta_0}\, \sigma^{-\frac{3+\lambda}{2}} \right)\quad \hbox{for}\,\,\,\sigma\ge 1 
\end{array}
\right.
\eeqn

\beqn
 \label{S2Remtinfty2}
g(t, x, 1)  = \left\{
\begin{array}{l} a_1\, t^{-\frac{1}{\lambda-1}}\, x^{-\frac{3}{2}}+
{\cal O}\left(t^{-\frac{1-2\varepsilon}{\lambda-1}}\, x^{-\frac{3}{2}+\varepsilon}+t^{-\frac{1}{\lambda-1}-\delta}\, x^{-\frac{3}{2}} \right)\quad \hbox{for}\,\,\,0<x<2\,t^{-\frac{2}{\lambda-1}}  \\ \\
a_2\, t^{-\frac{\lambda+1}{\lambda-1}}\, x^{-\frac{3+\lambda}{2}}+
{\cal O}\left(t^{-\frac{\lambda+1+2 \varepsilon}{\lambda-1}}\, x^{-\frac{3+\lambda}{2}-\varepsilon}+t^{-\frac{\lambda+1}{\lambda-1}-\delta}\, x^{-\frac{3}{2}} \right)\quad \hbox{for}\,\,\,x>2\,t^{-\frac{2}{\lambda-1}} 
\end{array}
\right.
\eeqn
\end{rem}
Our second result is about the regularity of the fundamental solutions of (\ref{e2.4bis}), (\ref{e2.5}).

\begin{rem}
\label{S2R1Bis}
Using the rescaling properties of the problem (\ref{e2.4bis}), (\ref{e2.5}),  
(\ref{S2ID}) has the self similar form (\ref{S2.SFSM}). It is then enough to restrict our analysis to the case $x_0=1$.
\end{rem}

Our strategy to solve the problem  (\ref{e2.4bis}), (\ref{e2.5}),  
(\ref{S2ID}) is to use Fourier analysis. The resulting problem is explicitly solvable by means of the Wiener Hopf method \cite{BZ}. Using the representation formula for the solution, we then prove Theorem \ref{S2T1-1} by deriving suitable a priori estimates. Similar arguments have been used in \cite{EMV}.

\subsection{Fourier variables.}
We make now a change of variables in order to have functions defined in all of the
real line $\RR$. To this end we define $x=e^X$, $X\in \RR$, as well as the Fourier transform
\bear
\widehat G(t, \xi)=\frac{1}{\sqrt{2 \pi}}\int_{\RR} e^{-i X \xi} G (t, X)dX,\quad G(t, X)=g(t, e^X)
\eear
Then, 
the problem  (\ref{e2.4bis}), (\ref{e2.5}),  (\ref{S2ID}) reads in terms of the new variables:
\bear
\label{e2.24}
&&{\partial \widehat G\over \partial t}(t, \xi )=\widehat G\left(t, \xi +{\lambda -1\over 2}i\right)\Phi \left(\xi +{\lambda
-1\over 2}i\right)\\
&&\widehat G(0, \xi)=\frac{1}{\sqrt{2 \pi}} \label{e2.24bis}
\eear
where the function $\Phi$ is given by:
\bear
\Phi (\xi ) & = & -\,\frac{2\sqrt \pi\, \,\Gamma(i\xi +1+{\lambda \over2})}{\Gamma(i\xi +\frac{\lambda+1}{2})}.
\label{e2.2490367}
\eear
as it is shown in Section  \ref{Appendix I}. The fact that the function $g(t, \cdot, 1)\in {\cal V}(3/2, (3+\lambda)/2)$ implies that the function $G(t, \cdot)$ is analytic in the strip:
$\mathcal S=\{\xi \in \CC;\,\,\Im m \xi \in (3/2, (3+\lambda)/2) \}$.

\section{The auxiliary function.}
\label{The auxiliary function}
\setcounter{equation}{0}
\setcounter{theo}{0}

The properties of the function $\Phi $, defined by (\ref{e2.2490367}) determine the properties of the equation
(\ref{e2.24}). The set of its zeros is:
\beqn
\label{e3.1}
\xi_z(n)= i\, \left(n+{1+\lambda \over 2}\right),\quad n=0, 1, \cdots
\eeqn
The singularities of the function $\Phi$ are given by the family of poles: 
\beqn
\label{e2.26}
\xi_p(n)= i\, \left(1+{\lambda \over 2}+n\right),\quad n=0, 1, \cdots
\eeqn
The asymptotic  behaviour of the function $\Phi$ as $\Re e (\xi) \to \pm \infty $ is the following:

\begin{prop}
\label{S3T1}
For all $M>0$ fixed:
\bean
\Phi (\xi ) & = & -\sqrt{2\pi}(1+iQ)\sqrt{Q\, \xi}-\frac{\sqrt{2\pi}(1+iQ)i}{\xi}\sqrt{Q\, \xi}
\left(\frac{1}{8}+\frac{\lambda}{4} \right)+{\cal O}\left( \frac{1}{|\xi|^{3/2}}\right)
\eean
as $\Re e(\xi)\to \infty$, uniformly on $\Im m(\xi)\in (-M, M)$ and where the function $Q$ is defined as:
\bear
\label{S3Etheta}
Q \equiv Q(\xi)= \hbox{sgn}(\Re e[\xi]).
\eear
\end{prop}

\begin{figure}[ht]
\centering
\includegraphics[width=1\textwidth,height=0.45\textheight]{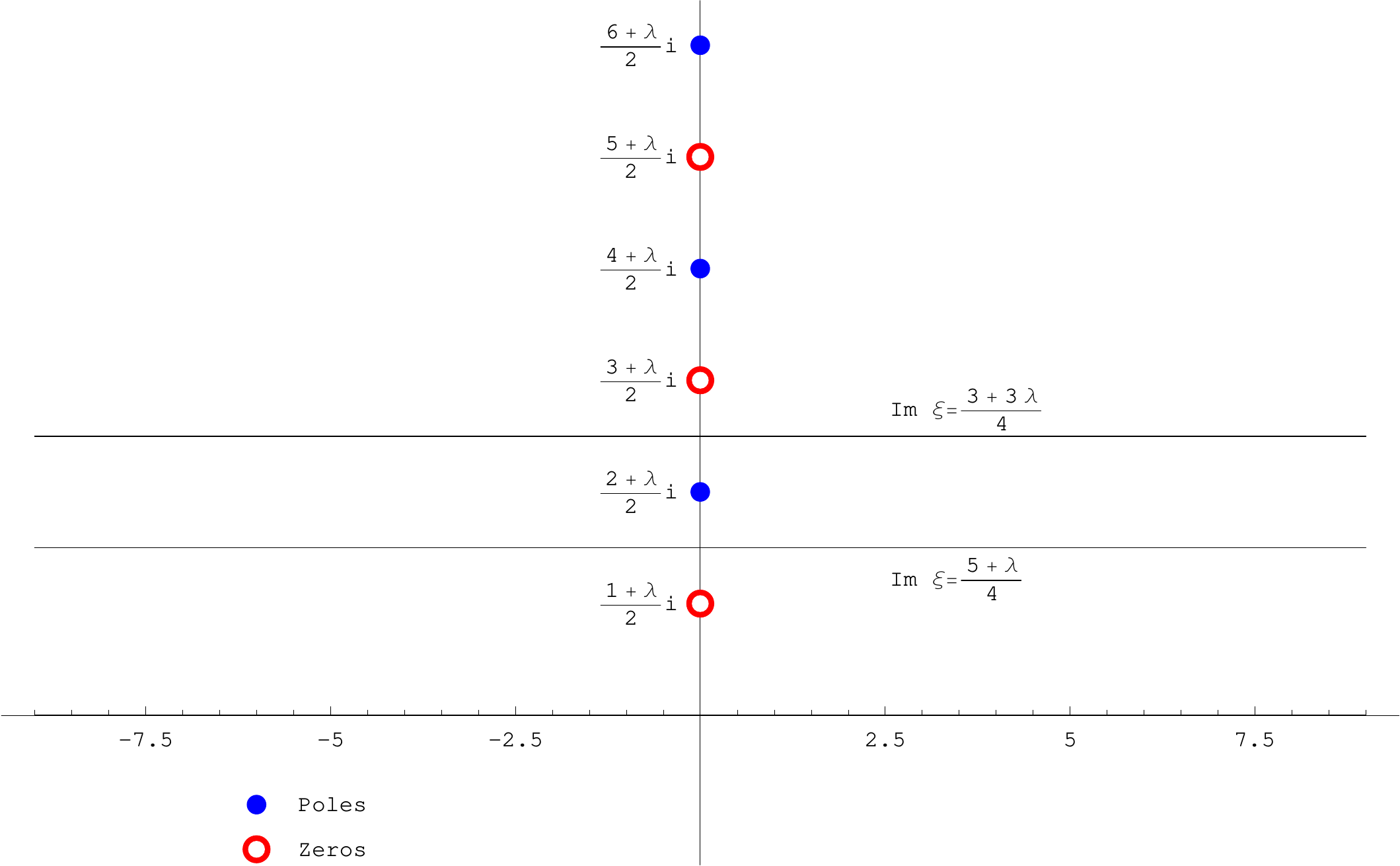}
\caption{Some relevant zeros and poles of the function $\Phi$.}\label{Cerosypolos}
\end{figure}

\section{Solving (\ref{e2.24})-(\ref{e2.24bis})}
\label{sole2}
\setcounter{equation}{0}
\setcounter{theo}{0}

Our goal is to solve the problem (\ref{e2.24})-(\ref{e2.24bis}) which we recall here:
\bear
&&\frac{\partial\widehat{G}}{\partial t}\left(  t,\xi\right)     =\widehat{G}\left(
t,\xi+\frac{\lambda-1}{2}i\right)  \Phi\left(  \xi+\frac{\lambda-1}
{2}i\right)  \label{E1}\\
&&\widehat{G}\left(  0^{+},\xi\right)     =\frac{1}{\sqrt{2\pi}}\label{E2}
\eear
where the function $\widehat{G}$ must be analytic in the strip $\mathcal S$. This
analyticity condition is a consequence of the decay required for $G\left(
t,X\right)  $ in order for the integral equation to make sense. Since we are
interested in deriving a solution $G\left(  t,X\right)  $ in the sense of distributions, we want to obtain boundedness of $\widehat{G}$ as $\left|
\Re e\left(  \xi\right)  \right|  \rightarrow\infty.$ In this case
we will obtain, for the particular solution $\widehat G$ constructed here, exponential
decay, something that means that $G\left(  t,\cdot\right)  \in C^{\infty}$ for $t>0.$

The following transformation allows to reduce (\ref{E1}), (\ref{E2}) to an
equation for a function depending only in one variable $\xi:$
\begin{equation}
\widehat{G}\left(  t,\xi\right)  =-\frac{\sqrt{2}}{\sqrt{\pi}i\left(
\lambda-1\right)  }\int_{\Im m\left(  y\right)  =\beta_{0}}
\frac{\mathcal{V}\left(  \xi\right)  }{\mathcal{V}\left(  y\right)  }
t^{\frac{2i\left(  \xi-y\right)  }{\lambda-1}}\Gamma\left(  -\frac{2i\left(
\xi-y\right)  }{\lambda-1}\right)  dy\label{E3}
\end{equation}
for some $\beta_{0}\in (3/2, 2)$. We have:

\begin{lem}
Suppose that $\mathcal{V}\left(  \eta\right)  $ is analytic in the strip
$\mathcal{S}$ and that $\mathcal{V}$
satisfies
\begin{equation}
\int_{\Im m\left(  y\right)  =\beta_{0}}\left|  \frac
{1}{\mathcal{V}\left(  y\right)  }\right|  e^{-\frac{\pi}{\left(
\lambda-1\right)  }\left|  y\right|  }\sqrt{\left|  y\right|  +1}\left|
dy\right|  <\infty\label{E4a}
\end{equation}
for any $\beta_{0}\in\left(  \frac{3}{2},2\right)  ,$ as well as:
\begin{equation}
\mathcal{V}\left(  \eta\right)  =-\mathcal{V}\left(  \eta+\frac{\lambda-1}
{2}i\right)  \Phi\left(  \eta+\frac{\lambda-1}{2}i\right)  \label{E5}
\end{equation}
for $\Im m\left(  \eta\right)  \in\left(  \frac{3}{2},2\right)  .$
Define $\widehat{G}\left(  t,\xi\right)  $ by means of (\ref{E3}) for
$\Im m\left(  \xi\right)  >\beta_{0}$. Then $\widehat{G}$ can be
extended analytically to $\mathcal{S}$ and it solves (\ref{E1}), (\ref{E2})
for $\Im m\left(  \eta\right)  \in\left(  \frac{3}{2},2\right)  .$
\end{lem}

\textbf{Proof.}
It just follows by direct computation. Indeed, notice that Stirling's formula,
that is uniformly valid for $\Gamma\left(  z\right)  ,\;\arg\left(  z\right)
\in\left(  -\pi+\varepsilon_{0},\pi-\varepsilon_{0}\right)  $ with
$\varepsilon_{0}>0$ (cf. \cite{AS}) implies:
\[
\left|  \Gamma\left(  -\frac{2i\left(  \xi-y\right)  }{\lambda-1}\right)
\right|  \leq C_{R}\frac{e^{-\frac{\pi}{\left(  \lambda-1\right)  }\left|
y\right|  }}{\sqrt{\left|  y\right|  +1}}\;\;
\]
for $\left|  \xi\right|  \leq R,$ $\Im m\left(  y\right)
=\beta_{0}.$ Therefore, the integral on the right-hand side of (\ref{E3})
converges for any $\xi\in\mathcal{S\cap}\left\{  \xi:\Im m\left(
\xi\right)  \in\left(  \beta_{0},\frac{3+\lambda}{2}\right)  \right\}  $ due
to (\ref{E4a}) and the function $\widehat{G}\left(  t,\xi\right)  $ satisfies:
\[
\left|  \widehat{G}\left(  t,\xi\right)  \right|  \leq C_{R}\int
_{\Im m\left(  y\right)  =\beta_{0}}\left|  \frac{1}
{\mathcal{V}\left(  y\right)  }\right|  e^{-\frac{\pi}{\left(  \lambda
-1\right)  }\left|  y\right|  }\frac{\left|  dy\right|  }{\sqrt{\left|
y\right|  +1}}<\infty\;\;,\;\;\left|  \xi\right|  \leq R
\]
Taking $\beta_{0}$ arbitrarily close to $\frac{3}{2}$ we obtain analyticity of
$\widehat{G}$ in $\mathcal{S.}$

Moreover, the derivative with respect to $t$ of $\widehat{G}$ in (\ref{E3}) can be
computed by means of:
\begin{equation}
\frac{\partial\widehat{G}}{\partial t}\left(  t,\xi\right)  =\frac{\sqrt{2}}
{\sqrt{\pi}i\left(  \lambda-1\right)  }\int_{\Im m\left(
y\right)  =\beta_{0}}\frac{\mathcal{V}\left(  \xi\right)  }{\mathcal{V}\left(
y\right)  }t^{\left[  \frac{2i\left(  \xi-y\right)  }{\lambda-1}-1\right]
}\Gamma\left(  -\frac{2i\left(  \xi-y\right)  }{\lambda-1}+1\right)
dy\label{E7}
\end{equation}
where we have used  $z\Gamma\left(  z\right)  =\Gamma\left(  z+1\right)  $. On the other hand, using (\ref{E3}) we obtain:
\[
\widehat{G}\left(  t,\xi+\frac{\left(  \lambda-1\right)  }{2}i\right)
=-\frac{\sqrt{2}}{\sqrt{\pi}i\left(  \lambda-1\right)  }\int
_{\Im m\left(  y\right)  =\beta_{0}}\frac{\mathcal{V}\left(
\xi+\frac{\left(  \lambda-1\right)  }{2}i\right)  }{\mathcal{V}\left(
y\right)  }t^{\frac{2i\left(  \xi-y\right)  }{\lambda-1}-1}\Gamma\left(
-\frac{2i\left(  \xi-y\right)  }{\lambda-1}+1\right)  dy
\]
and using (\ref{E5}) and (\ref{E7}), (\ref{E1}) follows.

It only remains to check (\ref{E2}). To this end we use contour deformation
and residue Theorem to transform (\ref{E3}) into:
\bean
&&\widehat{G}\left(  t,\xi\right)    =\frac{1}{\sqrt{2\pi}}-\widehat{G}_{r}\left(
t,\xi\right)  \\
&&\widehat{G}_{r}\left(  t,\xi\right)     =\frac{\sqrt{2}}{\sqrt{\pi}i\left(
\lambda-1\right)  }\int_{\Im m\left(  y\right)  =\beta_{1}}
\frac{\mathcal{V}\left(  \xi\right)  }{\mathcal{V}\left(  y\right)  }
t^{\frac{2i\left(  \xi-y\right)  }{\lambda-1}}\Gamma\left(  -\frac{2i\left(
\xi-y\right)  }{\lambda-1}\right)  dy
\eean
where $\beta_{1}>\Im m\left(  \xi\right)  .$ Using (\ref{E4a}) it
follows that:
\[
\left|  \widehat{G}_{r}\left(  t,\xi\right)  \right|  \leq C_{R}\,t^{\frac{2\left(
\beta_{1}-\Im m\left(  \xi\right)  \right)  }{\lambda-1}
}\;\;,\;\;\xi\in\mathcal{S}\cap\left\{  \left|  \xi\right|  \leq R\right\}
\]
and therefore $\widehat{G}_{r}$ converges to zero uniformly in bounded sets of
$\xi,$ whence (\ref{E2}) follows.
\qed
\begin{rem}
A heuristic explanation for the formula (\ref{E3}) can be given using Laplace transform.
Suppose that we define the Laplace transform of $\widehat{G}\left(  t,\xi\right)
$ in $t$ as:
\[
\widetilde{G}\left(  z,\xi\right)  =\int_{0}^{\infty}\widehat{G}\left(  t,\xi\right)
e^{-zt}dt
\]

Then, (\ref{E1}), (\ref{E2}) becomes:
\begin{equation}
z\widetilde{G}\left(  z,\xi\right)  =\widetilde{G}\left(  t,\xi+\frac{\lambda-1}
{2}i\right)  \Phi\left(  \xi+\frac{\lambda-1}{2}i\right)  +\frac{1}{\sqrt
{2\pi}}\label{E7a}
\end{equation}

The solution of this equation can be formally reduced to (\ref{E5}) by means
of the transformation:
\begin{equation}
\widetilde{G}\left(  z,\xi\right)  =\exp\left(  -\frac{2i}{\lambda-1}\log\left(
-z\right)  \xi\right)  \mathcal{V}\left(  \xi\right)  H\left(  z,\xi\right)
\label{E8}
\end{equation}

The reason for using $\log\left(  -z\right)  $ instead of $\log\left(
z\right)  $ is that $\widetilde{G}\left(  \cdot,\xi\right)  $ can be expected to
be analytic for $\Re e\left(  z\right)  >a$ for some
$a\in\mathbb{R}$ and in this form the function $\log\left(  -z\right)  $ can
be expected to be analytic in this region assuming that $\arg\left(  z\right)
\in\left(  -\pi,\pi\right)  .$ The transformation (\ref{E8}) brings
(\ref{E7a}) to:
\begin{equation}
H\left(  z,\xi\right)  -H\left(  z,\xi+\frac{\lambda-1}{2}i\right)  =\frac
{1}{\sqrt{2\pi}}\frac{e^{\frac{2i}{\lambda-1}\log\left(  -z\right)  \xi}
}{z\mathcal{V}\left(  \xi\right)  }\label{E9}
\end{equation}

Equation (\ref{E9}) can be solved using conformal mapping and Cauchy's
formula. Using the change of variables:
\begin{equation}
H\left(  z,\xi\right)  =h\left(  z,\zeta\right)  \;\;\;,\;\;\zeta
=e^{\frac{4\pi}{\lambda-1}\left(  \xi-\beta_{0}i\right)  }\;\label{B1}
\end{equation}
and where, for the sake of simplicity we will write, with some slight abuse of
notation
\[
\mathcal{V}\left(  \xi\right)  =\mathcal{V}\left(  \zeta\right)
\]
we obtain:
\begin{equation}
h\left(  z,\zeta+i0\right)  -h\left(  z,\zeta-i0\right)  =\frac{e^{\frac{4\pi
}{\lambda-1}\beta_{0}\alpha\left(  z\right)  }}{\sqrt{2\pi}}\frac
{\zeta^{\alpha\left(  z\right)  }}{z\mathcal{V}\left(  \zeta\right)
}\;\;,\;\;\zeta\in\mathbb{R}^{+}\label{B2}
\end{equation}
with $h$ analytic in $\mathbb{C\setminus R}^{+}$ and:
\[
\alpha\left(  z\right)  =\frac{1}{2\pi i}\log\left(  -z\right)
\]

The solution of (\ref{B2}) can be obtained, assuming that $\frac{\zeta
^{\alpha\left(  z\right)  }}{\mathcal{V}\left(  \zeta\right)  }$ satisfies
suitable boundedness estimates for small and large $\zeta,$ and using Cauchy's
formula:
\[
h\left(  z,\zeta\right)  =\frac{1}{2\pi i}\frac{1}{\sqrt{2\pi}}\frac
{e^{\frac{4\pi}{\lambda-1}\beta_{0}\alpha\left(  z\right)  }}{z}\int
_{0}^{\infty}\frac{s^{\alpha\left(  z\right)  }}{\mathcal{V}\left(  s\right)
}\frac{ds}{\left(  s-\zeta\right)  }
\]
and, using (\ref{B1}):
\begin{equation}
H\left(  z,\xi\right)  =\frac{1}{2\pi i}\frac{1}{\sqrt{2\pi}}\frac{1}{z}
\int_{-\infty}^{\infty}\frac{e^{\frac{4\pi\alpha\left(  z\right)  }{\lambda
-1}y}}{\mathcal{V}\left(  y\right)  }\frac{dy}{1-e^{\frac{4\pi}{\lambda
-1}\left(  \xi-y\right)  }}\label{B3}
\end{equation}
It then follows from (\ref{E8}) that:
\[
\widetilde{G}\left(  z,\xi\right)  =\frac{1}{2\pi i}\frac{1}{\sqrt{2\pi}}\frac
{1}{z}\mathcal{V}\left(  \xi\right)  \int_{-\infty}^{\infty}\frac
{e^{\frac{4\pi\alpha\left(  z\right)  }{\lambda-1}\left(  y-\xi\right)  }
}{\mathcal{V}\left(  y\right)  }\frac{dy}{1-e^{\frac{4\pi}{\lambda-1}\left(
\xi-y\right)  }}
\]
and inverting the Laplace transform we finally obtain (\ref{E3}).
\end{rem}

\section{On the solutions of (\ref{E5}).}
\label{solecinco}
 \setcounter{equation}{0}
\setcounter{theo}{0}

Equation (\ref{E5}) admits infinitely many solutions. Indeed, given any
solution $\mathcal{V}_{part}\left(  \xi\right)  $ we can obtain any other one
by means of:
\[
\mathcal{V}\left(  \xi\right)  =\mathcal{V}_{part}\left(  \xi\right)  P\left(
\xi\right)
\]
where:
\begin{equation}
P\left(  \xi\right)  =P\left(  \xi+\frac{\lambda-1}{2}i\right)  \label{C0a}
\end{equation}
Given such a non uniqueness a natural and essential question  is then how to chose one of them. We may state several sufficient conditions that would ensure that $\widehat G$ is the Fourier transform of a tempered distribution. First we want the function $\widehat G$ to be defined. This is guaranteed by the condition (\ref{E4a}) above. However, this condition is not sufficient to prove that $\widehat G(t, \xi)$ is globally bounded with respect to $\xi$. The difficulty comes from the fact that, if the behaviours of ${\cal V}(\xi)$  
are too disparate as $\Re e(\xi)$ tends to plus or minus infinity, the quotient 
$\frac{{\cal V}(\xi)}{{\cal V}(\xi+Y)}$ may be strongly increasing in some regions of the integral in (\ref{E3}). A sufficient condition to avoid this difficulty is to have:
\bear
\label{S5EassymptV1}
&&\left|{\cal V}(\xi)\right|\approx e^{B_{\pm} |\xi|},\quad
|B_{\pm}|\le \frac{\pi}{(\lambda-1)} 
\eear
as $\Re e (\xi)\to \pm \infty$. The decay rate of the Gamma function in (\ref{E3}) may then control the possible growth of the quotient $\frac{{\cal V}(\xi)}{{\cal V}(\xi+Y)}$ uniformly on $\xi$.

 Another requirement that we need for the function ${\cal V}$ comes from  the requirement that $\widehat G$ must be analytic in the strip $\mathcal S$. This is ensured by imposing also that ${\cal V}$ is also analytic in that same strip.\\
We will then construct a function ${\cal V}$ analytic in that strip, satisfying equation (\ref{E5}) for $\Im m(\xi) \in (3/2, 2)$, satisfying conditions (\ref{E4a}) and (\ref{S5EassymptV1}).

A particular solution of (\ref{E5}) 
can be easily obtained using Cauchy's formula. To this end we take the
logarithm of both sides of (\ref{E5}) to obtain:
\begin{equation}
\log\left(  \mathcal{V}\left(  \xi\right)  \right)  =\log\left(
\mathcal{V}\left(  \xi+\frac{\lambda-1}{2}i\right)  \right)  +\log\left(
-\Phi\left(  \xi+\frac{\lambda-1}{2}i\right)  \right)  \label{C1}
\end{equation}
or equivalently
\begin{equation}
\log\left(  \mathcal{V}\left(  \xi-\frac{\lambda-1}{2}i\right)  \right)  =\log\left(
\mathcal{V}\left(  \xi\right)  \right)  +\log\left(
-\Phi\left(  \xi\right)  \right).  \label{C1bis}
\end{equation}
Let us take any $\beta_1$ such that $\Phi\left(  \xi\right)
$ has no zeros nor poles along the line $\Im m\left(  \xi\right)
=\beta_{1}.$ We define:
\[
\psi\left(  \zeta\right)  =\log\left(  \mathcal{V}\left(  \xi\right)  \right)
\;\;,\;\;\zeta=e^{\frac{4\pi}{\lambda-1}\left(  \xi-\beta_{1}i\right)
}\;\;,\;\;Q\left(  \zeta\right)  =\log\left(  -\Phi\left(  \xi\right)  \right)
\]
Equation (\ref{C1bis}) then becomes
\begin{equation}
\psi\left(  \zeta+i0\right)  =\psi\left(  \zeta-i0\right)  +Q\left(
\zeta-i0\right)  \;\;,\;\;\zeta\in\mathbb{R}^{+}\label{C2}
\end{equation}
with $\psi$ analytic in $\mathbb{C\setminus R}^{+}.$ Taking into account that
$\left|  Q\left(  \zeta\right)  \right|  \leq C\left(  1+\left|  \log\left(
\zeta\right)  \right|  \right)  $ we can obtain a particular solution of
(\ref{C2}) as:
\[
\psi\left(  \zeta\right)  =\frac{1}{2\pi i}\int_{\mathbb{R}^{+}}Q\left(
s\right)  \left[  \frac{1}{s-\zeta}-\frac{1}{s+1}\right]  ds
\]
where the term $1/(s+1)$ has been added to the classical Cauchy integral in order to ensure the convergence of the integral. Then, returning to the variable $\xi$ we obtain:
\begin{equation}
\label{S6Tdefinu}
\mathcal{V}_{part, \beta_1}\left(  \xi\right)  =\exp\left(  \frac{2}{\left(
\lambda-1\right)  i}\int_{\Im m\left(  \eta\right)  =\beta_1
}\log\left(  -\Phi\left(  \eta\right)  \right)
\left[  \frac{1}{1-e^{\frac{4\pi}{\lambda-1}\left(  \xi-\eta\right)  }}
-\frac{1}{1+e^{-\frac{4\pi}{\lambda-1}\eta}}\right]  d\eta\right)  \label{C3}
\end{equation}
where $\Im m\left(  \xi\right)  \in\left(  \beta_{1}-\frac{\lambda-1}{2},\beta_1\right)  .$

Formula (\ref{C3}) provides a particular solution of (\ref{E5}). On the other
hand, we can obtain an infinite family of solutions of (\ref{C0a}) given by:
\begin{equation}
P\left(  \xi\right)  =e^{\frac{4\pi}{\lambda-1}\ell\xi}\;\;,\;\;\ell
\in\mathbb{Z}\label{C4}
\end{equation}
Let us define a family of solutions of (\ref{E5}):
\begin{equation}
\mathcal{V}\left(  \xi\right)  =e^{\frac{4\pi\ell}{\lambda-1}\xi
}\mathcal{V}_{part, \beta_1}\left(  \xi\right)  \label{C5}
\end{equation}

Actually, using Fourier series, it can be seen that any solution of
(\ref{C0a}) can be written as a infinite linear combination of the functions
$\mathcal{V}_{\ell}\left(  \xi\right)  $. 

The formula (\ref{C3}) does not define uniquely the function $\mathcal V_{part, \beta_1}$ unless we prescribe the value of $\beta_1$ and the argument of the function $\ln (-\Phi (\eta))$. The different possible choices of this argument just differ by a factor $2\pi\ell\, i$ and therefore the resulting functions $\mathcal V_{part, \beta_1}$ would differ by a multiplicative factor (\ref{C4}). Proposition \ref{S3T1} implies that $\arg(-\Phi(\eta))\to \pi/4 +2\, \pi\, \ell\, i$ as $\Re e(\eta) \to +\infty$. In order to avoid exponential factors in some of the forthcoming formulas, we determine uniquely the function $\ln (-\Phi (\eta))$ by choosing:
\bear
\label{A2ZZargu}
\lim_{\Re e(\eta)\to +\infty}\arg(-\Phi(\eta))=\frac{\pi}{4}
\eear

Notice that in the formula (\ref{C3}) there exists an infinite possibility of choices of the constant $\beta_1$. These functions may be extended analytically moving $\xi$ and simultaneously the contour of integration in such a way that the condition $-(\lambda-1)/2<\Im m(\xi-\eta)<0$ always holds.
The only true obstruction to extend analytically the functions $\mathcal{V}_{part, \beta_1}$ arises from crossing with the contour deformation the zeros or poles of the function $\Phi$. Suppose that $\xi_{sing}$ is a zero or a pole of $\Phi$ and $\beta_1$, $\beta_2$ are such that:
\bean
\xi_{sing}-\frac{1}{2}<\beta_1<\xi_{sing}<\beta_2<\xi_{sing}+\frac{1}{2}.
\eean
Then:
\bear
\label{S5EZZ1}
\frac{\mathcal{V}_{part, \beta_1}(\xi)}{\mathcal{V}_{part, \beta_2}(\xi)}=-\left(\frac{e^{\frac{4\pi}{\lambda-1}(\xi_{sing}-\xi)}-1}{1+e^{\frac{4\pi}{\lambda-1}\xi_{sing}}}\right)^{-n}
\eear
where:

\begin{equation}
n = \left\{
\begin{array}{l}1\quad \hbox{if}\,\,\,\xi_{sing}\,\,\,\hbox{is a zero}  \\ \\
-1\quad \hbox{if}\,\,\,\xi_{sing}\,\,\,\hbox{is a pole.} 
\end{array} 
\right.
\end{equation}
Combining (\ref{S5EZZ1}) with (\ref{C3}) we can then extend any function $\mathcal{V}_{part, \beta}$ to the whole complex plane as a meromorphic function. As it could be expected the different functions $\mathcal{V}_{part, \beta}$ can be related to each other by means of linear combinations of functions of the form given in (\ref{C5}).

In order to obtain  the function $\mathcal{V}(\xi)$ with the properties requested above, it is  sufficient to take
\bear
\label{S5EZZ2}
\mathcal{V}(\xi)=\mathcal{V}_{part, \beta_1}(\xi),\quad \Im m(\xi)\in \left(\beta_1-\frac{\lambda-1}{2}, \beta_1 \right),
\eear
with
\bear
\label{S5EZZ3}
\quad\beta_1\in \left(\frac{2+\lambda}{2}, \frac{3+\lambda}{2}\right)
\eear
Moving the contour of integration if needed, inside the strip $\Im m (\eta)\in (2+\lambda)/2, (3+\lambda)/2$ we obtain that the function 
$\mathcal V(\xi)$ has no zeros nor poles in the whole strip $\mathcal S$. The function $\mathcal V$ is then defined once and for all along the remainder of the paper.
It only remains to check that this function satisfies the two conditions (\ref{E4a}) and (\ref{S5EassymptV1}).
It follows from Proposition \ref{A1T101} in Appendix  \ref{Some technical propositions.} that:
\bear
\label{S5Tabcd1234}
C_\delta e^{-\frac{1}{2}(\frac{\pi}{\lambda-1}+\delta)|\xi|}\le 
\left|\mathcal{V}\left(  \xi\right)\right| \leq
C_\delta e^{-\frac{1}{2}(\frac{\pi}{\lambda-1}-\delta)|\xi|}\quad \hbox{for}\,\,\,\Im m(\xi)\in \left(\frac{3}{2}, \frac{3+\lambda}{2}\right)
\eear
for $\delta>0$ arbitrarily small and $C_\delta>0$ a constant depending on $\delta$. 
This behaviour implies both (\ref{E4a}) and (\ref{S5EassymptV1}).

Summarizing, we have shown:
\begin{prop}
\label{S5Tpropnu}
The function $\mathcal V (\xi)$ defined by means of (\ref{S6Tdefinu}), (\ref{S5EZZ2}) with $\beta_1$ as in (\ref{S5EZZ3}) can be extended analytically to the strip $\mathcal S$ and meromorphically to the whole complex plane. It satisfies the equation (\ref{E5}) as well as the estimates (\ref{S5Tabcd1234}). Morever, $\mathcal V (\xi)\not =$ in all the strip 
$\mathcal S$.
\end{prop}
\begin{prop}
\label{S5TpropG}
The function $\widehat G(t, \xi)$ defined by means of (\ref{E3})
with $\mathcal V$ defined in Proposition \ref{S5Tpropnu}, solves (\ref{E1}) (\ref{E2}).
Moreover we have the following representation formula:
\bear
\label{S5E2JK42-1516}
&&\frac{{\cal V}(\xi)}{{\cal V}(y)}=
\exp\left[\frac{2}{(\lambda-1)i}
\int_{Im\, \eta  =\beta_1}
\ln\left( -\Phi (\eta)\right)\times\right.  \nonumber \\
&&\hskip 4cm 
\left.
\left({1\over 1-e^{\frac{4\pi}{\lambda-1} (\xi-\eta) }} 
-{1\over 1-e^{\frac{4\pi}{\lambda-1} (y-\eta) }}
\right)d\eta\right] \label{S5E2JK42bis-1}
\eear
for $\xi$ and $y$  such that  $\beta_1-(\lambda-1)/2<\Im m(\xi)< \beta_1$ and $\beta_1-(\lambda-1)/2<\Im m(y)< \beta_1$.
\end{prop}
\section{Decay estimates for the function $\widehat G(t, \xi)$.}
\label{decayghat}
\setcounter{equation}{0}
\setcounter{theo}{0} 
Through all the following of this paper we shall use the following function $A(z)$ that we  define  by means of:
\bear
\label{S5Edefof A}
&&\Gamma (z)=\sqrt{2 \pi}e^{-z}z^{z-1/2}A(z).
\eear
Notice that by the Stirling's formula:
\bear
&&A(z)\to 1\label{S5E2JKLM2} 
\eear
uniformly as $|z|\to  \infty$ and $arg(z)\in (-\pi+\varepsilon_0, \pi-\varepsilon_0)$ for any $\varepsilon_0 >0$ small.
\subsection{The case $0<t<1$.}
\begin{lem}
\label{S5Tregularity10}
The function $\widehat G$ defined in (\ref{E3}) satisfies the following.
\bear
\label{S5Tregularity10E1}
|\widehat G(t, \xi)| & \le & \kappa e^{-a\sqrt{|\xi|}t}\\
\label{S5Tregularity10E2}
\left(1+|\xi|^{1/2}\right)\left|\frac{\partial}{\partial \xi}\widehat G(t, \xi)\right|+
\left(1+|\xi|^{3/2}\right)\left|\frac{\partial^2}{\partial \xi^2}\widehat G(t, \xi)\right|
& \le & \kappa\, t\,  e^{-a\sqrt{|\xi|}t}
\eear
for all $t\in (0, 1)$ and for all $\xi \in \{\xi \in \CC; \Im m(\xi)\in (3/2, (3+\lambda)/2)\}\cup \{\xi\in \CC;\,\, \Im m \xi \in [-L, L]\,, \,\,|\xi|>2\,L \}$
for any  $L>0$  sufficiently large and where $\kappa_1$ and $a$ are positive numerical numbers depending on $L$.
\end{lem}
The proof of this Lemma requires the following result:
\begin{lem}
\label{S5Elemagen}
For any fixed constant $B>0$ consider a function
\bear
\label{S5Ewdoble}
W(t, \xi)=\int_{\Im m(Y)=-\gamma_1}m(\xi, Y)e^{\Psi(\xi, Y, t)}\, dY
\eear
where the function $m(\xi, Y)$ is  analytic with respect to $Y$ in the domain:
\bean
\left|\Re e\left(Y+\frac{1}{8} sign\left(\Re e(\xi) \right)\xi \right)\right|\le B\, |\Im m(Y)|,\,\,
sign (\Re e(Y))=sign (\Re e(\xi))
\eean
and in the strip:
\bean
\Im m(Y)\in \left[-\gamma_1, \gamma_1+\frac{\lambda-1}{2}\right].
\eean
Then, for any $L>0$ sufficiently large, there exists $\xi_0>0$ sufficiently large, and depending on $L$, such that:
\begin{itemize}
\item If 
\bear
\label{S5EcondM1}
|m(\xi, Y)|\le C
\eear then, 
\bear
\label{S5EconcM1}
|W(t, \xi)|\le Ce^{-a|\xi|^{1/2}\, t}
\eear
 for all $t\in [0, 1]$ and all  $\xi$ such that $|\Re e(\xi)|\ge \xi_0$ and $\{\xi \in \CC; \Im m(\xi)\in (3/2, (3+\lambda)/2)\}\cup \{\xi\in \CC;\,\, \Im m \xi \in [-L, L]\,, \,\,|\xi|>2\,L \}$.
\item If 
\bear
\label{S5EcondM2}
|m(\xi, Y)|\le C(1+|Y|)\,\,\, \hbox{and}\quad m(\xi, 0)=0 
\eear
 then, 
\bear
\label{S5EconcM2}
|W(t, \xi)|\le C\,t\, |\xi|^{1/2}\, e^{-a|\xi|^{1/2}\, t}
\eear
 for all $t\in [0, 1]$  and all $\xi $ such that $|\Re e(\xi)|\ge \xi_0$ and $\{\xi \in \CC; \Im m(\xi)\in (3/2, (3+\lambda)/2)\}\cup \{\xi\in \CC;\,\, \Im m \xi \in [-L, L]\,, \,\,|\xi|>2\,L \}$.
\end{itemize}
\end{lem}
\textbf{Proof of Lemma \ref{S5Elemagen}.}
It is convenient to introduce the new variable $Z$ as: $Y=\sqrt{|\xi|}Z$. Then the  function $\Psi$ becomes:
\bear
&&\Psi (\xi, Y, t)=\Phi(\xi, Z, t)=\frac{2}{(\lambda-1)\, i}
\int_{Im\, \eta  =\beta_0+\frac{\lambda-1}{2}-\varepsilon}
\ln\left(-\Phi (\eta)\right)
\Theta(\eta-\xi, \sqrt{|\xi|} Z )d\eta-\nonumber\\
&&-\sqrt{|\xi|}\left(\frac{2 i Z}{\lambda-1}\ln (t)+\frac{2 i Z}{\lambda-1}
-\left(\frac{2 i Z}{\lambda-1}-\frac{1}{2  \sqrt{|\xi|}} \right)\ln \left(\frac{2 i Z}{\lambda-1} \right)-\right.\nonumber\\
&&\left.-\left(\frac{2 i Z}{\lambda-1}-\frac{1}{2 \sqrt{|\xi|}}\right)\ln |\xi|^{1/2}\right)
\label{S5E2JKLM3bis}
\eear
The possibility of extending the function $\Phi(\xi, Z, t)$ as a function of $Z$ to some suitable cones has been studied in Lemma \ref{S5TPhi}. Moreover in Lemma \ref{S5Lcriticalpt} the existence of a critical point $Z_c$ of the function $\Phi(\xi, Z, t)$ with he asymptotics (\ref{S5Ecriticalpt}) has been proved.

Using the analyticity properties of the functions $\Phi(\xi, Z, t)$ and $A(z)$ as well as the change of variables $Y=\sqrt{|\xi|\, Z}$ we can obtain a new representation formula of $\widehat G$ 
\bear
\label{S5E2JKLM4387Z}
W(t, \xi)=\sqrt{|\xi|}
\int_{{\cal C}_t}e^{\Phi (\xi, Z, t)} m(\xi, \sqrt{|\xi|} Z)dZ
\eear
with the new integration contour ${\cal C}_t$ given in Figure 2. Notice that we have moved the portion of the contour $\Im m Z= (\beta_0- \Im \xi)|\xi|^{-1/2}$ where $|\Re e(Z)|$   is large to the line $\Im m(Z)=\gamma_1$ since in that deformation we do not cross any singularity of the integrand. \\ 
We now consider separately the two different cases $|\xi|^2\, t\to +\infty$ and $|\xi|^2\, t$ bounded.
\begin{figure}
\begin{tikzpicture}
\draw[->] (-6.5,1) -- (6,1) node[right]{$\Re e Z$};
\draw[->] (-3,-3) -- (-3,3) node[above]{$\Im m Z$};
\draw[->] (-6.5, 2) -- (-5, 2);
\draw[->] (-5, 2) -- (-3.6, 2);
\draw[->] (-3.6, 2) -- (-3.6, 0);
\draw[->] (-3.6, 0) -- (-3.6, -2);
\draw[->] (-3.6, -2) -- ( -2, -2);
\draw[->] (-2, -2) -- ( 0, -2) node{$\times$};
\draw[->] (0, -2) -- ( 0, -2) node[below]{$Z_c$};
\draw[->] (0,-2) -- (2, -2);
\draw[->] (2, -2) -- (2, 0);
\draw[->] (2, 0) -- (2, 2);
\draw[->] (2,2)-- (4, 2);
\draw[->] (4,2)-- (6.5, 2) node[above]{$\Im m Z=\gamma_1\, t$};
\end{tikzpicture}
\caption{The curve ${\cal C}_t$}
\end{figure}
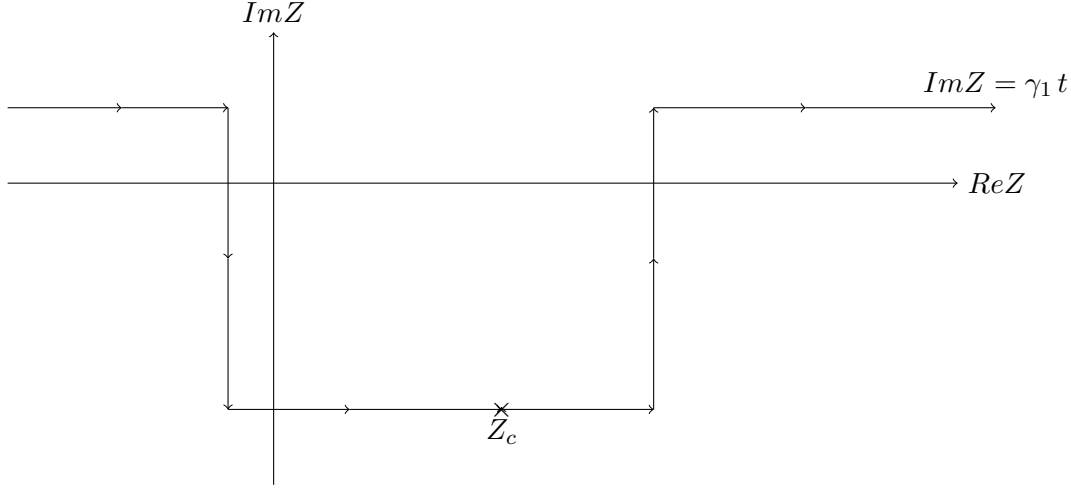
\vskip 0.4cm 
\noindent
\textbf {The estimate of $W(t, \xi)$ as $|\xi|\, t^2\to +\infty$.} We write the function $W(t, \xi)$ as follows:
\bear
\label{S5E2JKLMn27}
&&W(t, \xi)  = {\cal I}_1+{\cal I}_2\\
&&{\cal I}_1=\sqrt{|\xi|}
\int_{{\cal C}_t\cap\{Z;\, |Z-Z_c|\le \delta_0\, t\}}e^{\Phi (\xi, Z, t)} m(\xi, \sqrt{|\xi|} Z)dZ\label{S5E2JKLMn2746529}\\
&&{\cal I}_2 =\sqrt{|\xi|}
\int_{{\cal C}_t\cap\{Z;\, |Z-Z_c|\ge \delta_0\, t\}}e^{\Phi (\xi, Z, t)} m(\xi, \sqrt{|\xi|} Z)dZ
\label{S5E2JKLMn2746530}
\eear
for some positive constant $\delta_0$ to be fixed.

We estimate first the integral ${\cal I}_1$ using Lemmas \ref{S5TPhi}--\ref{S5Ttercerder} and Taylor's expansion we obtain first:
\bear
\label{S5Ekandl60k}
\Phi (\xi,  Z, t)= \Phi (\xi,  Z_c, t)+\frac{(Z-Z_c)^2}{2}\frac{\partial^2 \Phi}{\partial Z^2}(\xi, Z_c\, t)+
{\cal O}\left(\frac{\sqrt{|\xi|}}{t^2}(Z-Z_c)^3\right)
\eear
for $|Z-Z_c|\le \varepsilon_0 t$, and then
\bear
\label{S5Ekandl61k}
\Phi (\xi,  Z, t)= \Phi (\xi,  Z_c, t)-\frac{1}{2}\frac{\sqrt{|\xi|}}{\sqrt{2\pi}\, t\, (1+i\, Q)}\left(1+\delta(\xi, Z, t) \right)|Z-Z_c|^2,
\eear
where $|\delta(\xi, Z, t)|$ can be made arbitrarily small if $|\xi|t^2$ is large and $\varepsilon_0$ sufficiently small. Therefore:

\bean
{\cal I}_1= \sqrt {|\xi|}
e^{\Phi (\xi, Z_c, t)}
\int_{{\cal C}_t\cap\{Z;\, |Z-Z_c|\le \delta_0\, t\}}\hskip -2.5cm e^{-\frac{1}{2}\frac{\sqrt{|\xi|}}{\sqrt{2\pi}\, t\, (1+i\, Q)}\left(1+\delta(\xi, Z, t) \right)|Z-Z_c|^2} m(\xi, \sqrt{|\xi|} Z)dZ
\eean
which gives, in the case (\ref{S5EcondM1}) :
\bear
\label{S5EcorolcondM1}
\left|{\cal I}_1(t, \xi)\right|\le C\, 
e^{-a \sqrt{|\xi|}\,  t}\qquad\hbox{as}\,\,\,|\xi|\, t^2\to +\infty,\,\,\,0<t<1,
\eear
and in the case (\ref{S5EcondM2}):
\bear
\label{S5EcorolcondM2}
\left|{\cal I}_1(t, \xi)\right|\le C\, 
t\sqrt{|\xi|}\, e^{-a \sqrt{|\xi|}\,  t}\qquad\hbox{as}\,\,\,|\xi|\, t^2\to +\infty,\,\,\,0<t<1.
\eear

We must now estimate the integral ${\cal  I}_2$ given by (\ref{S5E2JKLMn2746530}).
\bean
{\cal I}_2 =\sqrt{|\xi|}
\int_{{\cal C}_t\cap\{Z;\, |Z-Z_c|\ge \delta_0\, t\}}e^{\Phi (\xi, Z, t)} m(\xi, \sqrt{|\xi|} Z)dZ
\eean
In order to estimate the integral ${\cal I}_2$ we use Lemma \ref{A10estimderivada}. To this end we split the integral as follows:
\bean
&&{\cal I}_2={\cal I}_{2,1}+{\cal I}_{2,2}\\
&&{\cal I}_{2,1}=\sqrt{|\xi|}
\int_{{\cal C}_t\cap\{Z;\, |Z-Z_c|\ge \delta_0\, t\}\cap \gamma(M, t)}e^{\Phi (\xi, Z, t)}m(\xi, \sqrt{|\xi|} Z)dZ\\
&&{\cal I}_{2,2}=\sqrt{|\xi|}
\int_{{\cal C}_t\setminus  \gamma(M)\cap \{Z;\,\,|Z|\le \varepsilon_1\sqrt{|\xi|}\}}e^{\Phi (\xi, Z, t)}m(\xi, \sqrt{|\xi|} Z)dZ.\\
&&{\cal I}_{2,3}=\sqrt{|\xi|}
\int_{{\cal C}_t\setminus  \gamma(M)\cap \{Z;\,\,|Z|\ge \varepsilon_1\sqrt{|\xi|}\}}e^{\Phi (\xi, Z, t)}m(\xi, \sqrt{|\xi|} Z)dZ.
\eean
where $\gamma(M, t)$ is the portion of the curve ${\mathcal C}_t$ along which $\Im m(Z)< \gamma_1\, t$. Using Lemma \ref{A10estimderivada} it follows that, in the case (\ref{S5EcondM1}):
\bear
\label{S5Ecorol2condM1}
\left|{\cal I}_{2,1}\right| & \le & C e^{\Phi (\xi, Z_c, t)}e^{-\varepsilon_0\sqrt{|\xi|}\, t}\sqrt{|\xi|}
\int_{{\cal C}_t\cap\{Z;\, |Z-Z_c| \ge  \delta_0\, t\}\cap \gamma(M, t)} m(\xi, \sqrt{|\xi|} Z)dZ \nonumber\\
& \le &  C e^{\Phi (\xi, Z_c, t)}e^{-\varepsilon_0\sqrt{|\xi|}\, t}\sqrt{|\xi|}\, t  \le  C e^{\Phi (\xi, Z_c, t)}e^{-\varepsilon_0/2\sqrt{|\xi|}\, t},
\eear
and in the case (\ref{S5EcondM2}) :
\bear
\label{S5Ecorol2condM2}
\left|{\cal I}_{2,1}\right| & \le &   C e^{\Phi (\xi, Z_c, t)}e^{-\varepsilon_0\sqrt{|\xi|}\, t}(\sqrt{|\xi|}\, t)^2  \le  C (\sqrt{|\xi|}\, t)\,e^{\Phi (\xi, Z_c, t)}e^{-\varepsilon_0/2\sqrt{|\xi|}\, t}.
\eear
The estimate of ${\cal I}_{2,2}$ follows using Lemma \ref{A10estimderivadamas}. In the case (\ref{S5EcondM1}) we obtain:
\bear
\label{S5Ecorol3condM1}
\left|{\cal I}_{2,2}\right|\le C\, \sqrt{|\xi|}\int_{|Z|\ge M\, t}e^{-a\sqrt{|\xi|} |Z|}\, dZ= Ce^{-a\, M\sqrt{|\xi|}t}.
\eear
and in case (\ref{S5EcondM2}) we have:
\bear
\label{S5Ecorol3condM2}
\left|{\cal I}_{2,2}\right|\le C\, \sqrt{|\xi|}
\int_{|Z|\ge M\, t}e^{-a\sqrt{|\xi|} |Z|}\, \sqrt{|\xi|}\, Z\, dZ\le C(\sqrt{|\xi|}\, t)e^{-a\, M\sqrt{|\xi|}t},
\eear
since $|\xi|^2\, t$ is large.

The third integral ${\cal I}_{2,3}$ is estimated using Proposition \ref{A1T101}. To this end we use the variable $Y$ and the identity:
\bean
e^{\Psi(\xi, Z, t)}=t^{-\frac{2\, i\, Y}{\lambda-1}}\frac{\mathcal V(\xi)}{\mathcal V (\xi+Y)}\frac{\Gamma \left( \frac{2\, i\, Y}{\lambda-1}\right)}{A \left( \frac{2\, i\, Y}{\lambda-1}\right)}
\eean
where $A$ is defined by formula (\ref{S5Edefof A}).
This  reduces the estimate of ${\cal I}_{2,3}$ to the estimate of the integral:
\bear
\label{S5Ejota}
J=\int_{\Im m(Y)=\gamma_1, |Y|\ge \varepsilon_1 |\xi|}\left|\frac{m(\xi, Y)}{A \left( \frac{2\, i\, Y}{\lambda-1}\right)} \right|
\left|{{\cal V}(\xi)\over {\cal V}(Y+\xi)}\right|
\left|t^{-\frac{2 iY}{\lambda-1}}\right|\, \left|\Gamma\left(\frac{2i Y}{\lambda-1} \right)\right|dY.
\eear
Using that $\gamma_1>0$, $0\le t \le 1$ and Stirling's formula,  it follows  that, in both cases (\ref{S5EcondM1}) and (\ref{S5EcondM2}):
\bean
J\le \int_{\Im m(Y)=\gamma_1, |Y|\ge \varepsilon_1 |\xi|}\left( 1+|Y|\right)
\left|{{\cal V}(\xi)\over {\cal V}(Y+\xi)}\right|\, e^{-\frac{\pi}{\lambda-1}|Y|}\, dY.
\eean
Proposition \ref{A1T101} gives the following bounds 
\bear
J & \le &  C\,\int_{\Im m(Y)=\gamma_1, |Y|\ge \varepsilon_1 |\xi|}\left( 1+|Y|\right) e^{\delta\, \pi\, |\xi|}\, e^{-\frac{\pi}{2(\lambda-1)}|Y|}dY \nonumber \\
& \le & Ce^{-\frac{\pi}{2(\lambda-1)}\varepsilon_1|\xi|}e^{\delta\, \pi\, |\xi|}.
\eear
with $\varepsilon_0>0$ if $\delta$ is sufficiently small. If we are in case (\ref{S5EcondM2}), we estimate:
\bean
e^{-\varepsilon_0|\xi|}\le e^{-\frac{\varepsilon_0}{2}|\xi|}e^{-\frac{\varepsilon_0}{2\, t^2}}\le 
e^{-\frac{\varepsilon_0}{2}\sqrt{|\xi|}\, t}e^{-\frac{\varepsilon_0}{2\, t^2}}
\eean
which is estimated by the right hand side of (\ref{S5EconcM2}).

This ends the proof of the estimate of $\mathcal I _2$ and then of the Lemma 
\ref{S5Tregularity10} in the domain where $|\xi|\, t^2$ large and $0<t<1$.

\textbf{The estimate of $W(t, \xi)$ for $|\xi|\, t^2$ bounded.} We suppose first that condition (\ref{S5EcondM1}) holds.
We split the integral in (\ref{S5E2JK43}) in two pieces:
\bear
&&W(t, \xi)=J_1+J_2 \label{S5EsplitW}\\
&&J_1(t, \xi)= 
\int_{Im\, Y =-\gamma_1\, t,\,\,|Y|\le \varepsilon_0|\xi|}\frac{m(\xi, Y)}{A \left( \frac{2\, i\, Y}{\lambda-1}\right)} 
{{\cal V}(\xi)\over {\cal V}(Y+\xi)}
t^{-\frac{2 iY}{\lambda-1}}\, \Gamma\left(\frac{2i Y}{\lambda-1} \right)dY.\nonumber\\
&&J_2(t, \xi)= \frac{\sqrt 2}{\sqrt \pi  i\,(\lambda-1)}
\int_{Im\, Y =-\gamma_1\, t,\,\,|Y|\ge \varepsilon_0|\xi|}\frac{m(\xi, Y)}{A \left( \frac{2\, i\, Y}{\lambda-1}\right)} 
{{\cal V}(\xi)\over {\cal V}(Y+\xi)}
t^{-\frac{2 iY}{\lambda-1}}\, \Gamma\left(\frac{2i Y}{\lambda-1} \right)dY\nonumber
\eear
The integral $J_2$ is estimated with the same argument used to bound the integral $J$ in (\ref{S5Ejota}). 
\bear
\label{S5E18640T}
|J_2|\le Ce^{-\varepsilon_0|\xi|}
\eear

We rewrite $J_1$ as:
\bean
J_1(t, \xi) & = & \sqrt{|\xi|}\, t
\int_{\Im m(\zeta)=-\frac{\gamma_1}{ \sqrt{|\xi|}}, |\zeta|\le \varepsilon_0 \frac{\sqrt{|\xi|}}{t}}e^{\Phi (\xi, \zeta\, t, t)} m(\xi, \sqrt{|\xi|}\, \zeta\, t)d\zeta
\eean
where
\bear
\Phi(\xi, \zeta t, t) &= &
-\sqrt{|\xi|}\, t\frac{2 i \zeta}{\lambda-1}\left[1-\ln \left(\frac{2 i \zeta }{\lambda-1} \right)
+\ln\left(2\sqrt{\pi}e^{iQ\frac{\pi}{4}} \right)\right]-\nonumber \\
&&-\frac{1}{2}\ln \left(t\, |\xi|^{1/2} \right)-\frac{1}{2}\ln \left(\frac{2\, i\, \zeta}{\lambda-1} \right)+h(\xi, \zeta t, t).
\eear
Notice that Lemma \ref{S5TPhidos} implies 
\bean
|h(\xi, \zeta t, t)|\le C\, \left(|\zeta|^2\, t^2+{\cal O}\left(\frac{1}{|\xi|} \right)\right)
\eean
for $t\, |\zeta|\le \varepsilon_0\sqrt{|\xi|}$, $\zeta t\in D(\xi, B)$, $|\xi|\ge \xi_0$ and  for some positive constants $C=C(B)$ and $\xi_0=\xi_0(B)$ sufficiently large. Notice that
\bean
\Re e (\Phi(\xi, \zeta t, t)) & \le & \sqrt{|\xi|}\, t\left(C-\frac{\pi}{\lambda-1}|\zeta|\right)-\frac{1}{2}\ln \left(t\, |\xi|^{1/2} \right)+
C\varepsilon_0\, \sqrt{|\xi|}\, t\, |\zeta|\\
& \le & \sqrt{|\xi|}\, t\left(C-\frac{\pi}{2(\lambda-1)}|\zeta|\right)-\frac{1}{2}\ln \left(t\, |\xi|^{1/2} \right)
\eean 
for some positive constant $C$, for $\zeta$ such that $\Im m(\zeta)=-\frac{\gamma_1}{ \sqrt{|\xi|}}$ and $|\zeta|\le \varepsilon_0 \frac{\sqrt{|\xi|}}{t}$, assuming that $\varepsilon_0$ is small. Using that $t\sqrt{|\xi|}$ is bounded it follows that, in the case (\ref{S5EcondM1}):
\bean
|J_1(t, \xi)|\le C \left(\sqrt{|\xi|}\,t\right)^{1/2} \int_{\Im m(\zeta)=-\frac{\gamma_1}{ \sqrt{|\xi|}}, |\zeta|\le \varepsilon_0 \frac{\sqrt{|\xi|}}{t}}
e^{-\frac{\pi}{\lambda-1}\sqrt{|\xi|}\, t|\zeta|}\frac{d|\zeta|}{\sqrt{|\zeta|}},
\eean
and computing the integral:
\bean
|J_1(t, \xi)|\le C.
\eean
Combining this estimate with (\ref{S5E18640T}),  estimate (\ref{S5EconcM1}) of the Lemma follows.\\
In the case (\ref{S5EcondM2}) we deform the contour integration in formula (\ref{S5Ewdoble})  using the analyticity properties of the function $m(\xi, Y)$ as well as the fact that $m(\xi, 0)=0$ to obtain:
\bear
\label{S5Ewdobleconpolo}
W(t, \xi)= - \frac{\pi\,(\lambda-1)}{  A(-1)}\, m\left(\xi, \frac{\lambda-1}{2}\, i\right) {{\cal V}(\xi)\over {\cal V}(\frac{\lambda-1}{2}\, i+\xi)}t\nonumber\\+
\int_{\Im m(Y)=\gamma_1+\frac{\lambda-1}{2}}m(\xi, Y)e^{\Psi(\xi, Y, t)}\, dY\nonumber \\
= \frac{\pi(\lambda-1)}{  A(-1)}\, m\left(\xi, \frac{\lambda-1}{2}\, i\right) \Phi\left(\frac{\lambda-1}{2}\, i+\xi\right)t\nonumber\\+
\int_{\Im m(Y)=\gamma_1+\frac{\lambda-1}{2}}m(\xi, Y)e^{\Psi(\xi, Y, t)}\, dY
= K_1+K_2.
\eear
using (\ref{E5}). By  (\ref{S5EcondM2}), $m(\xi, \frac{\lambda-1}{2}\, i)$ is uniformly bounded. Using then Proposition \ref{S3T1} we obtain that:
\bear
\label{S5Ewdobleconpolocota}
|K_1|\le C(1+|\xi|^{1/2})t.
\eear
We are then left with the integral $K_2$ which is formally very similar to (\ref{S5Ewdoble}) although the integral contour is different. 
We split this integral in two terms $J_1$ and $J_2$ like in (\ref{S5EsplitW}). The term $J_2$ is bounded as (\ref{S5E18640T}) by a similar argument as before with one small difference which that the term $m(\xi, Y)$ is now bounded as $(1+|Y|)$. The term $J_1$ can be written  as:
\bean
J_1(t, \xi)= 
\int_{Im\, Y =\gamma_1+\frac{\lambda-1}{2},\,\,|Y|\le \varepsilon_0|\xi|}\frac{m(\xi, Y)}{A \left( \frac{2\, i\, Y}{\lambda-1}\right)} 
{{\cal V}(\xi)\over {\cal V}(Y+\xi)}
t^{-\frac{2 iY}{\lambda-1}}\, \Gamma\left(\frac{2i Y}{\lambda-1} \right)dY.
\eean
Using Lemma \ref{S5TPhi}, we may estimate $|{{\cal V}(\xi)\over {\cal V}(Y+\xi)}|$ by $Ce^{\left(\frac{\pi}{2(\lambda-1)}+\delta_0 \right)|Y|}$ in the domain of integration. On the other hand, since $Im\, Y =\gamma_1+\frac{\lambda-1}{2}$ we may estimate
$t^{-\frac{2 iY}{\lambda-1}}$ by $t^{1+\frac{2\gamma_1}{\lambda-1}}$. Therefore, using the decay of the Gamma function:

\bear
\label{S5Ewdobleconpolocotacinco}
|J_1(t, \xi)|& \le & Ct^{1+\frac{2\gamma_1}{\lambda-1}}
\int_{\Im m(Y)=\gamma_1+\frac{\lambda-1}{2}}
e^{\left(\frac{\pi}{2(\lambda-1)}+\delta_0 \right)|Y|}\, e^{-\frac{\pi}{\lambda-1}|Y|}d|Y|\\
& \le &  Ct^{1+\frac{2\gamma_1}{\lambda-1}}.
\eear
Combining (\ref{S5Ewdobleconpolocota}) and (\ref {S5Ewdobleconpolocotacinco}) we deduce the estimate (\ref{S5EconcM2}) when
 $|\xi|\, t^2$ is bounded. This concludes the proof of the Lemma \ref{S5Elemagen}.\qed
\\

\noindent
\textbf{Proof of Lemma \ref{S5Tregularity10}.} 
Using the change of variables: $y-\xi=Y$ in (\ref{E3}) we obtain
\bear
\label{S5E2JK43}
\widehat G(t, \xi)=-\frac{\sqrt 2}{\sqrt \pi  i\,(\lambda-1)}
\int_{Im\, Y =-\gamma_1}
{{\cal V}(\xi)\over {\cal V}(Y+\xi)}
t^{-\frac{2 iY}{\lambda-1}}\, \Gamma\left(\frac{2i Y}{\lambda-1} \right)dY.
\eear
where $\gamma_1$ is a positive constant sufficiently small. 
We rewrite the function $\widehat G(t, \xi)$ as follows.
\bear
\label{S5E2JKLM4387}
\widehat G(t, \xi)=\frac{\sqrt 2}{\sqrt \pi  i\,(\lambda-1)}
\int_{\Im m(Y)=-\gamma_1 }e^{\Psi (\xi, Y, t)} A\left(\frac{2 i Y}{\lambda-1} \right)dY
\eear
 where
\bear
&&\Psi(\xi, Y, t)=\frac{2}{(\lambda-1)\, i}
\int_{Im\, \eta  =\beta_1}
\ln\left(-\Phi (\eta)\right)
\Theta(\eta-\xi, Y)d\eta- \nonumber\\
&&-\frac{2 i Y}{\lambda-1}\ln (t)-\frac{2 i Y}{\lambda-1}
+\left(\frac{2 i Y}{\lambda-1}-\frac{1}{2} \right)\ln \left(\frac{2 i Y}{\lambda-1} \right)
\label{S5E2JKLM3} \\
&&\Theta(\sigma, Y)=\frac{1}{1-e^{-\frac{4\pi}{\lambda-1} \sigma}} 
-\frac{1}{1-e^{\frac{4\pi}{\lambda-1} (-\sigma+Y) }}.\label{S5E2JKLM4}
\eear
 \\ \\
In order to obtain estimates (\ref{S5Tregularity10E1}) and (\ref{S5Tregularity10E2}) for bounded values of $\xi$ we use contour deformation. In particular, crossing the pole at $Y=0$ in integral (\ref{S5E2JKLM4387}) and using residue's Theorem we obtain:
\bear
\label{S5E2JK43residue}
&&\widehat G(t, \xi)=\frac{1}{\sqrt{2\pi}}+\widehat G_1(t, \xi)\\
&&\widehat G_1(t, \xi)=-\frac{\sqrt 2}{\sqrt \pi  i\,(\lambda-1)}
\int_{Im\, Y =\gamma_1}
{{\cal V}(\xi)\over {\cal V}(Y+\xi)}
t^{-\frac{2 iY}{\lambda-1}}\, \Gamma\left(\frac{2i Y}{\lambda-1} \right)dY.
\label{S5E2JK43tresillo}
\eear
with $\gamma_1>0$ sufficiently small. Using Proposition \ref{A1T101} it follows that 
\bean
|\widehat G_1(t, \xi)|\le C t^{\frac{2 \gamma_1}{\lambda-1}}
\eean
uniformly for $\xi$ in bounded sets. This yields estimate (\ref{S5Tregularity10E1}) for $\xi$ in bounded sets. If we differentiate in (\ref{S5E2JK43tresillo}) with respect to $\xi$ we obtain:
\bear
\label{S5Tregularity10E5}
&&\frac{\partial^{\ell} }{\partial \xi^{\ell}}\widehat G(t, \xi)=
-\frac{\sqrt 2}{\sqrt \pi  i\,(\lambda-1)}
\int_{Im\, Y =\gamma_1}
m_\ell(\xi, Y)
t^{-\frac{2 iY}{\lambda-1}}\, \Gamma\left(\frac{2i Y}{\lambda-1} \right)dY\\
\label{S5Tregularity10E6}
&&m_\ell(\xi, Y)=\frac{\partial ^{\ell} }{\partial \xi^{\ell}}\left({{\cal V}(\xi)\over {\cal V}(Y+\xi)}\right)
\eear
Using the analyticity properties of the functions $m_\ell(\xi, Y)$ we deform the integration contour in the integral
(\ref{S5Tregularity10E5}). The first singularity that is met is the pole of function $\Gamma\left(\frac{2i Y}{\lambda-1} \right)$ located at $Y=(\lambda-1)i/2$. (This point is below the first zero of the function $\mathcal V (\xi+Y)$ which is located at $Y=(2+(\lambda/2)-\xi)\, i$ and $2+(\lambda/2)-\xi>(\lambda-1)/2$.) We  deduce
\bear
\label{S5Tregularity10E7}
&&\frac{\partial^{\ell} }{\partial \xi^{\ell}}\widehat G(t, \xi)=
\frac{2\, \sqrt {2\, \pi}}{\sqrt \pi  i\,(\lambda-1)}t\, m_\ell(\xi, \frac{\lambda-1}{2}\, i)\\
&&-\frac{\sqrt 2}{\sqrt \pi  i\,(\lambda-1)}
\int_{Im\, Y =\gamma_1+\frac{\lambda-1}{2}}
m_\ell(\xi, Y)
t^{-\frac{2 iY}{\lambda-1}}\, \Gamma\left(\frac{2i Y}{\lambda-1} \right)dY.
\eear
using the analyticity properties of the functions $m_\ell(\xi, Y)$ we deduce estimate (\ref{S5Tregularity10E2}) for bounded values of $\xi$.

In order to complete the proof of Lemma \ref{S5Tregularity10} we apply the method of the stationary phase. To this  end we differentiate the expression in (\ref{S5E2JKLM4387}) with respect to $\xi$, obtain:
\bear
\label{S5E2JKLM4387der2}
\frac{\partial^\ell}{\partial \xi^\ell}\widehat G(t, \xi)=\frac{\sqrt 2}{\sqrt \pi  i\,(\lambda-1)}
\int_{Im\, Y =-\gamma_1}
\left[\frac{\partial^\ell \Psi}{\partial \xi^\ell}(\xi, Y, t)\, A\left(\frac{2 i Y}{\lambda-1} \right)\right]\, e^{\Psi (\xi, Y, t)}dY.\eear
for $\ell =1, 2$ and use Lemma \ref{S5Elemagen} with three choices for $m(t, \xi)$: 
\bear
\label{S5Tchoiceone}
&&m(\xi, Y)= \frac{\sqrt 2}{\sqrt \pi  i\,(\lambda-1)}\, A\left( \frac{2\, i\, Y}{\lambda-1}\right)\\
\label{S5Tchoicethree}
&&m(\xi, Y)= \frac{\sqrt 2}{\sqrt \pi  i\,(\lambda-1)}\, (|\xi|+1)^\ell\, \frac{\partial^\ell \psi}{\partial \xi^\ell}(\xi, Y, t)\, A\left( \frac{2\, i\, Y}{\lambda-1}\right).
\eear
for $\ell =1, 2$. (Notice that $\frac{\partial \psi}{\partial \xi}(\xi, Y, t)$ and $\frac{\partial^2 \psi}{\partial \xi^2}(\xi, Y, t)$ are independent of $t$). 

The function $m$ in (\ref{S5Tchoiceone}) satisfies condition (\ref{S5EcondM1}). On the other hand, using Lemma \ref{A11Tderivadapsi}, it follows that for the choices (\ref{S5Tchoicethree}), $\ell=1, 2$:
$
|m(\xi, Y)|\le C |Y|.
$
Moreover, by the definition of the function $\Psi$ (cf. (\ref{S5E2JKLM3}))
\bean
\frac{\partial \psi}{\partial \xi}(\xi, Y) 
 & = & \frac{\partial}{\partial \xi}\left(\ln \left(\frac{\mathcal V(\xi)}{\mathcal V(\xi+Y)} \right) \right)
\eean
and therefore $m(\xi, 0)=0$ with the choice (\ref{S5Tchoicethree}) and $\ell=1$. This follows by a similar argument for the choice (\ref{S5Tchoicethree}), $\ell=2$. Applying then Lemma \ref{S5Elemagen} the estimates (\ref{S5Tregularity10E1}) and (\ref{S5Tregularity10E1}) follow for $|\xi|$ sufficiently large and this concludes the proof of Lemma \ref{S5Tregularity10}.
\qed

\subsection{The case $t>1$.}
\begin{lem}
\label{S5Tregularity11}
The function $\widehat G$ defined in (\ref{E3}) satisfies that for any $\varepsilon_0>0$ arbitrarily small, there exists two positive constants $\kappa_1$ and $a$ such that
\bear
\label{S7Elemaseistres}
|\widehat G(t, \xi)|\le \kappa_1\, t^{-\frac{1}{\lambda-1}+\varepsilon_0} e^{-a\sqrt{|\xi|}}
\eear
for all $\xi$ such that $\Im m(\xi)\in (3/2, (3+\lambda)/2)$ and all $t>1$.
\end{lem}
\textbf{Proof.} We deform the integration contour in the expression (\ref{S5E2JK43}) to the line: $\Im m(Y)=(\lambda-1)/2-\varepsilon_0$ with $\varepsilon_0$ arbitrarily small in order to avoid the zero of the function $\mathcal V (\xi+Y)$ at $\xi+Y=1$. Then we perform the change of variables $Y=\sqrt{|\xi|}\, Z$, and perform a new deformation of the contour of integration in the $Z$ variable to the curve $\mathcal D$ in Figure 3.
\begin{figure}
\label{curveD}
\begin{tikzpicture}
\draw[->] (-6.5,1) -- (6,1) node[right]{$\Re e Z$};
\draw[->] (-3,-3) -- (-3,3) node[above]{$\Im m Z$};
\draw[->] (-6.5, 2) -- (-5, 2);
\draw[->] (-5, 2) -- (-3.6, 2);
\draw[->] (-3.6, 2) -- (-3.6, 0);
\draw[->] (-3.6, 0) -- (-3.6, -2);
\draw[->] (-3.6, -2) -- ( -2, -2);
\draw[->] (-2, -2) -- ( 0, -2) node{$\times$};
\draw[->] (0, -2) -- ( 0, -2) node[below]{$Z_c$};
\draw[->] (0,-2) -- (2, -2);
\draw[->] (2, -2) -- (2, 0);
\draw[->] (2, 0) -- (2, 2);
\draw[->] (2,2)-- (4, 2);
\draw[->] (4,2)-- (6.5, 2) node[above]{$\Im m Z=\gamma_1$};
\end{tikzpicture}
\caption{The curve ${\cal D}$}
\end{figure}
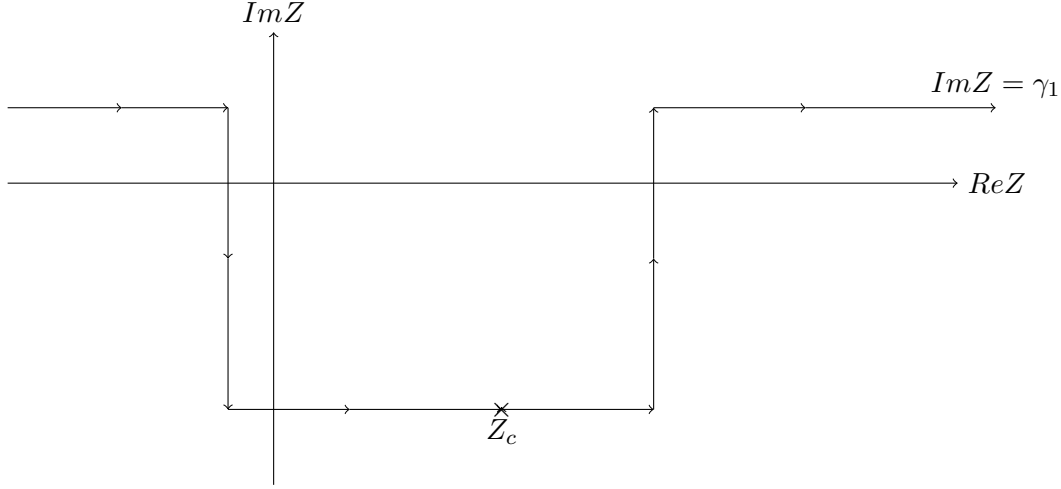
 
\noindent to obtain:
\bear
\label{S5E2JKLM4387Z3}
\widehat G(t, Z)=\frac{\sqrt 2 \sqrt{|\xi|}}{\sqrt \pi  i\,(\lambda-1)}
\int_{{\cal D}}t^{-\frac{2\, i}{\lambda-1} \sqrt{|\xi|}\, Z}e^{\Phi (\xi, Z, 1)} A\left(\frac{2 i \sqrt{ |\xi|} Z}{\lambda-1} \right)dZ.
\eear
Then,
\bean
|\widehat G(t, Z)|\le C\sqrt{|\xi|}\, t^{-\frac{1}{\lambda-1}+\varepsilon_0}
\int_{{\cal D}}\left|e^{\Phi (\xi, Z, 1)} A\left(\frac{2 i \sqrt{ |\xi|} Z}{\lambda-1} \right)\right|dZ.
\eean
We argue now in the same way as in the previous case with $t=1$ splitting the integral in the same pieces to obtain the estimate of the Lemma.\qed

\section{Estimates on the function $G(t, X)$.}
\label{decayG}
\setcounter{equation}{0}
\setcounter{theo}{0}  
We may now take the inverse Fourier transform of $\widehat G(t, \xi)$ to  obtain the function:
\bear
\label{S5E103}
G(t, X)=\frac{1}{\sqrt{2 \pi}}\int_{\Im m(\xi)=\beta_0}e^{i X \xi} \widehat G(t, \xi) d\xi
\eear
where $\beta_0$ is in the interval $(3/2, (3+\lambda)/2)$.
\begin{prop}
For all $t>0$ the function $G(t, \cdot)$ defined by (\ref{S5E103}) belongs to  $C^\infty(\RR)$ and,  for $\ell =0, 1, 2, \cdots$. Moreover, for any fixed $R>0$, it  satisfies:
\bear
\label{S7Ereguarity}
\left|\frac{\partial ^{\ell} G}{\partial X^{\ell}}(t, X)\right|\le \frac{C_{\ell, R}}{t^{2(1+\ell)}},\quad \hbox{for}\,\,\,0<t\le 1,\,\, |X|\le R.
\eear
and, for any $\varepsilon_0$ arbitrarily small and $R>0$, there exists a positive constant $C_{\ell, \varepsilon_0, R}$ such that
\bear
\label{S7Ereguaritydos}
\left|\frac{\partial ^{\ell} G}{\partial X^{\ell}}(t, X)\right|\le \frac{C_{\ell, \varepsilon_0, R}}{t^{\frac{1}{\lambda-1}-\varepsilon_0}},\quad \hbox{for}\,\,\,t\ge 1,\,\, |X|\le R
\eear

\end{prop}
\textbf{Proof.} This Proposition is just a consequence of (\ref{S5E103}) as well as from (\ref{S5Tregularity10E1}) for $t\in (0, 1)$ and (\ref{S7Elemaseistres}) for $t>1$.\qed

The estimates (\ref{S7Ereguarity}) and (\ref{S7Ereguaritydos}) do not give any detailed description of the function $G$. We now proceed to obtain detailed descriptions of the function $G$ on the different regions in $X$ and $t$.

\subsection{Behaviour as $t\to 0$, $0\le |X|\le 1$.}

We begin by showing that the function $G$ behaves like a mollifier of the Dirac measure when $t\to 0$ and for $X$ small. The asymptotic profile of the solution $G(t, X)$ can be heuristically guessed as follows.  If we assume that the function $g$, solution of (\ref{e2.4bis}) with $g(0, x)\sim \delta(x-1)$ is small, then equation (\ref{e2.4bis}) may be approximated, for $t$ and $x-1$ small, as
\bear
\label{S7Eheuristic1}
\frac{\partial g}{\partial t}= \int_0^{1/2}\frac{g(x-y)-g(x)}{y^{3/2}}\, dy, 
\qquad g(x, 0)=\delta(x-1) 
\eear
The equation in (\ref{S7Eheuristic1})  describes the probability distribution  for a system of particles whose size is initially equal to one and have a probability per unit of time  $\chi_{0\le y \le 1/2}\,\frac{dy}{y^{3/2}}$ of coalescing with a particle of size in the interval $(y, y+dy)$. If we write the function $g$ as:
\bean
g(t, x)=\frac{1}{\sqrt{2\pi}}\int_{\RR}\widehat g(t, x)\, e^{i\, k\, x}\, dk
\eean
Applying Fourier transform we obtain:
\bean
&&\frac{\partial \widehat g}{\partial t}(t, k)=m(k)\, \widehat g(t, k),\qquad \widehat g(k, 0)=\frac{e^{-i\, k}}{\sqrt {2\pi}}\\
&&m(k)=\int_0^{1/2}\frac{e^{-i\, k\, y}-1}{y^{3/2}}\, dy.
\eean
The multiplier $m(k)$ can be computed explicitly but its exact formula is not  needed to compute the asymptotics of the function $g(t, x)$. The only relevant information that we really need is
\bean
m(k)\approx -2\sqrt{\pi\, k\, i}, \quad \hbox{for}\,\,\,|k|\to +\infty.
\eean
Then,
\bean
\widehat g(t, k)\approx  \frac{e^{-i\, k}}{\sqrt {2\pi}}\, e^{-2\sqrt{\pi\, k\, i}\, t}, \quad \hbox{for}\,\,\,|k|\to +\infty
\eean
and inverting the Fourier transform we obtain
\bean
g(t, x)\approx
\frac{2}{t^2}\,\Psi\left(\frac{x-1}{t^2}\right)
\quad \hbox{as}\,\,\,t\to 0,
\eean
with
\beqn
\label{S7Epsichi}
\Psi(\chi)=\left\{
\begin{array}{l}
\frac{\pi}{\chi^{3/2}}\, e^{-\frac{\pi}{\chi}}\quad \hbox{for}\,\,\,\chi\ge 0,\\
 \\
0\quad \hbox{for}\,\,\,\chi<0.
\end{array}
\right.
\eeqn
Finally, since $x=e^X$, $x-1=e^X-1\sim X$ when $X\sim 0$ we obtain
\bear
\label{S7Frupokj}
G(t, X)\approx \frac{2}{t^2}\,\Psi\left(\frac{X}{t^2}\right) \quad \hbox{as}\,\,\,t\to 0.
\eear
The fact that the  support of $G(t, X)$ is contained in $\RR^+$ is a consequence of the interpretation of (\ref{S7Eheuristic1}) in terms of coagulation of particles given above.

\subsubsection{The region where $X={\cal O}(t^{2})$.}

We now prove that the fundamental solution $G(t, X)$ behaves in the self similar form (\ref{S7Frupokj}) in the region where  $X={\cal O}(t^{2})$ using the explicit representation formula given by (\ref{E3}), (\ref{S5E2JK42-1516}) and (\ref{S5E103}). 
We are interested in the limit $\lim_{t\to 0^+} t^2\,G(t, t^2\, \chi)=\Psi(\chi)$ in compact sets of $\chi$. To this end we rewrite (\ref{S5E103}) as follows:
\bear
\label{S6EGChi}
G(t, \chi)=\frac{1}{\sqrt{2 \pi}\, t^2}\int_{\Im m(\eta)=\beta_0\, t^2}e^{i \eta\, \chi} \widehat G\left(t, \frac{\eta}{t^2}\right) d\eta.
\eear
Formula (\ref{S6EGChi}) suggests that in order to compute $\Psi(\chi)$ we need to obtain  $\lim_{t\to 0^+}\widehat G\left(t, \frac{\eta}{t^2}\right)$ for $\eta$  such that $\Im m(\eta)=\beta_0\, t^2$. To this end we use the expression (\ref{S5E2JK43}) that implies:
\bear
\label{S5E2JK43et}
\widehat G(t, \eta/t^2)=-\frac{\sqrt 2}{\sqrt \pi  i\,(\lambda-1)}
\int_{Im\, Y =-\gamma_1}
{{\cal V}(\eta/t^2)\over {\cal V}(Y+\eta/t^2)}
t^{-\frac{2 iY}{\lambda-1}}\, \Gamma\left(\frac{2i Y}{\lambda-1} \right)dY.
\eear

\begin{prop}
\label{S5Eperfilautosem}
\bean
\lim_{t\to 0^+} \left(t^2\, G(t, t^2\, \chi) \right)= \psi(\chi)
\eean
unformly for $\chi$ in compact sets of $\RR$ where $\psi(\chi)$ is as in (\ref{S7Epsichi}).
\end{prop}
The proof of Proposition \ref{S5Eperfilautosem} requires several Lemmas.

\begin{lem} 
\label{S6lemFracnues}
For all $\varepsilon_0>0$ and $M>0$  there exists a function $h_{\varepsilon_0, M}(t)$ such that
\bear
\label{S6lemFracnuesE1}
\lim_{t\to 0^+}h_{\varepsilon_0, M}(t)=0
\eear
and 
\bear
\label{S6lemFracnuesE2}
\left|{{\cal V}(\eta/t^2)\over {\cal V}(Y+\eta/t^2)}t^{-\frac{2\, i\, Y}{\lambda-1}}- 
e^{-\frac{2\, i\, Y}{\lambda-1}\ln \left(2\, \sqrt \pi \sqrt{i\, \eta} \right)}
 \right|\le h_{\varepsilon_0, M}(t)
 \eear
for all $Y$ such that $|\Im m(Y)|\le 1/4$, $|\Re e (Y)|\le \frac{1}{t\, |\ln t|}$,  for $\Im m(\eta/t^2)$ in compact subsets of $(3/2, (3+\lambda)/2)$ and $\varepsilon_0\le |\eta|\le M$. Moreover there is $\delta_0>0$ sufficiently small (depending on $\varepsilon_0$ and $M$) such that:
\bear
\label{S6lemFracnuesE3}
\left|{{\cal V}(\eta/t^2)\over {\cal V}(Y+\eta/t^2)}t^{-\frac{2\, i\, Y}{\lambda-1}}
 \right|\le C\, e^{\frac{3\pi}{4}\frac{|Y|}{\lambda-1}}
 \eear
for all $Y$ such that $|\Im m(Y)|\le 1/4$, $|\Re e (Y)|\le \frac{\delta_0}{t^2}$,  for $\Im m(\eta/t^2)$ in compact subsets of $(3/2, (3+\lambda)/2)$ and $\varepsilon_0\le |\eta|\le M$.
\end{lem}
\begin{rem}
In the  Lemmas \ref{S6lemFracnues} until \ref{S5Lemexplicit} and in Proposition \ref{S5Eperfilautosem} we choose the branch of the function square root as follows:
\bean
\sqrt z=|z|^{1/2}\, e^{i\frac{\theta}{2}}\,\,\,\hbox{with}\,\,\theta\in (-\pi, \pi].
\eean
\end{rem}

\textbf{Proof.} Lemma \ref{S6lemFracnues} is a consequence of Lemma \ref{S5TPhi} where $\xi=\eta /t^2$, $Z=Y/\sqrt{|\xi|}$.  By formula (\ref{S5E2JK42bis-1}) we have:
\bean
\frac{{\cal V}(\eta/t^2)}{{\cal V}(\eta/t^2+Y)} & = & \exp\left[\frac{2}{(\lambda-1)\, i}
\int_{Im\, \eta  =\beta_1}
\ln\left(-\Phi (\eta)\right)
\Theta(\eta-\eta/t^2, Y )d\eta\right]\\
& = & \exp\left[F\left(\eta/t^2, t\, Y/\sqrt{|\eta|} \right)\right]
\eean
and using (\ref{S5TPhiE3}):
\bean
\left|F\left(\eta/t^2, t\, Y/\sqrt{|\eta|} \right) 
+\frac{2\, i\, Y}{\lambda-1}\ln \left(2\, \sqrt \pi \sqrt{i\, \eta} \right)
 \right|\le C\, t^2\frac{|Y|^2}{|\eta|}.
\eean
Therefore, in the region where $|\Re e(Y)|\le 1/(t|\ln t|)$ we obtain (\ref{S6lemFracnuesE2}). On the other hand, if 
$|\Re e(Y)|\le \delta_0 /t^2$ we deduce  (\ref{S6lemFracnuesE3}) using that:
\bean
\left|\Re e\left(\frac{2\, i\, Y}{\lambda-1}\ln \left(2\, \sqrt \pi \sqrt{i\, \eta} \right)\right)\right|\le \frac{\pi}{2(\lambda-1)}|Y|+C
\eean
as $|\eta|\to +\infty$ and for $\Im m(Y)=-\gamma_1$. \qed
\begin{lem}
\label{S6lemLimghat}
For all positive constants $M$, $\varepsilon_0$ such that $M>\varepsilon_0$:
\bean
\lim_{t\to 0^+}\widehat G\left(t, \frac{\eta}{t^2}\right)=-\frac{\sqrt 2}{\sqrt \pi  i\,(\lambda-1)}\int_{\Im m(Y)=
-\gamma_1}e^{-\frac{2\, i\, Y}{\lambda-1}\ln \left(2\, \sqrt \pi \sqrt{i\, \eta} \right)}\Gamma\left(\frac{2i Y}{\lambda-1} \right)dY,
\eean
uniformly for  $\Im m(\eta/t^2)$ in compact subsets of $(3/2, (3+\lambda)/2)$ and  $\varepsilon_0\le |\eta|\le M$.
\end{lem}
\textbf{Proof.} We split the integral in (\ref{S5E2JK43et}) as follows:
\bean
\widehat G\left(t, \frac{\eta}{t^2}\right) & = & -\frac{\sqrt 2}{\sqrt \pi  i\,(\lambda-1)}
\int_{Im\, Y =-\gamma_1,\, |\Re e(Y)|\le \frac{1}{t^2\,|\ln t|}}
{{\cal V}(\eta/t^2)\over {\cal V}(Y+\eta/t^2)}
t^{-\frac{2 iY}{\lambda-1}}\, \Gamma\left(\frac{2i Y}{\lambda-1} \right)dY\\
&-&\frac{\sqrt 2}{\sqrt \pi  i\,(\lambda-1)}
\int_{Im\, Y =-\gamma_1,\, \frac{1}{t^2\,|\ln t|}\le |\Re e(Y)|\le \frac{\delta_0}{t^2}}
{{\cal V}(\eta/t^2)\over {\cal V}(Y+\eta/t^2)}
t^{-\frac{2 iY}{\lambda-1}}\, \Gamma\left(\frac{2i Y}{\lambda-1} \right)dY\\
& + & \frac{\sqrt 2}{\sqrt \pi  i\,(\lambda-1)}
\int_{Im\, Y =-\gamma_1,\, |\Re e(Y)|\ge \frac{\delta_0}{t^2}}
{{\cal V}(\eta/t^2)\over {\cal V}(Y+\eta/t^2)}
t^{-\frac{2 iY}{\lambda-1}}\, \Gamma\left(\frac{2i Y}{\lambda-1} \right)dY\\
&=& J_1+J_2+J_3.
\eean

\bear
J_1 & = & -\frac{\sqrt 2}{\sqrt \pi  i\,(\lambda-1)}
\int_{Im\, Y =-\gamma_1,\, |\Re e(Y)|\le \frac{1}{t^2\,|\ln t|}}
\left({{\cal V}(\eta/t^2)\over {\cal V}(Y+\eta/t^2)}t^{-\frac{2\, i\, Y}{\lambda-1}}- 
e^{-\frac{2\, i\, Y}{\lambda-1}\ln \left(2\, \sqrt \pi \sqrt{i\, \eta} \right)}
 \right)\times \nonumber \\
&& \times \, \Gamma\left(\frac{2i Y}{\lambda-1} \right)dY +\frac{\sqrt 2}{\sqrt \pi  i\,(\lambda-1)}
\int_{Im\, Y =-\gamma_1,} 
e^{-\frac{2\, i\, Y}{\lambda-1}\ln \left(2\, \sqrt \pi \sqrt{i\, \eta} \right)}
\, \Gamma\left(\frac{2i Y}{\lambda-1} \right)dY \nonumber\\
&&-\frac{\sqrt 2}{\sqrt \pi  i\,(\lambda-1)}
\int_{Im\, Y =-\gamma_1,\, |\Re e(Y)|\ge \frac{1}{t^2\,|\ln t|}} 
e^{-\frac{2\, i\, Y}{\lambda-1}\ln \left(2\, \sqrt \pi \sqrt{i\, \eta} \right)}
\, \Gamma\left(\frac{2i Y}{\lambda-1} \right)dY.\label{S6lemLimghatE1}
 \eear
Using that:
 \bean
\left| e^{-\frac{2\, i\, Y}{\lambda-1}\ln \left(\sqrt {i\eta} \right)}\right|\le C\, e^{\frac{\pi}{2(\lambda-1)}|Y|}\quad \hbox{and}\quad
\left|\Gamma \left(\frac{2\, i\, Y}{\lambda-1}\right)\right|\le C\, e^{-\frac{\pi}{(\lambda-1)}|Y|}
 \eean
 the last term in (\ref{S6lemLimghatE1}) is estimated as:
 \bean
\left| \frac{\sqrt 2}{\sqrt \pi  i\,(\lambda-1)}
\int_{Im\, Y =-\gamma_1,\, |\Re e(Y)|\ge \frac{1}{t^2\,|\ln t|}} 
e^{-\frac{2\, i\, Y}{\lambda-1}\ln \left(2\, \sqrt \pi \sqrt{i\, \eta} \right)}
\, \Gamma\left(\frac{2i Y}{\lambda-1} \right)dY\right|\le Ce^{-\frac{a}{t^2\, |\ln t|}}
 \eean
 for some positive constant $a$. In the first term of (\ref{S6lemLimghatE1}), using (\ref{S6lemFracnuesE2}) in Lemma \ref{S6lemFracnues} we are led to estimate:
 \bean
h(t)\, \int_{Im\, Y =-\gamma_1}e^{-\frac{\pi}{(\lambda-1)}|Y|}|dY| 
 \eean
which tends to zero as $t\to 0^+$ by (\ref{S6lemFracnuesE1}).\\
We now consider the integral $J_3$. This term is estimated using the estimates of the function $\mathcal V$ proved in Proposition \ref{A1T101} and Stirling's formula:
\bean
|J_3| & \le &  C_\delta t^{-\frac{2\, \gamma_1}{\lambda-1}}e^{-\frac{1}{t^2}\left(\frac{\pi\, \delta_0}{\lambda-1}-2\delta M\right)}\le C_\delta t^{-\frac{2\, \gamma_1}{\lambda-1}} e^{-\frac{a}{t^2}}
\eean
for some positive constant $a$ choosing $\delta$ sufficiently small.\\
The integral $J_2$ is estimated using (\ref{S6lemFracnuesE3}) in Lemma \ref{S6lemFracnues}:
\bean
|J_2|\le C
\int_{Im\, Y =-\gamma_1,\, \frac{1}{t^2\,|\ln t|}\le |\Re e(Y)|\le \frac{\delta_0}{t^2}}e^{\frac{3\pi}{4(\lambda-1)}|Y|}e^{-\frac{\pi}{\lambda-1}|Y|}|d Y|\le
Ce^{-\frac{a}{t^2\, |\ln t|}}
\eean
for some positive constant $a$. \qed
\begin{lem}
\label{S5Lemexplicit}
\bean
\frac{-\sqrt 2}{\sqrt \pi  i\,(\lambda-1)}\int_{\Im m(Y)=
-\gamma_1}e^{-\frac{2\, i\, Y}{\lambda-1}\ln \left(2\, \sqrt \pi \sqrt{i\, \eta} \right)}\Gamma\left(\frac{2i Y}{\lambda-1} \right)dY=\sqrt{2\, \pi}\, e^{-2\sqrt{\pi \, \eta\, i }}
\eean
\end{lem}
\textbf{Proof.} We first perform the change of variable: $2iY/(\lambda-1)=s$:
\bean
\frac{-\sqrt 2}{\sqrt \pi  i\,(\lambda-1)}\int_{\Im m(Y)=
-\gamma_1}\!\!\!\!e^{-\frac{2\, i\, Y}{\lambda-1}\ln \left(2\, \sqrt \pi \sqrt{i\, \eta} \right)}\Gamma\left(\frac{2i Y}{\lambda-1} \right)dY=
\frac{1}{\sqrt{2\, \pi}}\int_{\Re e(s)=\gamma_1}e^{-s\, \ln(2\, \sqrt{\pi\, \eta\, i})}\, \Gamma(s)\, ds.
\eean
Then we deform the integration contour and reduce the integral to the sum of the residues of the function
$e^{-s\, \ln(2\, \sqrt{\pi\, \eta\, i})}$ at the poles $s=-n$ of the Gamma function with residues $\frac{(-1)^n}{n!}$.
\qed

\noindent
\textbf{Proof of Proposition \ref{S5Eperfilautosem}.} 
We split the integral in (\ref{S6EGChi})
\bean
t^2\, G(t, \chi)= I_1+I_2+I_3
\eean
where
\bean
I_1 & = & \frac{1}{\sqrt{2 \pi}}\int_{\Im m(\eta)=\beta_0\, t^2,\, |\eta|\le \varepsilon_0}e^{i \eta\, \chi} \widehat G\left(t, \frac{\eta}{t^2}\right) d\eta\\
I_2 & = & \frac{1}{\sqrt{2 \pi}}\int_{\Im m(\eta)=\beta_0\, t^2,\,  \varepsilon_0 \le |\eta|\le M}e^{i \eta\, \chi} \widehat G\left(t, \frac{\eta}{t^2}\right) d\eta\\
I_3 & = & \frac{1}{\sqrt{2 \pi}}\int_{\Im m(\eta)=\beta_0\, t^2,\, |\eta|\ge M}e^{i \eta\, \chi} \widehat G\left(t, \frac{\eta}{t^2}\right) d\eta.
\eean
Using Lemma \ref{S5Tregularity10}:
\bear
\label{S5Ecotajotaunodos}
|I_1|+I_3|\le C\int_{\Im m(\eta)=\beta_0\, t^2,\, |\eta|\le \varepsilon_0,\, |\eta|\ge M}e^{-a|\eta|}d\eta\le C \left(\varepsilon_0+e^{-a\, M} \right).
\eear
The integral $I_2$ is estimated using the previous Lemmas:
\bean
I_2=\frac{1}{\sqrt{2 \pi}}\int_{\Im m(\eta)=\beta_0\, t^2,\,  \varepsilon_0 \le |\eta|\le M}e^{i \eta\, \chi} (\widehat G\left(t, \frac{\eta}{t^2}\right)-\sqrt{2\, \pi}\, e^{-2\sqrt{\pi \, \eta\, i }}) d\eta+\\
\int_{\Im m(\eta)=\beta_0\, t^2,\,  \varepsilon_0 \le |\eta|\le M}e^{i \eta\, \chi}\, e^{-2\sqrt{\pi \, \eta\, i }} d\eta
= I_{2,1}+I_{2,2}
\eean
The first term is bounded as
\bean
|I_{2,1}|\le C\int_{\Im m(\eta)=\beta_0\, t^2,\,  \varepsilon_0 \le |\eta|\le M}\left|\widehat G\left(t, \frac{\eta}{t^2}\right)-\sqrt{2\, \pi}\, e^{-2\sqrt{\pi \, \eta\, i }}\right| |d\eta|
\eean
and tends to zero by Lemma \ref{S6lemLimghat} and Lemma \ref{S5Lemexplicit}. And,
\bean
\lim_{t\to 0^+}I_{2,1}=\int_{\Im m(\eta)=0,\,  \varepsilon_0 \le |\eta|\le M}e^{i \eta\, \chi}\, e^{-2\sqrt{\pi \, \eta\, i }} d\eta.
\eean
Now taking the limit as $\varepsilon_0\to 0$, $M\to +\infty$ and using (\ref{S5Ecotajotaunodos}) :
\bear
\label{S5perfilexplicitdos}
\lim_{t\to 0^+}t^2\, G(t, \chi)=\int_{\Im m(\eta)=0}e^{i \eta\, \chi}\, e^{-2\sqrt{\pi \, \eta\, i }} d\eta.
\eear
The integral in the right hand side of (\ref{S5perfilexplicitdos}) may be calculated explicitly using contour deformation. For $\chi<0$ the contour is sent to the region where $\Im m(\eta)\to -\infty$ and  the integral gives zero. When $\chi>0$ the deformation is made to the upper half plane in such a way that it avoids the cut along the half line $ \eta \in i\, \RR^+$ and the integral is reduced to :
\bean
2\,\int_0^\infty \sin\left( 2 \sqrt{\pi\, \lambda}\right)e^{-\chi\, \lambda}\, d\lambda=\frac{\pi}{\chi^{3/2}}\, e^{-\frac{\pi}{\chi}}.
\eean
\qed
\subsubsection{Estimates of $G(t, X)$ for  $t^{2}\le X\le 1$.}
We obtain an estimate in self similar form in the domain where $t^{2}\le X\le C$ that is required to show that $G(t, X)$ converges to the Dirac measure as $t\to 0$.
\begin{prop}
\label{S5Txt2peque} 
For any $\delta \in (0, 1/2)$, there exists a positive constant $C_\delta$ such that
\bean
|G(t, X)|\le C\frac{t^{1-2\, \delta}}{|X|^{\frac{3}{2}-\delta}}\quad \hbox{for}\,\,\,t^2\le |X|\le 1.
\eean
\end{prop}
\textbf{Proof.} We integrate by parts twice in formula (\ref{S5E103}) and obtain:
\bear
\label{S5E103byparts}
G(t, X)=\frac{1}{\sqrt{2 \pi}}\frac{1}{X^2}\int_{\Im m(\xi)=\beta_0}
\left(e^{i X \xi}-1 \right)\frac{\partial ^2}{\partial \xi^2}\widehat G(t, \xi) d\xi
\eear
Using that $\left|e^{i X \xi}-1 \right|\le C_\delta |X|^{1/2+\delta}|\xi|^{1/2+\delta}$ for $\delta\in [0, 1/2]$, as well as (\ref{S5Tregularity10E2}) we deduce:
\bean
|G(t, X| & \le & \frac{C_\delta}{X^2}\int_{\Im m(\xi)=\beta_0}\frac{t}{(1+|\xi|^{3/2})}|X|^{1/2+\delta}|\xi|^{1/2+\delta}e^{-t\, \sqrt{|\xi|}} |d\xi|\\
& \le & \frac{C_\delta\, t}{|X|^{\frac{3}{2}-\delta}}\int_{\Im m(\xi)=\beta_0}e^{-t\, \sqrt{|\xi|}}\frac{ |d\xi|}{|\xi|^{1-\delta}}
\eean
and the result follows.\qed

\subsection{Behaviour of $G(t, X)$ for $t\ge 1$.} The behaviour of the function $G$ as $t\to +\infty$ has a self similar structure. This is seen by writing the function $G(t, X)$, given in (\ref{E3}) and  (\ref{S5E103}), in terms of the variable
\bear
\label{S5Eteta}
\theta=X+\frac{2}{\lambda-1}\ln(t)
\eear
\bear
G(t, X)=\frac{i}{\pi (\lambda-1)}\int_{\Im m(\xi)=\beta_0} d\xi\, e^{i\xi\theta}
\int_{\Im m(y)=\beta_1}dy\, \frac{{\cal V}(\xi)}{{\cal V}(y)}t^{-\frac{2\, i\, y}{\lambda-1}} 
\, \Gamma\left(-\frac{2i }{\lambda-1}(\xi-y) \right)\nonumber
\\ \label{S5E104000001}
\eear
with $3/2<\beta_0<\beta_1<(3+\lambda)/2$ . \\
Moving the integration contour of $y$ downward in the expression of $G(t, X)$, the first singularity to be met in the integrand of of (\ref{S5E104000001}) is $y=i$ which is a zero of  ${\mathcal V}(y)$  (cf. Proposition \ref{SAT8-1K}). That gives:
\bear
&& G(t, X)  =  t^{\frac{2}{\lambda-1}}\, \Psi_1(\theta)+\nonumber \\
&&\frac{i}{\pi (\lambda-1)}\int_{\Im m(\xi)=\beta_0} d\xi\, e^{i\xi\theta}
\int_{\Im m(y)=\beta_2}dy\, \frac{{\cal V}(\xi)}{{\cal V}(y)}t^{-\frac{2\, i\, y}{\lambda-1}}\Gamma\left(-\frac{2i }{\lambda-1}(\xi-y) \right)\label{S5E104}\\
&&\Psi_1(\theta)  = 
\frac{2}{\lambda-1}\int_{\Im m(\xi)=\beta_0} d\xi\, e^{i\xi\theta}\frac{{\cal V}(\xi)}{{\cal V}'(i)}
\Gamma\left(-\frac{2i }{\lambda-1}(\xi-i) \right) \label{S5E104-BIS}
\eear
where now $\beta_2\in  ((3-\lambda)/2, 1)$. 

Notice that ${\mathcal V}'(i)\not =0$ by the following reason.  If we differentiate the formula (\ref{E5}) we obtain:
\bean
{\mathcal V}'(i)=-{\mathcal V} '(\frac{\lambda+1}{2}\, i) \Phi(\frac{\lambda+1}{2}\, i)-
{\mathcal V} (\frac{\lambda+1}{2}\, i) \Phi'(\frac{\lambda+1}{2}\, i)
\eean
At $\xi=\frac{\lambda+1}{2}\, i$ the function $\mathcal V$ is analytic and $\Phi$ has a zero. Therefore,
\bean
{\mathcal V}'(i)=-
{\mathcal V} (\frac{\lambda+1}{2}\, i) \Phi'(\frac{\lambda+1}{2}\, i)
\eean
By Proposition \ref{SAT8-1K}, ${\mathcal V} (\frac{\lambda+1}{2}\, i)\not =0$. Moreover, by Proposition \ref{e2.25}, 
\bean
\Phi(\xi)\sim-2\, \pi\,i (\xi-\frac{\lambda+1}{2}\, i)\quad \hbox{as}\,\,\,|\Re e (\xi)|\to +\infty
\eean
and therefore $\Phi'(\frac{\lambda+1}{2}\, i)=-2\pi\,i$. We then obtain the expression for $\Psi_1$:
\bear
\Psi_1(\theta)  = 
\frac{1}{\pi(\lambda-1)\, i\, {\mathcal V}(\frac{\lambda+1}{2}\, i)}\int_{\Im m(\xi)=\beta_0} d\xi\, e^{i\xi\theta}{\cal V}(\xi)
\Gamma\left(\frac{2 i}{\lambda-1}(i-\xi) \right) \label{S5E104-BIS-Ter}
\eear

We study now the function $\Psi(\theta)$ and give its behaviour as $\theta\to \pm \infty$. 
\begin{prop}
\label{S5T1}
The following estimates hold:
\bear
\Psi_1 (\theta) = C_1\,e^{-\frac{3}{2}\theta}+{\cal O}\left(e^{-\left(\frac{4-\lambda}{2}+\varepsilon\right)\theta} \right) \quad \hbox{as}\,\,\,\theta\to -\infty,
\label{S5T1E1}\\
\Psi_1 (\theta)= C_2\,e^{-\frac{3+\lambda}{2}\theta}+{\cal O}\left(e^{-\left(\lambda+1-\varepsilon\right)\theta} \right)\quad \hbox{as}\,\,\,\theta\to +\infty,
\label{S5T1E2}
\eear
for some positive constant $\varepsilon$ arbitrarily small and
where 
\bean
C_1=\frac{2\, i\,}{(\lambda-1)}\frac{{\cal V}((1+\frac{\lambda}{2})i)}{{\cal V}(\frac{\lambda+1}{2}i)}\;\;\hbox{and}\;\; C_2=-\frac{\Gamma\left(\frac{\lambda+1 }{\lambda-1}\right)}{2\pi\, i}
\frac{{\cal V}(2\,i)}{{\mathcal V}(\frac{\lambda+1}{2}\, i)}.
\eean
\end{prop}
\textbf{Proof of Proposition \ref{S5T1}.}  
We use again contour deformation. In order to obtain the behaviour as $\theta \to -\infty$ we deform the contour integration in $\Psi$ downward. The first singularity of the integrand that we meet is $\xi=3i/2$ which is a pole of $\mathcal V (\xi)$. 
Using (\ref{E5}) and (\ref{e2.25}) we obtain:
\bear
{\mathcal R}es \left({\cal V}, \xi=\frac{3\, i}{2} \right) & = & -i\, {\cal V}\left(\left(1+\frac{\lambda}{2}\right)\, i \right)
\label{S5Eresitroisdemis}\\
{\mathcal R}es \left({\cal V}, \xi=\frac{(3+\lambda)\, i}{2} \right) & = & -\frac{\mathcal V(2i)}{4\pi\, i}.
\label{S5Eresitroisdemislambda}
\eear
Therefore

\bear
\label{S5E2JK21}
\Psi_1(\theta) & = & \frac{2\, i\,}{(\lambda-1)}\frac{{\cal V}((1+\lambda/2)i)}{{\cal V}((\lambda+1)/2)i)}
\, e^{-\frac{3}{2}\theta}\nonumber \\
&&+\frac{1}{\pi(\lambda-1)\, i\, {\mathcal V}(\frac{\lambda+1}{2}\, i)}\int_{\Im m(\xi)=\beta_3} d\xi\, e^{i\xi\theta}{\cal V}(\xi)
\Gamma\left(-\frac{2i }{\lambda-1}(\xi-i) \right)
\eear
\bear
\label{S5E2JK22}
\Psi_1(\theta) & = & -\frac{\Gamma\left(\frac{\lambda+1 }{\lambda-1}\right)}{2\pi\, i}
\frac{{\cal V}(2\,i)}{{\mathcal V}(\frac{\lambda+1}{2}\, i)}\,e^{-\frac{3+\lambda}{2}\theta}  \nonumber \\ \nonumber \\
&&+\frac{1}{\pi(\lambda-1)\, i\, {\mathcal V}(\frac{\lambda+1}{2}\, i)}\int_{\Im m(\xi)=\beta_4} d\xi\, e^{i\xi\theta}{\cal V}(\xi)
\Gamma\left(-\frac{2i }{\lambda-1}(\xi-i) \right)
\eear
where $\beta_3\in ((4-\lambda)/2,3/2)$, $\beta_4\in ((3+\lambda)/2, 1+\lambda)$. We have derived (\ref{S5E2JK21}), (\ref{S5E2JK22}) deforming the contour of integration upward and downward respectively.\\
Proposition \ref{A1T101} ensures that the function
${\cal V}(\xi)\Gamma\left(-\frac{2i }{\lambda-1}(\xi-i) \right)$ is integrable and then, for $\Re e\, \theta \le 0$:
\bear
\label{S5E2JK223}
\left|\int_{\Im m (\xi)=\beta_3} d\xi\, e^{i\xi\theta}{\cal V}(\xi)
\Gamma\left(-\frac{2i }{\lambda-1}(\xi-i) \right)\right|\le
Ce^{-\beta_3\, \Re e(\theta)}.
\eear
while for $\theta \ge 0$:
\bear
\label{S5E2JK224}
\left|\int_{\Im m (\xi)=\beta_4} d\xi\,  e^{i\xi\theta}{\cal V}(\xi)
\Gamma\left(-\frac{2i }{\lambda-1}(\xi-i) \right)\right|\le
Ce^{-\beta_4\, \Re e(\theta)}.
\eear
Proposition (\ref{S5T1}) follows from (\ref{S5E2JK21}), (\ref{S5E2JK22}),  (\ref{S5E2JK223}) and (\ref{S5E2JK224}).
\qed

\subsubsection{Estimate of the remainder term in formula  (\ref{S5E104}).}
Let us call that term:
\bear
\label{S5EGr}
G_1(t, \theta) & = & \frac{i}{\pi (\lambda-1)}\int_{\Im m(\xi)=\beta_0} d\xi\, e^{i\xi\theta}
\int_{\Im m(y)=\beta_2}dy\, \frac{{\cal V}(\xi)}{{\cal V}(y)}t^{-\frac{2\, i\, y}{\lambda-1}}\Gamma\left(-\frac{2i }{\lambda-1}(\xi-y) \right) \nonumber \\
\eear
We have then the following Lemma.
\begin{lem}
\label{regularity1} For any positive constant $C_1$ 
there exist positive constants $A, \delta_0$ and $C_2$  such that, for all $t>1$:
\bear
&&|G_1(t, \theta)|\, e^{\frac{3 \theta}{2}}\le C_2\, t^{\frac{2}{\lambda-1}-\delta_0}\qquad \hbox{for all $\theta \le 0$}  \label{S5Reg2}\\
&&|G_1(t, \theta)|\, e^{\frac{3+\lambda}{2}\theta}\le C_2\, t^{\frac{2}{\lambda-1}-\delta_0}\qquad \hbox{for all $\theta \ge 0$}.  \label{S5Reg3}
\eear
\end{lem}
\textbf{Proof of Lemma \ref{regularity1}.} 
The function $G_1$ is may be written as:
\bear
G_1(t, \theta) & = & 
\frac{i}{\pi (\lambda-1)}\int_{\Im m(\xi)=\beta_0} d\xi\, e^{i\xi\theta}\, 
H_1\,(t, \xi) \label{S5EQR1}\\
H_1(t, \xi) & = &
\int_{\Im m(y)=\beta_2}dy\
\, \frac{{\cal V}(\xi)}{{\cal V}(y)}t^{-\frac{2\, i\, y}{\lambda-1}}\Gamma\left(-\frac{2i }{\lambda-1}(\xi-y) \right)
 \label{S5EQR2}
\eear
The function $H_1$ is estimated in the same way as the function $\widetilde G(t, \xi)$ in Lemma \ref{S5Tregularity11}. To this end we first perform the change of variables: $y=\xi+\sqrt{|\xi|}\, Z$ and obtain:

\bear
\label{S7EH1cambio}
H_1(t, \xi) & = &
\sqrt{|\xi|}\int_{\Im m(Z)=\frac{\beta_2-\beta_0}{\sqrt{|\xi|}}}dy\
\, \frac{{\cal V}(\xi)}{{\cal V}(\sqrt{|\xi|}Z+\xi)}t^{-\frac{2\, i\, (\sqrt{|\xi|}Z+\xi)}{\lambda-1}}\Gamma\left(\frac{2i }{\lambda-1}\sqrt{|\xi|}\, Z \right)
\eear
Then we deform the contour of integration in (\ref{S7EH1cambio}) to the new contour ${\cal D}_1$.
\begin{figure}
\begin{tikzpicture}
\draw[->] (-6.5,3) -- (6,3) node[right]{$\Re e Z$};
\draw[->] (-3,-3) -- (-3,4) node[above]{$\Im m Z$};
\draw[->] (-6.5, 2) -- (-5, 2);
\draw[->] (-5, 2) -- (-3.6, 2);
\draw[->] (-3.6, 2) -- (-3.6, 0);
\draw[->] (-3.6, 0) -- (-3.6, -2);
\draw[->] (-3.6, -2) -- ( -2, -2);
\draw[->] (-2, -2) -- ( 0, -2) node{$\times$};
\draw[->] (0, -2) -- ( 0, -2) node[below]{$Z_c$};
\draw[->] (0,-2) -- (2, -2);
\draw[->] (2, -2) -- (2, 0);
\draw[->] (2, 0) -- (2, 2);
\draw[->] (2,2)-- (4, 2);
\draw[->] (4,2)-- (6.5, 2) node[below]{$\Im m Z=\frac{\beta_2-\beta_0}{\sqrt{|\xi|}}$};
\end{tikzpicture}
\caption{The curve ${\cal D}_1$}
\end{figure}
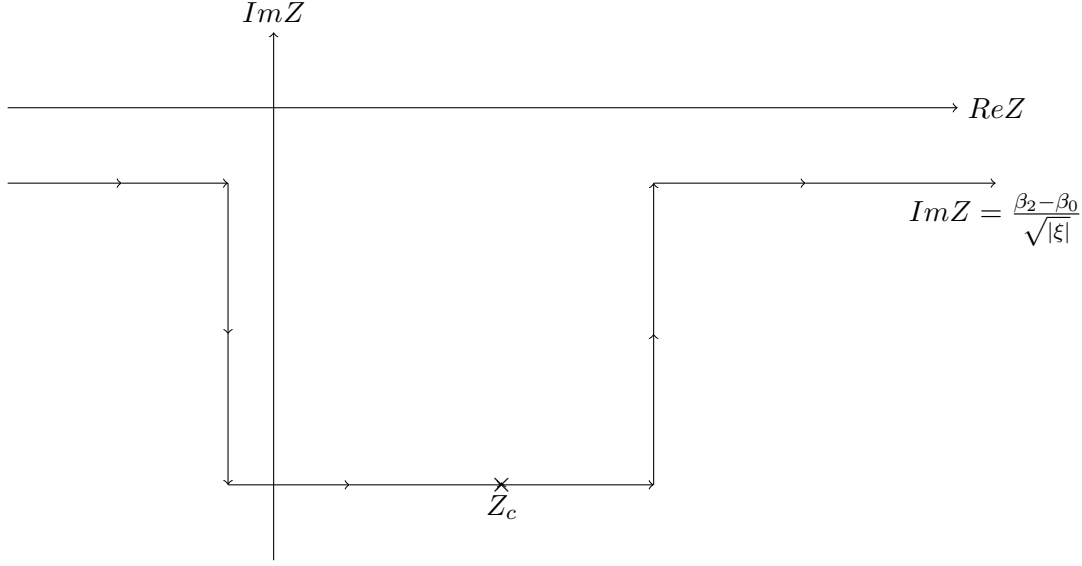
Since along this new contour $\Im m(Z)=\frac{\beta_2-\beta_0}{\sqrt{|\xi|}}$, we have  
$
\Im m (\sqrt{|\xi|}Z+\xi)\le \beta_2
$
and then
\bean
|t^{-\frac{2\, i\,y}{\lambda-1}}|\le C t^{\frac{2\, \beta_2}{\lambda-1}}=t^{\frac{2}{\lambda-1}-\delta}.
\eean
We are then left with the term
\bean
\sqrt{|\xi|}\int_{{\mathcal D}_1}\left|
\, \frac{{\cal V}(\xi)}{{\cal V}(\sqrt{|\xi|}Z+\xi)}t^{-\frac{2\, i\, (\sqrt{|\xi|}Z+\xi)}{\lambda-1}}\Gamma\left(\frac{2i }{\lambda-1}\sqrt{|\xi|}\, Z \right)\right||dy|
\eean
that may be estimated following the same arguments as in the proof of Lemma \ref{S5Tregularity11}. We obtain then the bound on $H_1(t, \xi)$:
\bear
\label{cotaHuno}
|H_1(t, \xi)|\le Ct^{\frac{2}{\lambda-1}-\delta}e^{-a\sqrt{|\xi|}}.
\eear
Using (\ref{S5E103}) we  deduce that, for any positive constant $R$ there exists a constant $C_R>0$ such that, for all $|\theta|\le R$:
\bean
|G_1(t, \theta)| & \le & C_R\,t^{\frac{2}{\lambda-1}-\delta} \eean 
for some positive constant $C_R$. This shows (\ref{S5Reg2}) and (\ref{S5Reg3}) for $\theta$ bounded. 

In order to prove (\ref{S5Reg2}) and (\ref{S5Reg3}) we
deform the contour of the integral with respect to $\xi$ in formula (\ref{S5EGr}). To obtain the estimate (\ref{S5Reg3}) as $\theta \to +\infty$ we deform the contour upward. The first singularity of $\widehat G_1(t, \xi)$ is located at $\xi = (3+\lambda)\, i/2$ (see Proposition \ref{SAT8-1K}). Therefore:
\bear
\label{S5EGrQr}
 G_1(t, \theta) & = & b_2(t)\, e^{-\frac{3+\lambda}{2}\theta}+ Q_1(t, \theta)
\eear
\bear
\label{S5EQr}
 Q_1(t, \theta) & = &  \frac{i}{\pi (\lambda-1)}\int_{\Im m(\xi)=\beta_5} d\xi\, e^{i\xi\theta}
\int_{\Im m(y)=\beta_1}dy\, \frac{{\cal V}(\xi)}{{\cal V}(y)}t^{-\frac{2\, i\, y}{\lambda-1}} 
\times \Gamma\left(-\frac{2i }{\lambda-1}(\xi-y) \right)\nonumber \\
\eear
where $\beta_5\in ((3+\lambda)/2, (4+\lambda)/2)$, 
and
\bean
b_2(t) 
& = & \frac{{\cal V}(2\, i)}{2\pi\, i(\lambda-1)}
\int_{\Im m(y)=\beta_1}dy\, \frac{t^{-\frac{2\, i\, y}{\lambda-1}} }{{\cal V}(y)}
\Gamma\left(-\frac{2i }{\lambda-1}\left(\frac{3+\lambda}{2}\, i-y\right) \right).
\eean
Using that again $\beta_1\in ((3-\lambda)/2, 1)$ and so in particular $\beta_1<1$, we deduce:
\bear
|b_2(t)| & \le &C t^{\frac{2}{\lambda-1}-\delta_0},\quad \hbox{for}\,\,\,t>1. \label{S5Esuno}
\eear
 We are then left with the term $Q_1(t, \theta)$ that we write:
 \bear
Q_1(t, \theta) & = &  \int_{\Im m(\xi)=\beta_5} d\xi\, e^{i\xi \theta}\, 
H_2(t, \xi)
\label{S5EQsubrtheta}\\
H_2& = & \frac{i}{\pi (\lambda-1)} \int_{\Im m(y)=\beta_1}dy\, \frac{{\cal V}(\xi)}{{\cal V}(y)}
t^{-\frac{2\, i\,y}{\lambda-1}} \Gamma\left(-\frac{2i }{\lambda-1}(\xi-y) \right).\label{S5EQsubrji}
\eear
The function $H_2$ is estimated in the same way as the function $\widehat G(t, \xi)$ in Lemma \ref{S5Tregularity11}. 
We change of variables: $y=\xi+\sqrt{|\xi|}\, Z$ and obtain:
\bear
\label{S7EH2cambio}
H_2(t, \xi)  = 
\sqrt{|\xi|}\int_{\Im m(Z)=\frac{\beta_1-\beta_0}{\sqrt{|\xi|}}}dy\
\, \frac{{\cal V}(\xi)}{{\cal V}(\sqrt{|\xi|}Z+\xi)}t^{-\frac{2\, i\, (\sqrt{|\xi|}Z+\xi)}{\lambda-1}}\Gamma\left(\frac{2i }{\lambda-1}\sqrt{|\xi|}\, Z \right)
\eear
Then we deform the integration contour in (\ref{S7EH2cambio})  to $\mathcal D _2$ 
\begin{figure}
\begin{tikzpicture}
\draw[->] (-6.5,3) -- (6,3) node[right]{$\Re e Z$};
\draw[->] (-3,-3) -- (-3,4) node[above]{$\Im m Z$};
\draw[->] (-6.5, 2) -- (-5, 2);
\draw[->] (-5, 2) -- (-3.6, 2);
\draw[->] (-3.6, 2) -- (-3.6, 0);
\draw[->] (-3.6, 0) -- (-3.6, -2);
\draw[->] (-3.6, -2) -- ( -2, -2);
\draw[->] (-2, -2) -- ( 0, -2) node{$\times$};
\draw[->] (0, -2) -- ( 0, -2) node[below]{$Z_c$};
\draw[->] (0,-2) -- (2, -2);
\draw[->] (2, -2) -- (2, 0);
\draw[->] (2, 0) -- (2, 2);
\draw[->] (2,2)-- (4, 2);
\draw[->] (4,2)-- (6.5, 2) node[below]{$\Im m Z=\frac{\beta_1-\beta_0}{\sqrt{|\xi|}}$};
\end{tikzpicture}
\caption{The curve ${\cal D}_2$}
\end{figure}
Since along this new contour $\Im m(Z)=\frac{\beta_1-\beta_0}{\sqrt{|\xi|}}$, we have  
$
\Im m (\sqrt{|\xi|}Z+\xi)\le \beta_1
$
and then
\bean
|t^{-\frac{2\, i\,y}{\lambda-1}}|\le C t^{\frac{2\, \beta_1}{\lambda-1}}=t^{\frac{2}{\lambda-1}-\delta}.
\eean
We are then left with the term
\bean
\sqrt{|\xi|}\int_{{\mathcal D}_1}\left|
\, \frac{{\cal V}(\xi)}{{\cal V}(\sqrt{|\xi|}Z+\xi)}t^{-\frac{2\, i\, (\sqrt{|\xi|}Z+\xi)}{\lambda-1}}\Gamma\left(\frac{2i }{\lambda-1}\sqrt{|\xi|}\, Z \right)\right||dy|
\eean
that may be estimated following the same arguments as in the proof of Lemma \ref{S5Tregularity11}. We obtain then the bound on $H_2(t, \xi)$:
\bear
\label{cotaHuno}
|H_2(t, \xi)|\le Ct^{\frac{2}{\lambda-1}-\delta}e^{-a\sqrt{|\xi|}}
\eear

Using now (\ref{S5EQsubrtheta}) we deduce that:
\bear
|Q_1(t, \theta)|\le Ct^{\frac{2}{\lambda-1}-\delta}e^{-\beta_5 \theta}.
\eear
Combining this with (\ref{S5Esuno}), estimate (\ref{S5Reg3}) follows.

The estimate  (\ref{S5Reg2}) for $\theta\to -\infty$ is obtained in a very similar way. We deform downward the contour of the integral (\ref{S5EQR2}) and we continue the proof as for (\ref{S5Reg3}). This concludes the proof of Lemma
\ref{regularity1}.
\qed

\subsection{Behaviour as $0\le t\le 1$, $|X|\to +\infty$.}
For the small values of time the solution is described in the $(t, X)$ variables as follows. 
\begin{lem}
\label{S5Ttpeq} For $0\leq t\leq 1\;$the following estimates hold: 
\begin{equation}
\label{S5TtpeqE1}
G(t,X)=\left\{ 
\begin{array}{l}
e^{-\frac{3}{2}X}\,t+\mathcal{O}\left( e^{-\left( \frac{3}{2}%
-\varepsilon \right) X}\,t\right) \quad \hbox{as}\,\,\,X\rightarrow -\infty 
\\ 
\\ 
C_{1}\,e^{-\frac{3+\lambda }{2}X}\,t+\mathcal{O}\left( e^{-\left( \frac{%
3+\lambda }{2}+\varepsilon \right) X}\,t\right) \quad \hbox{as}%
\,\,\,X\rightarrow +\infty .
\end{array}
\right. 
\end{equation}
where:
\bean
C_1=\frac{{\mathcal V}(2i)}{4\, \pi\, {\mathcal V}
\left(\left(1+\frac{\lambda}{2}\right)\, i \right)}
\eean
\end{lem}
\textbf{Proof of Lemma \ref{S5Ttpeq}.}
Using (\ref{S5E103}) we obtain: 
\begin{eqnarray*}
G(t,X) &=&\frac{1}{\sqrt{2\pi }}\int_{\Im m(\xi )=\beta _{0}}e^{i\xi X}%
\widehat{G}\left( t,\xi \right) d\xi = \\
&=&\frac{i}{\pi (\lambda -1)}\int_{\Im m(\xi )=\beta _{0}}e^{i\xi X}d\xi
\int_{\Im m(Y)=-\gamma _{1}}dy\frac{\mathcal{V}(\xi )}{\mathcal{V}(\xi +Y)}%
t^{-\frac{2\,i}{\lambda -1}Y}\,\Gamma \left( \frac{2iY}{\lambda -1}\right) 
\end{eqnarray*}
where $\beta _{0}\in \left( \frac{3}{2},\frac{3+\lambda }{2}\right) $ and $%
\gamma _{1}$ small. Lemma \ref{S5Tregularity10} implies that $\widehat{G}\left( t,\xi \right) $
is exponentially bounded on $\left| \xi \right| $ for large $\left| \xi
\right| $ and this yields convergence of the integral. Moreover, such
exponential decay holds also for $\xi \in \left( \frac{3}{2}-\delta _{0},%
\frac{3+\lambda }{2}+\delta _{0}\right) $ with $\delta _{0}>0$.

We can now deform the contour integration on $\xi $ crossing the poles of $%
\widehat{G}\left( t,\xi \right) $ that are due to the poles of $\mathcal{V}(\xi
).$ The closest poles are at $\xi =\frac{3}{2}i,\;\xi =\frac{3+\lambda }{2}i$
respectively. We deform the contour upwards if $X>0$ and downwards if $X<0.$
We then obtain using (\ref{S5Eresitroisdemis}), (\ref{S5Eresitroisdemislambda}): 
\begin{eqnarray}
&&G(t,X)=\nonumber\\
&&-2\pi\left( \frac{i}{\pi (\lambda -1)}\right)\, {\cal V}\left(\left(1+\frac{\lambda}{2}\right)\, i \right)\, e^{-\frac{3}{2}X} \int_{\Im m(Y)=-\gamma
_{1}}dy\frac{t^{-\frac{2\,i}{\lambda -1}Y}}{\mathcal{V}(\frac{3}{2}i+Y)}%
\,\Gamma \left( \frac{2iY}{\lambda -1}\right) +  \nonumber \\
&&+\frac{i}{\pi (\lambda -1)}\int_{\Im m(\xi )=\beta _{1}}e^{i\xi X}d\xi
\int_{\Im m(Y)=-\gamma _{1}}dy\frac{\mathcal{V}(\xi )}{\mathcal{V}(\xi +Y)}%
t^{-\frac{2\,i}{\lambda -1}Y}\,\Gamma \left( \frac{2iY}{\lambda -1}\right)  
\nonumber \\
&&\equiv J_{1}+J_{2}  \label{W1E1}
\end{eqnarray}
\begin{eqnarray}
&&G(t,X) = -\frac{\mathcal V(2i)\, i}{2\pi (\lambda -1)}e^{-\frac{3+\lambda}{2}X}\!\!\! \int_{\Im m(Y)=-\gamma
_{1}}\!\!\!dy\frac{t^{-\frac{2\,i}{\lambda -1}Y}}{\mathcal{V}(\frac{3+\lambda }{2}%
i+Y)}\,\Gamma \left( \frac{2iY}{\lambda -1}\right) +  \nonumber \\
&&+\frac{i}{\pi (\lambda -1)}\int_{\Im m(\xi )=\beta _{2}}e^{i\xi X}d\xi
\int_{\Im m(Y)=-\gamma _{1}}dy\frac{\mathcal{V}(\xi )}{\mathcal{V}(\xi +Y)}%
t^{-\frac{2\,i}{\lambda -1}Y}\,\Gamma \left( \frac{2iY}{\lambda -1}\right)  
\nonumber \\
&&\equiv J_{1}+J_{2}  \label{W1E2}
\end{eqnarray}
where $\beta _{1}=\frac{3}{2}-\delta \;,\;\beta _{2}=\frac{3+\lambda }{2}%
+\delta $ with $\delta >0$ small. 

The terms $J_2$ in  (\ref{W1E1}), (\ref{W1E2}) can be estimated easily. Indeed, they can be written in the form:
\bean
G(t,X)=\frac{1}{\sqrt{2\pi }}\int_{\Im m(\xi )=\beta _{\ell }}e^{i\xi X}%
\widehat{G}\left( t,\xi \right) d\xi \;\;,\;\;\ell =1,2
\eean
Integrating by parts we obtain:
\bean
G(t,X)=-\frac{1}{\sqrt{2\pi }X^{2}}\int_{\Im m(\xi )=\beta _{\ell }}d\xi
e^{i\xi X}\frac{\partial ^{2}}{\partial \xi ^{2}}\left( \widehat{G}\left( t,\xi
\right) \right) 
\eean

We can now estimate $\frac{\partial ^{2}}{\partial \xi ^{2}}\left( \widehat{G}%
\left( t,\xi \right) \right) $ using  Lemma \ref{S5Tregularity10}. It then follows that:
\bean
\left| J_{2}\right| \leq \frac{C}{X^{2}}\int_{\Im m(\xi )=\beta _{\ell
}}\left| e^{i\xi X}\right| \frac{t}{\left( 1+\left| \xi \right|
^{3/2}\right) }e^{-a\sqrt{\left| \xi \right| }t}\left| d\xi \right| 
\leq 
\frac{Ce^{\beta _{\ell }X}}{X^{2}}\int_{\Im m(\xi )=\beta _{\ell }}\frac{t}{%
\left( 1+\left| \xi \right| ^{3/2}\right) }e^{-a\sqrt{\left| \xi \right| }%
t}\left| d\xi \right|   \label{W1E3}
\eean

In order to estimate the integral we split it as follows:
\bean
\int_{\Im m(\xi )=\beta _{\ell }}\frac{t}{\left( 1+\left| \xi \right|
^{3/2}\right) }e^{-a\sqrt{\left| \xi \right| }t}\left| d\xi \right|
& = & \int_{\Im m(\xi )=\beta _{\ell },\;\left| \xi \right| \leq \frac{1}{t^{2}}}%
\left[ ...\right] \left| d\xi \right| +\int_{\Im m(\xi )=\beta _{\ell
},\;\left| \xi \right| \geq \frac{1}{t^{2}}}\left[ ...\right] \left| d\xi
\right|\\ &\leq & Ct+Ct^{2}
\eean

Then:
\bean
\left| J_{2}\right| \leq Cte^{\beta _{\ell }X}\;\;\hbox{for\ \ }\left|
X\right| \geq 1
\eean
where $\ell=1$ for $X<0$ and $\ell =2$ for $X>0$.
In order to compute the terms $J_{1}$ in (\ref{W1E1}), (\ref{W1E2}) we
deform the contour on $Y$ upwards. We cross the first pole of the function $%
\Gamma \left( \frac{2iY}{\lambda -1}\right) $ for $Y=0.$ However, this point
is not a pole of the integrand, because the functions $\mathcal{V}(\frac{3}{2%
}i+Y),\;\mathcal{V}(\frac{3+\lambda }{2}i+Y)$ have also a pole at $Y=0.$
Therefore the first pole of the integrand that is found is the one at $Y=%
\frac{\lambda -1}{2}i.$ Notice that $\mathcal{V}(\frac{3}{2}i+Y)$ does not
have a zero before, since $\frac{3}{2}+\frac{\lambda -1}{2}=1+\frac{\lambda 
}{2}<\frac{3}{2}+\frac{\lambda }{2}$. On the other hand $\frac{3+\lambda }{2}%
+\frac{\lambda -1}{2}=1+\lambda $ while the first zero of $\mathcal{V}\left(
\eta \right) $ is at $\eta =2+\frac{\lambda }{2}$ and $1+\lambda <2+\frac{%
\lambda }{2}.$ Then, after deforming the integral contour as indicated,
 the terms $J_{1}$ can be written  as:
\bean
J_1
& = & e^{-\frac{3}{2}X}\, t -\frac{2\, i}{\lambda-1}{\mathcal V}\left(\left(1+\frac{\lambda}{2}\right)\, i \right) e^{-\frac{3}{2}X}
\int_{\Im m(Y)=\gamma_{2}}dy\frac{t^{-\frac{2\,i}{\lambda -1}Y}}{\mathcal{V}(\frac{3}{2}i+Y)}%
\,\Gamma \left( \frac{2iY}{\lambda -1}\right)\\
& = &e^{-\frac{3}{2}X}\left(t+J_3\right)
\eean
and
\bean
J_1 &= & e^{-\frac{3+\lambda}{2}X}
\frac{{\mathcal V}(2i)\, t}{4\, \pi\, {\mathcal V}
\left(\left(1+\frac{\lambda}{2}\right)\, i \right)}
-\\
&&\hskip 2.5cm -\frac{i\, {\mathcal V}(2i)}{2\, \pi\, (\lambda-1)} e^{-\frac{3+\lambda}{2}X}
\int_{\Im m(Y)=\gamma_{2}}dy\frac{t^{-\frac{2\,i}{\lambda -1}Y}}{\mathcal{V}(\frac{3+\lambda}{2}i+Y)}%
\,\Gamma \left( \frac{2iY}{\lambda -1}\right)\\
& = &e^{-\frac{3+\lambda}{2}X}\left(\frac{{\mathcal V}(2i)\, t}{4\, \pi\, {\mathcal V}
\left(\left(1+\frac{\lambda}{2}\right)\, i \right)}+J_3\right)
\eean
where $\gamma_2>(\lambda-1)/2$ is such that $\gamma_2-(\lambda-1)/2$ is small. In both cases there exist positive constants $\delta$ and $C_\delta$  such that for all $t>1$:
\bean
|J_3| & \le & C_\delta t^{1+\delta}.
\\ 
\left| J_{1}\right| & \leq & Cte^{\beta _{\ell }X}\;\;\hbox{for\ \ }\left|
X\right| \geq 1
\eean
where $\ell=1$ for $X<0$ and $\ell =2$ for $X>0$.

It then follows that:
\bean
\left| J_{1}\right| +\left| J_{2}\right| \leq Cte^{\beta _{\ell }X}\;\;\hbox{%
for\ \ }\left| X\right| \geq 1
\eean
where, as before,  $\ell=1$ for $X<0$ and $\ell =2$ for $X>0$, and Lemma \ref{S5Ttpeq} follows.

\section{The initial value problem.}
\label{invalpbm}
\setcounter{equation}{0}
\setcounter{theo}{0}

Using the fundamental solution $g$ obtained in Theorem \ref{S2T1-1} we can
obtain a solution of the initial value problem
\begin{eqnarray}
\frac{\partial h}{\partial t} &=&L\left[ h\right]   \label{IN1} \\
h\left( 0,x\right)  &=&h_{0}\left( x\right)   \label{IN2}
\end{eqnarray}
with $L\left[\, \cdot\, \right] $ defined in (\ref{e2.5}). Assuming
that there are not difficulties with the integrals written below, we would
expect, due to the linearity of the problem (\ref{IN1}), (\ref{IN2}) the
following representation formula for their solutions (cf. Theorem \ref{S2T1-1}):
\begin{equation}
h\left( t,x\right) =\int_{0}^{\infty }h_{0}\left( y\right) g\left( ty^{\frac{%
\lambda -1}{2}},\frac{x}{y},1\right) \frac{dy}{y}  \label{IN3}
\end{equation}
We first precise sufficient conditions on $h_{0}$ that allow to define $%
h\left( t,x\right) $ in (\ref{IN3}). 

\begin{theo}
\label{S8TCauchy}
Suppose that the function $h_{0}\in C\left( \mathbb{R}^{+}\right) $ satisfies
\begin{equation}
\int_{0}^{1}\left| h_{0}\left( y\right) \right| y^{\lambda
}dy+\int_{1}^{\infty }\left| h_{0}\left( y\right) \right| dy<\infty 
\label{IN4}
\end{equation}

Then the function $h\left( t,x\right) $ defined for $t\geq 0,\;x>0$ by means
of (\ref{IN3}) solves the initial value problem (\ref{IN1}), (\ref{IN2}).
\end{theo}
The proof of this Theorem reduces to a detailed analysis of the conditions on $h_0$ yielding integrability of the right hand side of (\ref{IN3}).

Under more stringent assumptions on $h_{0}$ it is possible to use
Theorem \ref{S2T1-1} to derive more detailed information on the asymptotics of
the solutions of (\ref{IN1}), (\ref{IN2}) for $x\rightarrow 0$ and $%
x\rightarrow \infty .$ The meaning of this asymptotics will be explained in
the next Section.

\begin{theo}
\label{S8Tcauchy2}
Suppose that 
\begin{eqnarray}
\left| h_{0}\left( x\right) \right|  &\leq &Cx^{-\frac{3}{2}+\varepsilon
}\;\;,\;\;0<x\leq 1\;\;,\;\;\varepsilon >0  \label{H1} \\
\left| h_{0}\left( x\right) \right|  &\leq &Cx^{-\left( 1+\varepsilon
\right) }\;\;,\;\;x\geq 1\;\;,\;\;\varepsilon >0  \label{H2}
\end{eqnarray}

Then the function $h\left( t,x\right) $ given in  (\ref{IN3}) satisfies for
any $t>0$
\bear
\left| h\left( t,x\right) -A_{-}\left( t\right) x^{-\frac{3}{2}}\right| 
&\leq &B_{-}\left( t\right) x^{-\frac{3}{2}+\varepsilon }\;\;\hbox{for \ }%
0<x\leq 1 \label{S8T2Amenos}\\
\left| h\left( t,x\right) -A_{+}\left( t\right) x^{-\frac{3+\lambda }{2}%
}\right|  &\leq &B_{+}\left( t\right) x^{-\frac{3+\lambda }{2}-\varepsilon
}\;\;\hbox{for\ \ }x\geq 1 \label{S8T2Amas}
\eear
for suitable functions $A_{-}\left( t\right) ,\;A_{+}\left( t\right)
,\;B_{-}\left( t\right) ,\;B_{+}\left( t\right) .$
\end{theo}
Detailed proofs of these two results will be given in \cite{EV}.

\section{Particle fluxes for singular solutions of the coagulation equation.}
\label{flux}

\subsection{Computing particle fluxes for the nonlinear equation.}

It is well known that smooth solutions of the coagulation equation
(\ref{e1.1}), (\ref{e1.1bis}) preserve the value of the quantity $\int_{0}^{\infty
}xf\left( x,t\right) dx.$ This could be expected given that the equation can
be interpreted as a description of the coalescence of particles with sizes $%
x,\;y$ to form a particle of size $\left( x+y\right) $ in the time interval $%
\left[ t,t+dt\right] $ taking place with a probability $K\left( x,y\right)
dt $ for each particle pair $\left( x,y\right) .$

Notice that, in spite of this conservation property, we cannot expect to be
able to rewrite (\ref{e1.1}), (\ref{e1.1bis}) as a conservation law with the form 
\bean
\frac{\partial }{\partial t}\left( xf\right) +\frac{\partial }{\partial x}%
\left( j\right) =0
\eean
with $j$ depending only on local properties of $f$ at the point $(t, x)$ due to the fact that the particles can change their size an amount of order
one in an infinitesimal interval of time $\left[ t,t+dt\right] .$ However,
we can expect to have equivalent ways of finding formulas with the form: 
\begin{equation}
\frac{d}{dt}\left( \int_{R_{1}}^{R_{2}}xf\left( x,t\right) dx\right)
=J_{R_{1},R_{2}}^{+}-J_{R_{1},R_{2}}^{-}  \label{Q1F1}
\end{equation}
for arbitrary values $0\leq R_{1}<R_{2},$ where $J_{R_{1},R_{2}}^{+}$
denotes the number of monomers that enter in the interval $x\in \left[
R_{1},R_{2}\right] $ coming from particles with sizes in the region $\mathbb{%
R}^{+}\setminus \left[ R_{1},R_{2}\right] $ for unit of time and $%
J_{R_{1},R_{2}}^{-}$ is the number of monomers that leave the set $x\in %
\left[ R_{1},R_{2}\right] $ for unit of time. 

A careful counting of the number of particles entering and leaving the interval $[R_1, R_2]$ yields

\begin{eqnarray}
J_{R_{1},R_{2}}^{+} &=&\frac{1}{2}\int_{D_{1;R_{1},R_{2}}^{+}}f\left(
y\right) f\left( z\right) K\left( y,z\right) \left( y+z\right)
dydz+\nonumber \\
&&+\int_{D_{2;R_{1},R_{2}}^{+}}f\left( y\right) f\left( z\right) K\left(
y,z\right) ydydz  \label{Q1F3} \\
D_{1;R_{1},R_{2}}^{+} &=&\left\{ \left( y,z\right) :y\leq R_{1},\;z\leq
R_{1},\;R_{1}\leq \left( y+z\right) \leq R_{2}\right\}   \nonumber \\
D_{2;R_{1},R_{2}}^{+} &=&\left\{ \left( y,z\right) :y\leq R_{1},\;R_{1}\leq
z\leq R_{2},\;R_{1}\leq \left( y+z\right) \leq R_{2}\right\}   \nonumber
\end{eqnarray}

\begin{eqnarray}
&&J_{R_{1},R_{2}}^{-}   = \frac{1}{2}\int_{D_{1;R_{1},R_{2}}^{-}}f\left(
y\right) f\left( z\right) K\left( y,z\right) \left( y+z\right)
dydz+\nonumber \\
&&+\int_{D_{2;R_{1},R_{2}}^{-}}f\left( y\right) f\left( z\right) K\left(
y,z\right) ydydz+ \int_{D_{3;R_{1},R_{2}}^{-}}f\left( y\right) f\left( z\right) K\left(
y,z\right) ydydz   \label{Q1F2} \\
D_{1;R_{1},R_{2}}^{-} &=&\left\{ \left( y,z\right) :R_{1}\leq y\leq
R_{2},\;R_{1}\leq z\leq R_{2},\;R_{2}\leq \left( y+z\right) \right\}   \nonumber
\\
D_{2;R_{1},R_{2}}^{-} &=&\left\{ \left( y,z\right) :R_{1}\leq y\leq
R_{2},\;R_{2}\leq z,\;R_{2}\leq \left( y+z\right) \right\}   \nonumber \\
D_{3;R_{1},R_{2}}^{-} &=&\left\{ \left( y,z\right) :R_{1}\leq y\leq
R_{2},\;z\leq R_{1},\;R_{2}\leq \left( y+z\right) \right\}   \nonumber
\end{eqnarray}

There are two particular cases of (\ref{Q1F1}) that will be particularly
relevant for our purposes, namely the cases 
\begin{equation}
R_{1}=0\;\;,\;R_{2}=R\in \left( 0,\infty \right)   \label{Q1V1}
\end{equation}
and 
\begin{equation}
R_{1}=R\in \left( 0,\infty \right) \;\;,\;\;R_{2}=\infty   \label{Q1V2}
\end{equation}

In the case (\ref{Q1V1}),  (\ref{Q1F1}) reduces to:
\begin{equation}
\frac{d}{dt}\left( \int_{0}^{R}xf\left( x,t\right) dx\right) =-J_{R}^{-}%
\left[ f\right]   \label{Q1V3}
\end{equation}
\bear
J_{R}^{-}\left[ f\right]  &=&\frac{1}{2}\int_{D_{1;R_{1},R_{2}}^{-}}f\left(
y\right) f\left( z\right) K\left( y,z\right) \left( y+z\right)
dydz+\int_{D_{2;R_{1},R_{2}}^{-}}f\left( y\right) f\left( z\right) K\left(
y,z\right) ydydz \nonumber \\
D_{1}^{-}(R) &=&\left\{ \left( y,z\right) :y\leq R,\;z\leq R,\;R\leq \left(
y+z\right) \right\}  \label{S9ED1menos} \\
D_{2}^{-}\left( R\right)  &=&\left\{ \left( y,z\right) :y\leq R,\;R\leq
z,\;R\leq \left( y+z\right) \right\} \label{S9ED2menos}
\eear
In the case (\ref{Q1V2}), equation (\ref{Q1F1}) becomes:
\begin{equation}
\frac{d}{dt}\left( \int_{R}^{\infty }xf\left( x,t\right) dx\right) =J_{R}^{+}%
\left[ f\right]   \label{Q1V4}
\end{equation}
\begin{eqnarray*}
J_{R}^{+}\left[ f\right]  &=&\frac{1}{2}\int_{D_{1}^{+}\left( R\right)
}f\left( y\right) f\left( z\right) K\left( y,z\right) \left( y+z\right)
dydz+\int_{D_{2}^{+}\left( R\right) }f\left( y\right) f\left( z\right)
K\left( y,z\right) ydydz \\
D_{1}^{+}\left( R\right)  &=&\left\{ \left( y,z\right) :y\leq R\;,\;z\leq
R\;,\;R\leq \left( y+z\right) \right\}  \\
D_{2}^{+}\left(R \right) &=&\left\{ \left( y,z\right) :y\leq R,\;R\leq
z,\;R\leq \left( y+z\right) \right\} 
\end{eqnarray*}
Using the formula (\ref{Q1V3}) we can easily prove that for the
kernel $K\left( y,z\right) =\left( yz\right) ^{\frac{\lambda }{2}}$ the
meaning of the solutions of (\ref{e1.1}), (\ref{e1.1bis}) having the asymptotics $%
f\left( x\right) \sim \frac{A}{x^{\frac{3+\lambda }{2}}}$ is that there is a
flux of particles escaping to infinity. Indeed, it is readily seen that for
this kernel:
\bean
\lim_{R\rightarrow \infty }J_{R}^{-}\left[ f\right] =2\pi A^{2}
\eean
Notice that  the  function $f\left( x\right) =%
\frac{A}{x^{\frac{3+\lambda }{2}}}$ can be thought as a stationary solution
of (\ref{e1.1}), (\ref{e1.1bis}) characterized by a constant flux of particles
propagating from the smaller to the larger values of $x.$

\subsection{Particle fluxes for the linearized equation.}

Since equation (\ref{e2.4bis}), (\ref{e2.5}) has been obtained linearizing (\ref{e1.1}), (\ref{e1.1bis}) we can derive formulas for the particle fluxes associated to
(\ref{e2.4bis}), (\ref{e2.5}) linearizing (\ref{Q1F1}). We recall that (\ref{e2.4bis}), (\ref{e2.5}) has been obtained using that 
\begin{equation}
f\left( x,t\right) =x^{-\frac{3+\lambda }{2}}+g\left( x,t\right) 
\label{R1E1}
\end{equation}
in (\ref{e1.1}), (\ref{e1.1bis}) and keeping just linear terms on $g.$ Plugging (\ref{R1E1}) into (\ref{Q1V1}) and keeping just the linear
term on $g$ we obtain:

\bear
\label{SFE5}
\frac{d}{dt}\left( \int_{0}^{R}xg\left( t,x\right) dx\right) =-J_{R,lin}^{-}
\eear
where $J_{R,lin}^{-}$ is: 

\bear
\label{SFE4}
J_{R,lin}^{-} & = & \int_{D^-_1(R)}\frac{g(z)}{y^{\frac{3+\lambda}{2}}}(y z)^{\frac{\lambda}{2}}(y+z)dydz+
\int_{D^-_1(R)}\frac{g(y)}{z^{\frac{3+\lambda}{2}}}(y z)^{\frac{\lambda}{2}}(y+z)dydz\nonumber \\
&&+\int_{D^-_2(R)}\frac{g(z)}{y^{\frac{3+\lambda}{2}}}(y z)^{\frac{\lambda}{2}}zdydz+
\int_{D^-_2(R)}\frac{g(y)}{z^{\frac{3+\lambda}{2}}}(y z)^{\frac{\lambda}{2}}z dydz \nonumber \\
& = & \int_{D^-_3(R)}\frac{z^{\frac{\lambda}{2}}}{y^{\frac{3}{2}}}(y+z) g(z)dydz+
\int_{D^-_2(R)}\frac{z^{\frac{\lambda}{2}+1}}{y^{\frac{3}{2}}}g(z) dydz +\int_{D^-_2(R)}\frac{y^{\frac{\lambda}{2}}}{z^{\frac{1}{2}}}g(y) dydz \nonumber  \\
& = & I_1(t, R)+I_2(t, R)+I_3(t, R) \label{SFE4bis}
\\
D^-_3(R) & = & \left\{0\le z\le R, \,\,0\le y \le R,\, R\le (y+z)\right\}. 
\label{SFE4ter}
\eear
and $D_{1}^{-}\left( R\right) ,D_{2}^{-}\left( R\right)$ defined by (\ref{S9ED1menos}) and (\ref{S9ED2menos}) respectively. Integrating (\ref{SFE5}) we obtain:
\bear
\label{SFE6}
\int_0^{R}x\, g(0, x)dx=\int_0^t J_{R,lin}^{-}(s)ds+\int_0^{R}x\, g(t, x)dx
\eear
The solution $g\left( t,x\right) $ obtained in Theorem \ref{S2T1-1} satisfies 
\bear
\label{SFE3}
g(t, x)\sim a(t)\, x^{-(3+\lambda)/2}\quad \hbox{as}\,\,x\to +\infty.
\eear

Taking the limit $R\rightarrow \infty $ in (\ref{SFE6}) and using the fact
that the terms $I_{1}\left( t,R\right) ,\;I_{3}\left( t,R\right) $ tend to
zero as $R\rightarrow \infty $ and that $\lim_{R\rightarrow \infty
}I_{2}\left( t,R\right) =a\left( t\right) $ , it follows that:

\bear
\label{SFE8}
\int_0^{\infty}x\, g(0, x)dx=\int_0^t a(s)\,ds+\int_0^{\infty}x\, g(t, x)dx.
\eear
The self-similar asymptotics (\ref{S2Ttinfty1})-(\ref{S2Ttinfty3}) implies that $%
\lim_{t\rightarrow \infty }\int_{0}^{\infty }xg\left( t,x\right) dx=0.$
Then:
\bear
\label{SFE9}
\int_0^{\infty}x\, g(0, x)dx=\int_0^\infty a(s)\,ds.
\eear
The left hand side of (\ref{SFE9}) is the initial total mass of the perturbation. The right hand side of (\ref{SFE9}) is the total amount of particles contained in clusters of infinite size. Equation (\ref{SFE9}) means that all the excess of particles initially introduced in the system move as $t\to +\infty$ to an infinitely large cluster.
\section{Appendix I.}
\label{Appendix I}
 \setcounter{equation}{0}
\setcounter{theo}{0}

We take now the Mellin transform on both hands of the equation. We recall that the Mellin transform of a function $g(y)$ is defined as:
\beqn
\label{e2.6}
{\mathcal M}(h)(s )=\int_0^{\infty }y^{s -1}h(y)dy.
\eeqn
Taking the Mellin transform of the right hand side of (\ref{e2.4bis}), (\ref{e2.5}) we obtain after  straightforward calculations:

\bear
\label{e2.11}
\frac{\partial}{\partial s}{\mathcal M}(g)(s) & = & {\mathcal M}(g)\left(s+{\lambda -1\over 2}\right)M\left(-s+\frac{3}{2}\right).
\\
\label{e2.11bisbis}
M(s) & = & 
\int_2^{\infty }\theta
^{1/2-s}\left((\theta -1)^{-3/2}-\theta ^{-3/2} \right)\, d\theta +\nonumber \\
\nonumber \\
& & +\int_{1/2}^1(1-\theta )^{-3/2}\left(\theta ^{s-1}-1\right)d \theta+{2^{-s}\over s} -2\sqrt 2 \label{e2.113200}\\
& = & I_1(s)+I_2(s)+I_3(s)+I_4.\label{e2.113241}
\eear
 \begin{rem}
\label{S2R2}
If the function $g$ satisfies the estimates (\ref{S2Est}), (\ref{S2Est2}) for some $r>1$ and $\rho<2$ such that $\rho < r$  we will have that its Mellin transform is well defined in the strip $\Re e{s} \in (\lambda/2 +\rho, \lambda/2 +r)$.
\end{rem}
\subsection{The function $M(s)$.}
We consider in this Section the auxiliary function $M$ obtained by taking the Mellin transform of the equation (\ref{e2.4bis}), (\ref{e2.5}) and first rewrite it in terms of Gamma functions.
\begin{prop}
\label{e2.15prop}
The function $M(s)$ defined in (\ref{e2.113200}) can be written as 
\bear
\label{e2.15bisprop}
M(s) & = -\,\frac{2\sqrt \pi \,\Gamma(s)}{\Gamma(s-1/2)}.
\eear
\end{prop}

\noindent
\textbf{Proof.}
We can write the term  $I_1$ of (\ref{e2.113241}) as
\bean
I_1(s) & = &
\int_2^{\infty}\theta^{1/2-s}\theta^{-3/2}\left\{
\left(1-\frac{1}{\theta} \right)^{-3/2}-1
\right\}d\theta 
\eean 
Using Binomial Theorem to expand $\left(1-\frac{1}{\theta} \right)^{-3/2}$ and integrating each term of the resulting series we obtain:
\bean
I_1=  2^{-s}\sum _{n=1}^{\infty }\pmatrix {-3/2 \cr n}(-1)^n\frac{ 2^{-n}}{s+n}.
\eean
Then,
\bear
I_1+I_3=2^{-s}\sum _{n=0}^{\infty }\pmatrix {-3/2 \cr n}(-1)^n\frac{ 2^{-n}}{s+n}.\label{A0E1}
\eear
Integrating by parts we obtain:
\bear
I_2(s) & = & 2\sqrt 2-2\sqrt 2 (1/2)^{s-1}-2(s-1)\int_{1/2}^1 (1-\theta )^{-1/2}\theta ^{s-2}d\theta. \label{A0E2}
\eear
In order to compute the last term in (\ref{A0E2}) we write:
\bear
\int_{1/2}^1 (1-\theta )^{-1/2}\theta ^{s-2}d\theta=\int_0^1
(1-\theta )^{-1/2}\theta ^{s-2}d\theta-\int_0^{1/2}(1-\theta )^{-1/2}\theta ^{s-2}d\theta,\label{A0E3}
\eear
Expanding $(1-\theta )^{-1/2}$ using the Binomial Theorem and integrating each term of the resulting series:
\bear
\int_0^{1/2}(1-\theta )^{-1/2}\theta ^{s-2}d\theta & = & 2^{-s}\left\{ \frac{2}{s-1}+\sum_{\ell=0}^{\infty}
\pmatrix {-1/2 \cr \ell +1}\frac{(-1)^{\ell+1}}{\ell+s}2^{-\ell}\right\} 
\label{A0E4}
\eear
Moreover,
\bear
\frac{2}{s-1}+\sum_{n=0}^{\infty}\pmatrix {-1/2 \cr n +1}\frac{(-1)^{n+1}}{n+s}2^{-n}  
= \frac{2(1-1/2)^{-1/2}}{s-1}  -  \frac{1}{2(s-1)}\sum_{n=0}^{\infty}\pmatrix {-3/2 \cr n +1}\frac{(-1)^{n}}{n+s}2^{-n}
\label{A0E5}
\eear
where we have used that
$
\pmatrix {-1/2 \cr n +1}=-\frac{1}{2(n+1)}\pmatrix {-3/2 \cr n }
$
as well as the Binomial Theorem.
Combining  (\ref{A0E4}) and (\ref{A0E5}) we deduce
\bear
\label{A0E7}
\int_0^{1/2}(1-\theta )^{-1/2}\theta ^{s-2}d\theta  = 
2^{-s}\left\{
\frac{2\sqrt 2}{s-1}
-\frac{1}{2(s-1)}\sum_{n=0}^{\infty}\pmatrix {-3/2 \cr n +1}\frac{(-1)^{n}}{n+s}2^{-n}\right\}.
\eear
Therefore, using (\ref{e2.11bisbis}), (\ref{A0E1}), (\ref{A0E2}), (\ref{A0E3})  and (\ref{A0E7}), 
\bear
M(s)  =  -2(s-1)\int_0^1(1-\theta)^{-1/2}\theta^{s-2}d\theta
 =  -\frac{2\,\sqrt 2 \,\Gamma(s)}{\Gamma(s-1/2)}
.\label{A0E8}
\eear
\qed

\subsection{Zeros and Poles of $M(s)$}
The function $M(s)$ is meromorphic on the whole complex plane. Its zeros are given by the poles of the function $\Gamma(s-1/2)$:
\bear
\label{A0E9}
s_z(n)=\frac{1}{2}-n, \quad n=0, 1, 2, \cdots
\eear
The second zero, $M(-1/2)=0$ corresponds to the solution $x^{-(3+\lambda )/2}$.
\\ 
On the other hand, the poles of $M(s)$ are those of the function $\Gamma(s)$, i. e. 
\bean
\label{e2.17}
s_p(n)= -n,\quad n=0, 1, 2, \cdots
\eean

\subsection{Fourier variables.}
We make now a change of variables in order to have functions defined in all of the
real line $\RR$. To this end we define:

\bean
\label{e2.18}
{\mathcal M}(g)(s)=\int_0^{\infty }g(y)y^{s-1}ds  =  \int_{-\infty }^{\infty
}g(e^{X})e^{-iX\xi }dX
\equiv \sqrt{2\, \pi}\widehat G(\xi )
\eean
i.e., ${\mathcal M}(g)(s)=\sqrt{2\pi} \widehat G(\xi )$ with $s=-i\xi $, then, 
\beqn
\label{e2.19}
{\mathcal M}(g)\left(s+{\lambda -1\over
2}\right)= \sqrt{2\, \pi}\widehat G\left(\xi + {\lambda -1\over2}i\right).
\eeqn
If we re-write the equation (\ref{e2.11}) in these new
variables we obtain,

\beqn
\label{e2.21}{\partial G\over \partial t}(\xi )=G\left(\xi +{\lambda -1\over 2}i\right)M\left(i\xi
+{3\over 2}\right).
\eeqn
It is more convenient to use instead of $M$ the function $\Phi$ defined as:
\bear
\label{e2.23}
 \Phi (\xi
) = M\left(i\xi +1+{\lambda \over2}\right).
\eear
Using (\ref{e2.15bisprop}), we obtain
$\Phi$:
\bear
\Phi (\xi ) & = & -\,\frac{2\sqrt \pi \,\Gamma(i\xi +1+{\lambda \over2})}{\Gamma(i\xi +\frac{\lambda+1}{2})}.
\label{e2.25}
\eear
The equation (\ref{e2.21}) reads then as equation 
(\ref{e2.24}).
The set of zeros of $\Phi$ are:
\bear
\label{A0E10}
\xi_z(n)=i\, \left(n+{\lambda +1\over2}\right)\quad n=0, 1, 2, \cdots
\eear
The set of poles is
\bear
\label{A0E11}
\xi_p(n)= i\, \left( n+1+{\lambda \over2}\right)\quad n=0, 1, 2, \cdots
\eear
Notice that by definition,
\bean
M\left(-{1\over 2}\right)=\Phi \left({3+\lambda \over 2}i\right)=0,\quad M\left({1\over 2}\right)=\Phi
\left({1+\lambda
\over 2}\, i\right)=0.
\eean

\subsection{Behaviour of $\Phi$ at infinity.}

\begin{theo}
\label{S10Th10.3}
Let us define $\Theta \equiv \Theta \left( s\right)
=sgn\left( \Im m\left( s\right) \right) .$ For any fixed $L>0,$ the
following asymptotic formulas hold:
\begin{eqnarray}
M\left( s\right)  &=&-2\sqrt{\pi s}\left( 1-\frac{3}{8s}+O\left( \frac{1}{%
s^{2}}\right) \right) \;\;\hbox{as\ \ }\left| \Im m\left( s\right)
\right| \rightarrow \infty   \label{F1E1} \\
M\left( s\right)  &=&-\sqrt{2\pi }\left( 1+i\Theta \right) \sqrt{-is\Theta }-%
\frac{3}{8}\frac{\sqrt{2\pi }\left( 1+i\Theta \right) \sqrt{-is\Theta }}{s}%
+O\left( \frac{1}{s^{3/2}}\right) \;\;\hbox{as\ \ }\left| \Im m\left(
s\right) \right| \rightarrow \infty\nonumber \\   \label{A1E117}
\end{eqnarray}
uniformly in sets where $\arg \left( s\right) \in \left( -\pi
+\varepsilon _{0},\pi -\varepsilon _{0}\right) $ for any $\varepsilon _{0}>0$.
\end{theo} 
\textbf{Proof }\bigskip\ Formula (\ref{F1E1}) is a consequence of
(\ref{e2.15bisprop}) as well as the asymptotic formula:
\begin{equation}
\Gamma \left( z\right) \sim \sqrt{2\pi }\left( z\right) ^{z-\frac{1}{2}%
}e^{-z}\left( 1+\frac{1}{12z}+O\left( \frac{1}{z^{2}}\right) \right) \;\;%
\hbox{as\ \ }\left| z\right| \rightarrow \infty   \label{F1E2}
\end{equation}
that is uniformly valid in sets $\arg \left( z\right) \in \left( -\pi
+\varepsilon _{0},\pi -\varepsilon _{0}\right) $ for any $\varepsilon _{0}>0.
$ 
\\
Then:
\begin{eqnarray*}
M\left( s\right)  =-\frac{2\sqrt{%
\pi }\left( s\right) ^{s-\frac{1}{2}}\left( s-\frac{1}{2}\right) ^{\frac{1}{2%
}}}{\left( s-\frac{1}{2}\right) ^{s-\frac{1}{2}}e^{\frac{1}{2}}}\left(
1+O\left( \frac{1}{s^{2}}\right) \right) \;\; \hbox{as\ \ } \left|s\right|\to \infty,
\end{eqnarray*}
uniformly in sets $\arg \left( s\right) \in \left( -\pi
+\varepsilon _{0},\pi -\varepsilon _{0}\right) $  and $\arg \left( s-1/2\right) \in \left( -\pi
+\varepsilon _{0},\pi -\varepsilon _{0}\right) $  for any $\varepsilon _{0}>0.$
Notice that:
\begin{eqnarray*}
\frac{\left( s\right) ^{s-\frac{1}{2}}}{\left( s-\frac{1}{2}\right) ^{s-%
\frac{1}{2}}} &=&\sqrt{e}\left[ 1-\frac{1}{8\left( s-\frac{1}{2}\right) }+O\left( 
\frac{1}{s^{2}}\right) \right]  \;\; \hbox{as\ \ } \left|s\right|\to \infty,
\end{eqnarray*}
uniformly in the same sets as above,
whence:
\begin{eqnarray*}
M\left( s\right)  &=&-2\sqrt{\pi s}\left( 1-\frac{3}{8s}+O\left( \frac{1}{s^{2}}\right)
\right)  \;\; \hbox{as\ \ } \left|s\right|\to \infty,
\end{eqnarray*}
uniformly in sets $\arg \left( s\right) \in \left( -\pi
+\varepsilon _{0},\pi -\varepsilon _{0}\right) $  for any $\varepsilon _{0}>0.$
and (\ref{F1E1}) follows. Formula (\ref{A1E117}) is a consequence of (\ref{F1E1}).
\qed

\section{Appendix II: Some technical propositions.}
\label{Some technical propositions.}
\setcounter{equation}{0}
\setcounter{theo}{0}
\noindent
We prove now some auxiliary results used to prove the results of the paper.
\begin{lem}
\label{A1T100}
Suppose that $f$ and $h$ are two analytic functions in the cone
\bean
C(2\varepsilon_0)=\left\{\zeta \in \CC; \, \zeta=|\zeta|e^{i\theta}, \,\theta\in (-2\varepsilon_0, 2\varepsilon_0)  \right\}
\eean
for some $\varepsilon_0>0$ and real valued in $\RR^+$. Let us also assume that
\bear
&&\int_0^{\infty}\frac{|f(re^{i\theta})|+|h(re^{i\theta})|}{1+r^2}dr<+\infty, \,\,\,\hbox{for any}\,\,\,\,\theta\in(-2\varepsilon_0, 2\varepsilon_0) \label{A1E118}\\
&&\lim_{|\zeta|\to 0}f(\zeta)=\theta_1\qquad\hbox{and}\quad \lim_{|\zeta|\to \infty}f(\zeta)=\theta_2 \label{A1E119}\\
&&|f'(\zeta)|=o\left(\frac{1}{|\zeta|} \right) \quad \hbox{as}\quad |\zeta|\to 0,\, \, |\zeta |\to \infty,\,\,\zeta \in C(2 \varepsilon_0). 
\label{A1E120}
\eear
Then, the function
\bear
\label{A1E121}
F(\zeta)=\frac{1}{2\pi i}\int_0^{\infty}(h(s)+if(s))\left(\frac{1}{s-\zeta}-\frac{1}{s+1} \right)ds
\eear
is analytic in the domain:
\bear
\label{A1E121bis}
D(\varepsilon_0)=\left\{\zeta \in \CC; \, \zeta=|\zeta|e^{i\theta}, \,\theta\in (-\varepsilon_0, 2\pi+\varepsilon_0)  \right\}.
\eear
 Moreover,
\bear
F(\zeta)=-\frac{\theta_1}{2\pi}\ln \zeta+i H(\eta)+o(\ln|\zeta|),\quad \hbox{as}\,\,\, \zeta \to 0, \,\,\zeta \in D(\varepsilon_0),\label{A1E122}\\
F(\zeta)=-\frac{\theta_2}{2\pi}\ln \zeta+i H(\eta)+o(\ln|\zeta|),\quad \hbox{as}\,\,\, |\zeta| \to +\infty, \,\,\zeta \in D(\varepsilon_0).\label{A1E123}
\eear
where the function $H(\zeta)$ is a real valued function defined by
\bear
\label{A1E123}
H(\zeta)=-\frac{1}{2\pi }\int_0^{\infty}h(s)\left(\frac{1}{s-\zeta}-\frac{1}{s+1} \right)ds.
\eear
\end{lem}
\textbf{Proof of Lemma \ref{A1T100}.} Using the Lemma C.2 of \cite{EMV} with the function $f$ we obtain (\ref{A1E122}) and (\ref{A1E123}). 
On the other hand, the condition $(\ref{A1E118})$ ensures that the function $F$ is well defined and of course real valued. \qed
\begin{prop}
\label{A1T101}
Let ${\cal V}(\xi)$ defined by (\ref{C3}) and (\ref{S5E103}). Then, for any $\delta>0$ arbitrarily small and all $M>0$ arbitrarily large, there exist two  positive constants $C_{1, \delta, M}, C_{2, \delta, M}$ such that
\bear
C_{1, \delta, M}e^{-(\frac{\pi}{2}\frac{1}{\lambda-1}+\delta)|\xi|}\le|{\cal V}(\xi)|\le C_{2, \delta, M}e^{-(\frac{\pi}{2}\frac{1}{\lambda-1}-\delta)|\xi|}\label{A1E130}
\eear
uniformly for $\Im m (\xi)$ in compact sets of $(3/2, (3+\lambda)/2)$
as well as for all $\xi$ such that $|\Re e(\xi)|\ge 1$, $|\Im m(\xi)|\le M$.
\end{prop}
\textbf{Proof of Proposition \ref{A1T101}}  Given $\xi$ such that $\Im m(\xi) \in (3/2, (3+\lambda)/2)$ we can represent the function $\mathcal V$ by (\ref{C3}) and (\ref{S5E103}) for $\beta_1$ such that $\beta_1-(\lambda-1)/2 <\xi \le \beta_1$. In order to simplify some of the calculations we use the following change of variables.
\bear
\label{A1ZZ1}
\zeta=e^{\frac{4 \pi}{\lambda-1}(\xi-\beta_1)}\\
\nu (\zeta)=\mathcal V(\xi) \label{A1ZZ2}\\
\varphi(\zeta)=\Phi(\xi) \label{A1ZZ3}
\eear
The function $\ln\left(-\varphi (s )\right)$ may be written as
\bear
\label{A1E132}
\ln\left(-\varphi (s )\right)=\ln\left(\left|\varphi (s )\right| \right)+i\arg\left(-\varphi (s )\right).
\eear
The functions $\ln\left(\left|\varphi (s )\right| \right)$ and $\arg\left(\varphi (s )\right)$ satisfy the hypothesis required to $h$ and $f$ respectively  in Lemma \ref{A1T101}. In particular, by Proposition \ref{S3T1} and the fact that $\Re e (-\Phi(y)) >0$ for all $y$ such that $\Im m(y)\in ((2+\lambda)/2, (3+\lambda)/2)$ we may normalize the argument of the function $\ln\left(-\varphi (s )\right)$ such that:
\bear
\label{A1E133}
\lim_{\zeta \to 0}\arg\left(-\varphi (\zeta)\right)=-\frac{ \pi}{4},\qquad
\lim_{\zeta \to \infty}\arg\left(-\varphi (\zeta)\right)=\frac{\pi}{4}.
\eear
Applying Lemma \ref{A1T100} it follows that:
\bear
{1\over 2\pi i } \int_0^{\infty }\ln\left(-\varphi (s ) \right)
\left( {1\over s -\zeta }-{1\over s +1}\right)\, ds & = & \frac{1}{8}\ln (-\zeta)+i H(\eta)+o(\ln|\zeta|)\label{A1E133bis}\\ 
&&\hbox{as}\,\,\, \zeta \to 0, \,\,\zeta \in D(\varepsilon_0), \nonumber\\
{1\over 2\pi i } \int_0^{\infty }\ln\left(-\varphi (s ) \right)
\left( {1\over s -\zeta }-{1\over s +1}\right)\, ds & = & -\frac{1}{8}
\ln (-\zeta)+i H(\eta)+o(\ln|\zeta|) \label{A1E134}\\
&&\quad \hbox{as}\,\,\, \zeta \to \infty, \,\,\zeta \in D(\varepsilon_0).\nonumber 
\eear
The two estimates in  (\ref{A1E130}) follow, for $\Im m (\xi)$ in compact sets of $(3/2, (3+\lambda)/2)$,
by taking exponentials in both sides of (\ref{A1E133bis}) and (\ref{A1E134}) and inverting the change of variables (\ref{A1ZZ1})-(\ref{A1ZZ3}).\\
In order to prove the estimate for  $\xi$ in the region  $|\Re e(\xi)|\ge 1$ and  $|\Im m(\xi)|\le M$, we extend analytically the function $\mathcal V (\xi)$ to such regions using (\ref{E5}) as well as the fact that, by Proposition \ref{S3T1}, we have for some positive constants $C_1$ and $C_2$:
\bean
C_1|\xi|^{1/2}\le |\Phi(\xi)|\le C_2|\xi|^{1/2}
\eean
for $|\Re e(\xi)|\ge 1$ and  $|\Im m(\xi)|\le M$.
\qed

\begin{prop}
\label{SAT8-1K}
The zeros of the function ${\cal V}$ are:
\bear
&&\left(1+\frac{\lambda}{2}+n+k\frac{\lambda-1}{2}\right)i, \quad  n=1,2, \cdots,\,\,\,k=0, 1, \cdots \label{SAT8-1Kzeros1}\\
&&\left(\frac{1+\lambda}{2}-k\frac{\lambda-1}{2}\right) i,  \quad k=1, 2, \cdots \label{SAT8-1Kzeros2}
\eear
The poles of the function ${\cal V}$ are:
\bear
&&\left(\frac{1+\lambda}{2}+n+k\frac{\lambda-1}{2} \right)i,\quad n=1, 2, \cdots, \quad k=0, 1,Ê\cdots \label{SAT8-1Kpoles1} \\
&&\left(1+\frac{\lambda}{2}-k\frac{\lambda-1}{2}\right)i,\quad k=1, 2,\cdots
\eear
\end{prop}
\textbf{Proof.} The proof follows from (\ref{E5}) and the distribution of zeros and poles of $\Phi$ given in (\ref{e3.1}) and (\ref{e2.26}). Consider first any $\eta \in \CC$ such that $\Im m(\eta)>\beta_0+(\lambda-1)/2$ and let $k$ be the first natural integer such that 
$$\Im m(\eta)-k\frac{\lambda-1}{2}\in \left(\beta_0, \beta_0+\frac{\lambda-1}{2}\right).$$
By (\ref{E5}),
\bear
\label{SA3E19-260K}
{\cal V}(\eta)=(-1)^j\frac{{\cal V}(\eta-j\frac{\lambda-1}{2}i)}
{\Phi(\eta)\Phi(\eta-\frac{\lambda-1}{2}i)\cdots \Phi(\eta-j \frac{\lambda-1}{2}i)}
\eear
as far as none of the points $\eta-\ell \frac{\lambda-1}{2}i$, $\ell =0, \cdots j$ is a zero nor a pole f $\Phi$. Therefore, if 
\bear
\label{SA3E19-261K}
\forall j=0, 1, \cdots k:\quad \eta-j \frac{\lambda-1}{2}i\, \not \in\, \{\xi_z(n), \xi_p(n),\,\,n=0, 1, \cdots \}
\eear
then
\bear
\label{SA3E19-261K}
{\cal V}(\eta)=(-1)^k\frac{{\cal V}(\eta-k\frac{\lambda-1}{2}i)}
{\Phi(\eta)\Phi(\eta-\frac{\lambda-1}{2}i)\cdots \Phi(\eta-k \frac{\lambda-1}{2}i)}
\eear
and therefore ${\cal V}(\eta)$ is well defined and not zero. If for some $ j\in \{0, 1, \cdots k \}$ and $n\in \{0, 1, \cdots \}$ we have 
$\eta-k\frac{\lambda-1}{2}i=\xi_z(n)$, then, by (\ref{SA3E19-260K}), the point $\eta$ would be a pole of ${\cal V}$. Moreover, that is the only way for a given $\eta$ to be a pole of ${\cal V}$. Therefore, the poles $\eta$ of ${\cal V}$ such that $\Im m(\eta)>\beta_0+(\lambda-1)/2$ are:
\bear
\label{SA3E19-262K}
\left(n+\frac{1+\lambda}{2}+k\frac{\lambda-1}{2}\right) i, \quad n=1, 2, \cdots, \,\, k=0, 1, \cdots
\eear
If, on the other hand, for some $ j\in \{0, 1, \cdots k \}$ and $n\in \{0, 1, \cdots \}$ we had
$\eta-k\frac{\lambda-1}{2}i=\xi_p(n)$, then, by (\ref{SA3E19-260K}), the point $\eta$ would be a zero of ${\cal V}$. And this would again be the only way  for $\eta$ to be a zero of ${\cal V}$. We deduce that the zeros $\eta$ of ${\cal V}$ such that $\Im m(\eta)>\beta_0+(\lambda-1)/2$ are:
\bear
\label{SA3E19-263K}
\left(1+n+\frac{\lambda}{2}+k\frac{\lambda-1}{2}\right) i, \quad n=1, 2, \cdots, \,\, k=0, 1, \cdots
\eear
We have thus proved that, above the strip $\Im m(\eta) \in (\beta_0, \beta_0+(\lambda-1) /2)$, the zeros and poles of ${\cal V}$ are given by (\ref{SAT8-1Kzeros1}) and (\ref{SAT8-1Kpoles1}) respectively.\\
A similar argument may be done when $\Im m(\eta)<\beta_0$. If $k$ is the least natural integer such that 
$$\Im m(\eta)+k\frac{\lambda-1}{2}\in \left(\beta_0, \beta_0+\frac{\lambda-1}{2}\right)$$
we have by (\ref{E5}),
\bear
\label{SA3E19-260K}
{\cal V}(\eta)=(-1)^j\frac{{\cal V}(\eta+j\frac{\lambda-1}{2}i)}
{\Phi(\eta)\Phi(\eta+\frac{\lambda-1}{2}i)\cdots \Phi(\eta+j \frac{\lambda-1}{2}i)}
\eear
as far as none of the points $\eta+\ell \frac{\lambda-1}{2}i$, $\ell =0, \cdots j$ is a zero nor a pole f $\Phi$. Notice that by the definition of $k$ and the choice of $\beta_0$, the only zero of $\Phi$ that my be founded is $\xi_z(0))=(1+\lambda)i/2$ and the only pole is $\xi_p(0)=(2+\lambda) i/2$. The proof then follows as before. \qed

\section{Appendix III: Stationary phase.}  
 \setcounter{equation}{0}
\setcounter{theo}{0}

We  must estimate in Sections \ref{decayghat} and \ref{decayG}  several integral expressions  of the form
\label{Stationary phase}
\bean
\int_{Im\, Y =\beta_0-\Im m\xi}e^{\Psi (\xi, Y, t)} A\left(\frac{2 i Y}{\lambda-1} \right)dY
\eean
for a given function $\Psi$ but different functions $A$. This is done using the stationary phase method. We collect in this Section some technical results about the function $\Psi (\xi, Y, t)$ and its critical points.

\subsection{The critical point.} We first compute the critical points of the function  $\Psi(\xi, Y, t)$ defined by (\ref{S5E2JKLM3}) or equivalently, that of $\Phi(\xi, Z, t)$ defined in (\ref{S5E2JKLM3bis}). 
Define:
\bear
 \label{S5Eanalyticlambda}
D(\xi, B)=\left\{Z\in \CC;\,\,\Im m Z <0,\,|\Im m Z|\le B \left|\Re e Z+\frac{Q}{8}\sqrt{|\xi|}\,\right|, \,\, 
sign (\Re e Z)=sign(\Re e \xi)\right\}
 \eear
 where $Q=sign(\Re e (\xi))$.

 \begin{figure}
\begin{tikzpicture}
\draw[->] (-6.5,0) -- (3,0) node[right]{$\Re e Z$};
\draw[->] (-1,-2) -- (-1,2.6) node[above]{$\Im m Z$};
\draw[->] (-5,0.01) -- (3,0.01);
\draw[->] (-5, 0.01) -- (2.5,-2.49);
\draw[->] (-5, 0) -- (2.5,-2.5);
\draw (-5, 0) node{$\times$};
\draw (-5, -0) node[below]{$\frac{\sqrt{|\xi|}}{8}$};
\end{tikzpicture}
\caption{$D(\xi, B)$ when $\Re e(\xi)>0$ and then $Q=1$.}
\end{figure}
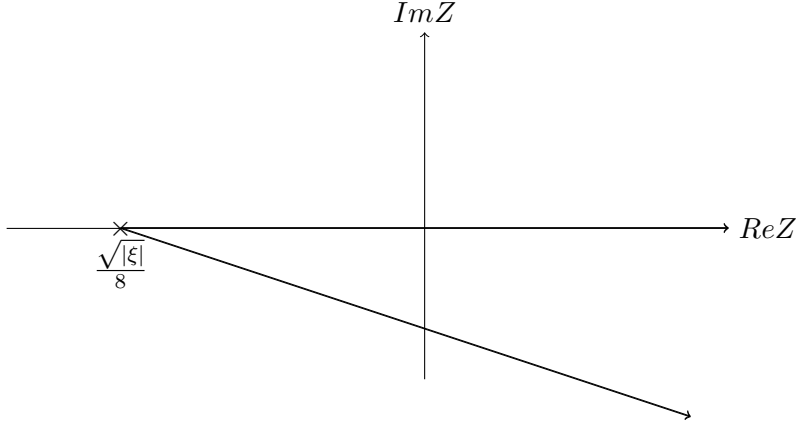
\vskip 1cm

\subsubsection{The case $0<t<1$.}

 We start with the following:
\begin{lem}
\label{S5TPhi}
Consider the function $F(\xi, Z)$ defined by means of
\bear
\label{S5TPhiE1}
F(\xi, Z)=\frac{2}{(\lambda-1)\, i}
\int_{\Im m(\eta)=\beta_1}
\ln\left(-\Phi (\eta)\right)
\Theta(\eta-\xi, \sqrt{|\xi|} Z )d\eta
\eear
for $\beta_1\in (3/2, (3+\lambda)/2)$, $\Im m(\xi)\in (3/2, \beta_1)$ and  
\bear
\label{S5TPhiE2}
\frac{1}{\sqrt{|\xi|}} (\beta_1-\Im m(\xi))-\frac{\lambda-1}{2}<\Im m(Z)\le \frac{(\beta_1-\Im m(\xi))}{\sqrt{|\xi|}}.
\eear
 For any constant $B>0$ the function $F(\xi, \cdot)$ can be extended analytically on the variable $Z$ to the domain  $|Z|\le \frac{\sqrt{|\xi|}}{8}$, $Z\in D(\xi, B)$ if $|\xi|\ge \xi_0$ for  $\xi_0=\xi_0(B)$ sufficiently large. Moreover, there exists a positive constant $C=C(B)$ such that
\bear
\label{S5TPhiE3}
\left|F(\xi, Z)+\frac{2\, i}{(\lambda-1)}
 \ln\left(-\Phi (\xi)\right)\, \sqrt{|\xi|}\, Z \right|\le C\, \left(Z^2+{\cal O}\left(\frac{1}{|\xi|} \right)\right)
\eear
for $|Z|\le \frac{\sqrt{|\xi|}}{8}$, $Z\in D(\xi, B)$ and $|\xi|\ge \xi_0$. 

\end{lem}
\textbf{Proof.} The function $F$ is well defined in  (\ref{S5TPhiE2}) since in that domain the variable $Z$ is in the region where the  function $\Theta $ is analytic. The function $F(\xi, \cdot)$ given by (\ref{S5TPhiE1}) is then analytic in the  strip 
$
|\Im m Z| \le \frac{\delta_0(\Im m \xi)}{\sqrt{|\xi|} }
$
 for some $\delta_0(\Im m \xi)$ sufficiently small. We now claim that
 for any fixed constant $B>0$, and $\Im m \xi \in (\beta_0, \beta_0+(\lambda-1)/2)$, if $\Re e \xi$ is sufficiently large (depending on $B$), this function $F(\xi, \cdot)$ may be extended  analytically  to the region$D(\xi, B)$.
To prove this we derive new representation formulas of the function $F$ performing suitable contour deformations in the variable $\eta$ in (\ref{S5TPhiE1}). \\
Notice that the singularities of the integrand in (\ref{S5TPhiE1}) are contained in the set $\Re e \eta =0$, $\eta= \xi +(\lambda-1)\ell /2$ and  $\eta= \xi+ \sqrt{|\xi|}\, Z +(\lambda-1)\ell /2$ for $\ell\in  \ZZ$. Given $Z_0$ in the region above, let be $\widehat Z_0$ such that:
\bean
\Re e \widehat Z_0=\Re e Z_0, \quad |\Im m \widehat Z_0|\le  \frac{\delta_0(\Im m \xi)}{\sqrt{|\xi|} }
\eean
and the integral curve $Im\, \eta  =\beta_0+\frac{\lambda-1}{2}-\varepsilon$ lies between the two points $\xi+ \widehat Z_0\sqrt{|\xi|}$ and $\xi+ \widehat Z_0\sqrt{|\xi|}+(\lambda-1)i/2$. (Notice that this is possible since $\delta_0(\Im m \xi)$ may be made as small as we need and $\Im m \eta >\Im m \xi$.)

 Consider the vertical segment of the complex plane connecting $Z_0$ and $\widehat Z_0$: $Z_{\theta}=(1-\theta)\widehat Z_0+\theta Z_0$, $\theta \in [0, 1]$. We then obtain an analytic extension of $F(\xi, \cdot)$ varying $\theta$ continuously from $0$ to one and deforming continuously the contour
$ Im\, \eta  =\beta_0+\frac{\theta-1}{2}-\varepsilon$ in such a way that:
\begin{itemize}
\item it always passes between $\xi+ \widehat Z_\theta\sqrt{|\xi|}$ and $\xi+ \widehat Z_\theta\sqrt{|\xi|}+(\lambda-1)i/2$ \\
\item we do not change the original integration contour for 
\bean
\left|\Re e \left(\eta-\xi-Z_0 \sqrt{|\xi|}\right)\right|\ge \frac{|\Re e (Z_0)\sqrt{|\xi|}}{2}
\eean
 \end{itemize}
 
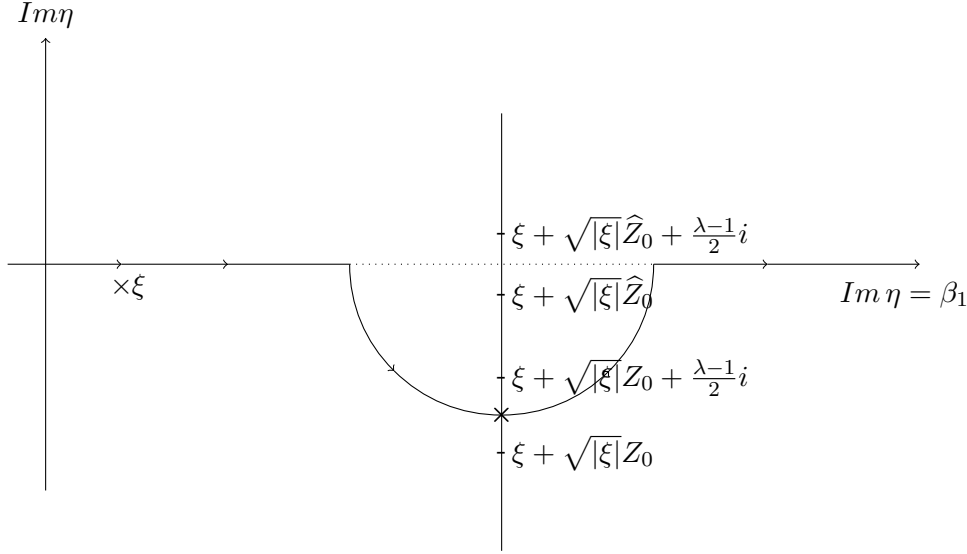
\begin{figure}
\label{figuracuatro}
\begin{tikzpicture}
\draw[->] (-6,-3) -- (-6,3) node[above]{$\Im m \eta$};
\draw[->] (-6.5, 0) -- (-5, 0);
\draw[->] (-5, 0) -- (-3.6, 0);
\draw (-3.6, 0) -- (-2, 0);
\draw[dotted] (-2,0) -- (2,0);
\draw[->] (-2,0) arc(180:225:2cm);
\draw(225:2cm) arc(225:270:2cm) node{$\times$};
\draw (-5, -0.3) node{$\times$};
\draw (-5, -0.3) node[right]{$\xi$};
\draw (0, 0.4) node{-};
\draw (0, 0.4) node[right]{$\xi+\sqrt{|\xi|}\widehat Z_0+\frac{\lambda-1}{2}i$};
\draw (0, -0.4) node{-};
\draw (0, -0.4) node[right]{$\xi+\sqrt{|\xi|}\widehat Z_0$};
\draw (0, -3.8) -- (0, 2);
\draw (0, -1.5) node{-};
\draw (0, -2.5) node{-};
\draw (0, -1.5) node[right]{$\xi+\sqrt{|\xi|} Z_0+\frac{\lambda-1}{2}i$};
\draw (0, -2.5) node[right]{$\xi+\sqrt{|\xi|} Z_0$};
\draw (270:2cm) node{$\times$};
\draw [->](270:2cm) arc(270:315:2cm);
\draw (315:2cm) arc(315:360:2cm);
\draw[->] (2,0) -- (3.5,0);
\draw[->] (3.5,0) -- (5.5,0);
\draw (5.3,-0.4) node{$Im\, \eta  =\beta_1$};
\end{tikzpicture}
\caption{The curve ${\cal C}_2$}
\end{figure}

 The first condition ensures that the integration contour never crosses any of the singularities of the function $\left(1-e^{-\frac{4 \pi}{\lambda-1}(\eta-\xi-Z\sqrt{|\xi|})}\right)^{-1}$ . The second one ensures that it does not cross neither any of the singularities of $\left(1-e^{\frac{4 \pi}{\lambda-1}(\eta-\xi)}\right)^{-1}$.\\ \\
Finally, since $sign (\Re e Z_0)=sign (\Re e \xi)$, the new integration contour never crosses the line $\Re e \eta =0$ where the singularities of $\ln \left(-\Phi(\eta)\right)$ are located.\\ \\
To estimate this integral we write
\bear
\label{S5E2JKLMn15}
&&\frac{2}{(\lambda-1)\, i}
\int_{{\cal C}_2}
\ln\left(-\Phi (\eta)\right)
\Theta(\eta-\xi, \sqrt{|\xi|} Z )d\eta = \frac{2\, \ln\left(-\Phi (\xi)\right)}{(\lambda-1)\, i}
\int_{{\cal C}_2}
\Theta(\eta-\xi, \sqrt{|\xi|}  Z)d\eta+\nonumber \\
&&+
\frac{2}{(\lambda-1)\, i}
\int_{{\cal C}_2}
\ln\left(\frac{\Phi (\eta)}{\Phi (\xi)}\right)
\Theta(\eta-\xi, \sqrt{|\xi|}t\,  Z )d\eta=I_1+I_2.
\eear
The first integral, $I_1$ is computed explicitly :
\bear
\label{S5E2JKLMn16}
I_1=\frac{2\, \ln\left(-\Phi (\xi)\right)}{(\lambda-1)\, i}
\sqrt{|\xi|}\, Z
=-\frac{2\, i}{(\lambda-1)}
 \ln\left(-\Phi (\xi)\right)\, \sqrt{|\xi|}\, Z.
\eear
In order to compute the second integral we have to distinguish the cases $\Re e \xi \to -\infty$ and $\Re e \xi \to +\infty$. Since both may be treated using similar arguments let us treat only the case $\Re e \xi \to +\infty$. In that case we decompose $I_2$ as follows
\bear
\label{S5E2JKLMn17}
I_2 & = & \frac{2}{(\lambda-1)\, i}
\int_{{\cal C}_2, \Re e \eta >0, |\eta -\xi|\le \frac{|\xi|}{4}}
[\cdots]\, d\eta+\int_{{\cal C}_2, \Re e \eta >0, |\eta -\xi|> \frac{|\xi|}{4}}
[\cdots]\, d\eta \nonumber \\
& + & \frac{2}{(\lambda-1)\, i}
\int_{{\cal C}_2, \Re e \eta <0}
[\cdots]\, d\eta
=I_{2,1}+I_{2,2}+I_{2, 3}
\eear
In $I_{2, 1}$ we use now Proposition \ref{S3T1} and Taylor's expansion to obtain:
\bean
\label{S5E2JKLMn18}
&&\frac{\Phi (\eta)}{\Phi (\xi)}
=\frac{\sqrt{\eta}\left(1+\frac{(2\lambda+1) i}{8\eta}+O(|\eta|^{-2}\right)}{
\sqrt{\xi}\left(1+\frac{(2\lambda+1) i}{8\xi}+O(|\xi|^{-2}\right)}
\eean
using now $\eta=\xi+(\eta-\xi)$ we can write:
\bean
&&\frac{\Phi (\eta)}{\Phi (\xi)}=1+{\cal O}\left(\frac{\eta-\xi}{\, \xi}\right)+{\cal O}\left(\frac{1}{\xi^2} \right),
\,\,\,  \hbox{as}\,\,\Re e \xi >1, \,\,|\eta-\xi|\le \frac{|\xi|}{4}.
\eean
Therefore:
\bean
\ln \left(\frac{\Phi (\eta)}{\Phi (\xi)} \right)={\cal O}\left(\frac{\eta-\xi}{\, \xi}\right)+{\cal O}\left(\frac{1}{\xi^2} \right),
\,\,\,  \hbox{as}\,\,\Re e \xi >1, \,\,|\eta-\xi|\le \frac{|\xi|}{4}.
\eean
We now estimate the two following integrals for all $Z\in D(\xi, B)$ and $|Z|\le \frac{\sqrt{|\xi|}}{8}$. The first one:
\bean
&&\int_{{\cal C}_2, \Re e \eta >0, |\eta -\xi|\le \frac{|\xi|}{4}}
\left(\frac{|\eta-\xi|}{|\xi|}\right)
\Theta(\eta-\xi, \sqrt{|\xi|} Z )d\eta=\frac{1}{| \xi|}\int_{\widetilde {\cal C}_2, \Re e \sigma >-\Re e \xi,\,  |\sigma|\le \frac{|\xi|}{4}}\left| \sigma
\Theta(\sigma, \sqrt{|\xi|} Z )\right|d\sigma\nonumber\\
&&\hskip 3cm \le
\frac{1}{| \xi|}\int_{\widetilde {\cal C}_2, \Re e \sigma >-\Re e \xi,\,  |\sigma|\le 2  |\xi|^{1/2} Z}\left| \sigma
\Theta(\sigma, \sqrt{|\xi|} Z )\right|d\sigma + \nonumber \\
&&\hskip 5cm +
\frac{1}{| \xi|}\int_{\widetilde {\cal C}_2, \Re e \sigma >-\Re e \xi,\,  |\sigma|\ge 2  |\xi|^{1/2} Z}\left| \sigma
\Theta(\sigma, \sqrt{|\xi|} Z )\right|d\sigma
\nonumber \\
&&\hskip 3cm\le C\, Z^2+C\frac{e^{-a|\xi|^{1/2} |Z|}}{|\xi|}\,\,\,
\hbox{for}\,\,\Re e \xi > \xi_0(B)
\eean
the second one:
\bean
&&\int_{{\cal C}_2, \Re e \eta >0, |\eta -\xi|\le \frac{|\xi|}{4}}
\frac{1}{|\xi|^2}
\left|\Theta(\eta-\xi, \sqrt{|\xi|} Z )\right|d\eta=\frac{1}{|\xi|^2}\int_{\widetilde{\cal C}_2, \Re e \sigma >-\Re e \xi, |\sigma|\le \frac{|\xi|}{4}}
\left|\Theta(\sigma, \sqrt{|\xi|} Z )\right|d\sigma \nonumber \\
&&\hskip 2cm\le\frac{1}{|\xi|^2}\int_{\widetilde{\cal C}_2, \Re e \sigma >-\Re e \xi, |\sigma|\le 2|\xi|^{1/2}Z}
\left|\Theta(\sigma, \sqrt{|\xi|} Z )\right|d\sigma +\nonumber \\
&&\hskip 5cm +\frac{1}{|\xi|^2}\int_{\widetilde{\cal C}_2, \Re e \sigma >-\Re e \xi, |\sigma|\ge 2|\xi|^{1/2}Z}
\left|\Theta(\sigma, \sqrt{|\xi|} Z )\right|d\sigma\nonumber \\
&&\hskip 2cm\le C\frac{|Z|}{|\xi|^{3/2}}+C \frac{e^{-a|\xi|^{1/2}|Z|}}{|\xi|^2}\le C\left(|Z|^2+\frac{1}{|\xi|^{3}}\right)+C \frac{e^{-a|\xi|^{1/2}|Z|}}{|\xi|^2}\,\,\,
\hbox{for}\,\,\Re e \xi >\xi_0(B)
\eean
where in both cases $\widetilde {\cal C}_2={\cal C}_2-\xi$, $\xi_0(B)$ is a positive constant sufficiently large, depending on $B$ and  $a$ and $C$  are positive constants which may depend on $B$ but are independent on $\Re e \xi$ and $Z\in D(\xi, B)$.
From where we deduce that, for all $\Re e \xi >\xi_0(B)$:
\bear
\label{S5E2JKLMn20}
|I_{2, 1}|\le C\left(|Z|^2+\frac{1}{|\xi|^{3}}\right)+C\frac{e^{-a|\xi|^{1/2} |Z|}}{|\xi|}
\eear

In the integral $I_{2, 2}$ we use the fact that when $|\eta-\xi|>\frac{|\xi|}{4}$ and  $\Re e \xi> \xi_0(B)$, the function $\Theta(\eta-\xi, Y)$ has an exponential decay $Ce^{-a|\xi|}$ with $C$ and $a$ as above as well as the inequality
$|\ln (-\Phi (\eta))|\le C|\ln (|\eta-\xi|+|\xi|)|\le C|(\ln (|\eta-\xi|)|
+|\ln (|\xi|)|)$ for $\eta$ is large. For $\eta$ of order one we use that $\ln (-\Phi (\eta))$ is of order one to derive a similar estimate. Then
\bear
\label{S5E2JKLMn21}
|I_{2,2}|\le 
\int_{|\eta -\xi|> \frac{|\xi|}{4}}
e^{-a|\eta-\xi|}\left( |\ln (|\eta-\xi|)|
+|\ln (|\xi|)|\right)\, d\eta={\cal O}\left(e^{-a |\xi|} \right).
\eear
Finally, the estimate of $I_{2, 3}$ follows using the same cut off properties of the function $\Theta$ since $\Re e \eta<0$ and $\Re e \xi \to +\infty$ implies that $\Re e (\eta-\xi)>C|\xi|$.  The final estimate of $I_2$, by (\ref{S5E2JKLMn17}),  is then,
\bear
\label{S5E2JKLMn22}
|I_2|\le  C\left(|Z|^2+\frac{1}{|\xi|^{3}}\right)+C\frac{e^{-a|\xi|^{1/2} |Z|}}{|\xi|}.
\eear
Using Proposition \ref{S3T1}, (\ref{S5E2JKLM3bis}), (\ref{S5E2JKLMn15}), (\ref{S5E2JKLMn16}) and (\ref{S5E2JKLMn22}) the Lemma follows.
\qed
\begin{lem}
\label{S5TPhidos}
For any $B>0$, the function $h$ defined by means of:
\bear
\label{S5E2JKLMn25}
\Phi(\xi, Z, t) &= &
-\sqrt{|\xi|}\frac{2 i Z}{\lambda-1}\left[1+\ln t-\ln \left(\frac{2 i Z}{\lambda-1} \right)
+\ln\left(2\sqrt{\pi}e^{iQ\frac{\pi}{4}} \right)\right]-\nonumber \\
&&-\frac{1}{2}\ln \left(|\xi|^{1/2} \right)-\frac{1}{2}\ln \left(\frac{2\, i\, Z}{\lambda-1} \right)+h(\xi, Z, t)\eear
satisfies 
\bean
|h(\xi, Z, t)|\le C\, \left(Z^2+{\cal O}\left(\frac{1}{|\xi|} \right)\right)
\eean
for $|Z|\le \frac{\sqrt{|\xi|}}{8}$, $Z\in D(\xi, B)$ and $|\xi|\ge \xi_0$ for some positive constants $C=C(B)$ and $\xi_0=\xi_0(B)$ sufficiently large.

\end{lem}
\textbf{Proof.} This Lemma is a direct consequence of Lemma (\ref{S5TPhi}). The only difficulty in order to estimate the function $\Phi$ defined in (\ref{S5E2JKLM3bis})
comes from the term ${\cal V}(\xi)/{\cal V}(\xi+Y)$ which corresponds to the integral term. This term, given by formula (\ref{S5E2JK42bis-1}) may also be written as:
\bean
\label{S5E2JKLMn10mnbis}
 \Lambda (\xi , Z)  =  \frac{{\cal V}(\xi)}{{\cal V}(\xi+\sqrt{|\xi|} Z)}
=\exp\left[\frac{2}{(\lambda-1)\, i}
\int_{Im\, \eta  =\beta_1}
\ln\left(-\Phi (\eta)\right)
\Theta(\eta-\xi, \sqrt{|\xi|} Z )d\eta\right]. 
\eean
\qed

As for the critical point $Z_c$ of $\Phi(\xi, \cdot, t)$, the precise result is the following.
\begin{lem}
\label{S5Lcriticalpt}
Suppose that $B>2 \sqrt \pi$. Let us define $\delta_0=(\lambda-1)\sqrt \pi /4$.
For all $0\le t\le 1$ and $\Im m \xi \in (\beta_0, \beta_0+(\lambda-1)/2)$, $|\xi|\,t^2\ge \xi_0(B)$ with $\xi_0(B)$ a positive constant sufficiently large, there exists a unique point $Z_c\in D(\xi, B)\setminus B_{\delta_0\, t}(0)$ such that  $\partial \Phi(\xi, Z_c, t)/\partial Z =0$. Moreover the following asymptotics holds uniformly for $0\le t \le 1$:
\bear
\frac{2\, i \, Z_c}{\lambda-1} & = & \sqrt{2 \pi}t(1+i Q))\left(1
+{\cal O}\left(\frac{1}{\sqrt{|\xi|}\, t}\right)\right)\,\,\,\hbox{as}\,\,\,|\Re e \xi|\, t^2 \to \infty.
\label{S5Ecriticalpt}
\eear
\end{lem}
\textbf{Proof.} Computing the derivative of the function $\Phi$ given by (\ref{S5E2JKLM3bis}) gives for all $Z\in D(\xi, B)$:
\bear
\label{S5Ederivadafi}
&&\frac{\partial \Phi}{\partial Z}(\xi, Z, t)=-
\frac{8\pi\sqrt{|\xi|}\,i}{(\lambda-1)^2}
\int_{{\cal C}_2}\ln\left(-\Phi (\eta)\right)\frac{e^{\frac{4 \pi}{\lambda-1}(\sqrt{|\xi|}Z-\eta+\xi)}}
{\left(1-e^{\frac{4\pi}{\lambda-1}(\sqrt{|\xi|}Z-\eta+\xi)}\right)^2}d\eta-
\nonumber \\
&&+\sqrt{|\xi|}\left(+
\left(\frac{2 i }{\lambda-1}\right)\ln \left( \frac{2 i (Z/t)}{\lambda-1}\right)
-\frac{1}{2  \sqrt{|\xi|}Z} +\frac{2\, i}{\lambda-1}\ln |\xi|^{1/2}\right).
\eear
We compute the leading term of the integral in the right hand side of (\ref{S5Ederivadafi}) as $|\xi|\to \infty$:
\bean
&&I(\xi, Z)=
\int_{{\cal C}_2}
\ln\left(-\Phi (\eta)\right)\frac{e^{\frac{4 \pi}{\lambda-1}(\sqrt{|\xi|}Z-\eta+\xi)}}
{\left(1-e^{\frac{4\pi}{\lambda-1}(\sqrt{|\xi|}Z-\eta+\xi)}\right)^2}d\eta= \nonumber \\
&&
\int_{\widehat {\cal C}_2}
\ln\left(-\Phi (\sigma+\xi+\sqrt{|\xi|}Z)\right)\frac{e^{-\frac{4 \pi}{\lambda-1}\sigma}}
{\left(1-e^{-\frac{4\pi}{\lambda-1}\sigma}\right)^2}d\sigma
\eean
where $\widehat {\cal C}_2=C_2-\xi-\sqrt{|\xi|}\, Z$. Using Proposition \ref{S3T1} we have that, uniformly for $Z\in D(\xi, B)$, $|Z|\le B$:
\bean
\ln\left(-\Phi (\sigma+\xi+\sqrt{|\xi|}Z)\right)=\ln\left(-\Phi (\xi+\sqrt{|\xi|}Z)\right)+A(\xi)\frac{\sigma}{|\xi|}+
{\cal O}\left( \frac{\sigma^2}{|\xi|^2}\right)
\eean
for $|\xi|\to +\infty$, where $A(\xi)$ is a bounded function of $sign (\Re e \xi)$. It then follows:

\bear
\label{S5Taabbcc}
&&I(\xi, Z)=
\ln\left(-\Phi (\xi+\sqrt{|\xi|}Z)\right)
\int_{\widehat {\cal C}_2}
\frac{e^{-\frac{4 \pi}{\lambda-1}\sigma}}
{\left(1-e^{-\frac{4\pi}{\lambda-1}\sigma}\right)^2}d\sigma+ \nonumber\\
&&+\frac{A(\xi)}{|\xi|}
\int_{\widehat {\cal C}_2}
\sigma \frac{e^{-\frac{4 \pi}{\lambda-1}\sigma}}
{\left(1-e^{-\frac{4\pi}{\lambda-1}\sigma}\right)^2}d\sigma+
\int_{\widehat {\cal C}_2}
{\cal O}\left( \frac{\sigma^2}{|\xi|^2}\right)\frac{e^{-\frac{4 \pi}{\lambda-1}\sigma}}
{\left(1-e^{-\frac{4\pi}{\lambda-1}\sigma}\right)^2}d\sigma\nonumber \\
&&+{\cal O}\left(e^{-a|\xi|}\right) \,\,\hbox{as}\,\,|\xi|\to +\infty
\eear
uniformly for $Z\in D(\xi, B)$, $|Z|\le B$, where $a>0$ is independent on $\xi$, $Z$.
Using:
\bean
\frac{d}{d\sigma}\left(\frac{1}{1-e^{-\frac{4\pi}{\lambda-1}\sigma}} \right)=
-\frac{4\pi}{\lambda-1}\frac{e^{-\frac{4 \pi}{\lambda-1}\sigma}}
{\left(1-e^{-\frac{4\pi}{\lambda-1}\sigma}\right)^2},
\eean
we obtain:
\bean
\int_{\widehat {\cal C}_2}
 \frac{e^{-\frac{4 \pi}{\lambda-1}\sigma}}
{\left(1-e^{-\frac{4\pi}{\lambda-1}\sigma}\right)^2}d\sigma=-\frac{\lambda-1}{4 \pi}.
\eean
On the other hand, in order to estimate the second term in the right hand side of (\ref{S5Taabbcc}) we deform the integration contour $\widehat {\cal C}_2$ to a horizontal line at a bounded distance of the real axis. The resulting integral con then be bounded by a positive constant independent of $\xi$, $Z$. The third term in the right hand side of (\ref{S5Taabbcc}) can be bounded using the specific form of $\widehat {\cal C}_2$ as:
\bean
C\left( \frac{|Z|^2}{|\xi|}+\frac{|Z|^3}{|\xi|^{1/2}}\right).
\eean
We notice that by Proposition \ref{S3T1}, we have:
\bean
\ln\left(-\Phi (\xi+\sqrt{|\xi|}Z)\right)=\ln\left(-\Phi (\xi)\right)+{\cal O}\left(\frac{|Z|}{\sqrt{|\xi|}} \right),
\,\,\hbox{as}\,\,|\xi|\to +\infty
\eean
uniformly for $Z\in D(\xi, B)$, $|Z|\le B$. 
Combining everything we deduce:
\bear
\label{S5Eabvt10}
&&I(\xi, Z)=-\frac{\lambda-1}{4 \pi}\ln\left(-\Phi (\xi)\right)
+{\cal O}\left(\frac{|Z|}{|\xi|^{1/2}}\right),
 \,\,\hbox{as}\,\,|\xi|\to +\infty
\eear
uniformly for $Z\in D(\xi, B)$, $|Z|\le B$. Combining (\ref{S5Ederivadafi}) and (\ref{S5Eabvt10}) it follows:
\bean
\frac{\partial \Phi}{\partial Z}(\xi, Z, t) & = & 
\frac{2i\sqrt{|\xi|}}{(\lambda-1)}\left(\ln\left(-\Phi (\xi)\right)+
\ln \left( \frac{2 i (Z/t)}{\lambda-1}\right)
+\frac{(\lambda-1)i}{4  \sqrt{|\xi|}Z} +\ln \left(|\xi|^{1/2}\right)+\right.\\
&&\hskip 6.5cm \left.
+{\cal O}\left(\frac{|Z|}{|\xi|^{1/2}}\right) \right) \,\,\hbox{as}\,\,|\xi|\to +\infty
\eean
uniformly for $Z\in D(\xi, B)$, $|Z|\le B$. 

Using Rouch\'e's Theorem it then follows that for $|\xi| t^2$ sufficiently large, $\xi \in Franja$, $0\le t \le 1$,  there exists a unique root of 
$({\partial \Phi}/{\partial Z})(\xi, Z, t)=0$ in $Z\in D(\xi, B)\setminus B_{\delta_0 t}$ satisfying the asymtotics (\ref{S5Lcriticalpt}) and the Lemma follows. \qed

We now derive estimates for higher order derivatives of $\Phi$. 
\begin{lem} 
\label{S5Tsecondder}
Suppose that $Z_c$ is as in Lemma \ref{S5Lcriticalpt}. Then the following asymptotics holds:
\bean
\frac{\partial ^2 \Phi}{\partial Z^2}(\xi, Z_c, t)=\frac{2\, i\, \sqrt{|\xi|}}{(\lambda-1)\, Z_c}\left(1+{\cal O}\left(\frac{1}{\sqrt{|\xi|}t}\right)\right)
\,\,\,\hbox{as}\,\,\,|\xi|\, t^2\to +\infty.
\eean
uniformly in $0\le t \le 1$.
\end{lem}

\textbf{Proof.} The second derivative of $\Phi$ with respect to $Z$ is:
\bean
&&\frac{\partial ^2 \Phi}{\partial Z^2}=-\frac{8 \pi |\xi| i}{(\lambda-1)^2}
\int_{{\cal C}_2}
[\ln\left(-\Phi (\eta)\right)]'\frac{e^{\frac{4 \pi}{\lambda-1}(Z\sqrt {|\xi|}-\eta+\xi)}}
{\left(1-e^{\frac{4\pi}{\lambda-1}(Z\sqrt {|\xi|}-\eta+\xi)}\right)^2}d\eta+
\frac{2 \sqrt{|\xi|}\, i}{(\lambda-1)\, Z}+\frac{1}{2 Z^2}\nonumber \\
&&=-\frac{8 \pi |\xi| i}{(\lambda-1)^2}
\int_{{\cal C}_2}
\frac{\Phi' (\eta)}{\Phi (\eta)}
\frac{e^{\frac{4 \pi}{\lambda-1}(Z\sqrt {|\xi|}-\eta+\xi)}}
{\left(1-e^{\frac{4\pi}{\lambda-1}(Z\sqrt {|\xi|}-\eta+\xi)}\right)^2}d\eta+\frac{2\, i \sqrt{|\xi|}}{(\lambda-1)\, Z}+\frac{1}{2 Z^2}.
\eean
Taking $Z=Z_c$ and using Lemma \ref{S5Lcriticalpt}, we deduce, uniformly in $0\le t \le 1$
\bear
\label{S5Enumeroahi}
&&\frac{\partial ^2 \Phi}{\partial Z^2}(\xi, Z_c, t)=\frac{2\, i\, \sqrt{|\xi|}}{(\lambda-1)\, Z_c}
-\frac{8 \pi |\xi| i}{(\lambda-1)^2}
\int_{{\cal C}_2}
\frac{\Phi' (\eta)}{\Phi (\eta)}
\frac{e^{\frac{4 \pi}{\lambda-1}(Z_c\sqrt {|\xi|}-\eta+\xi)}}
{\left(1-e^{\frac{4\pi}{\lambda-1}(Z_c\sqrt {|\xi|}-\eta+\xi)}\right)^2}d\eta+{\cal O}\left( \frac{1}{t}\right),\nonumber\\
\eear
as $|\xi| t^2\to +\infty$. Using Proposition \ref{S3T1}:
\bear
\label{S5Elajota}
&&J(\xi)=\int_{{\cal C}_2}
\frac{\Phi' (\eta)}{\Phi (\eta)}
\frac{e^{\frac{4 \pi}{\lambda-1}(Z_c\sqrt {|\xi|}-\eta+\xi)}}
{\left(1-e^{\frac{4\pi}{\lambda-1}(Z_c\sqrt {|\xi|}-\eta+\xi)}\right)^2}d\eta=
\frac{1}{2}\int_{{\cal C}_2}
\frac{1}{\eta}
\frac{e^{\frac{4 \pi}{\lambda-1}(Z_c\sqrt {|\xi|}-\eta+\xi)}}
{\left(1-e^{\frac{4\pi}{\lambda-1}(Z_c\sqrt {|\xi|}-\eta+\xi)}\right)^2}d\eta\nonumber \\
&&+\int_{{\cal C}_2}
{\cal O}\left(\frac{1}{|\xi|^2}\right)
\frac{e^{\frac{4 \pi}{\lambda-1}(Z_c\sqrt {|\xi|}-\eta+\xi)}}
{\left(1-e^{\frac{4\pi}{\lambda-1}(Z_c\sqrt {|\xi|}-\eta+\xi)}\right)^2}d\eta+{\cal O}\left(e^{-a|\xi|}\right),\,\,
\hbox{as}\,\,|\xi|\to +\infty
\eear
where $a$ is a positive constant independent on $\xi$. The first term in the right hand side is estimated as:
\bean
&&\int_{{\cal C}_2}
\frac{1}{\eta}
\frac{e^{\frac{4 \pi}{\lambda-1}(Z_c\sqrt {|\xi|}-\eta+\xi)}}
{\left(1-e^{\frac{4\pi}{\lambda-1}(Z_c\sqrt {|\xi|}-\eta+\xi)}\right)^2}d\eta=
\frac{1}{\xi}\int_{{\cal C}_2}
\frac{e^{\frac{4 \pi}{\lambda-1}(Z_c\sqrt {|\xi|}-\eta+\xi)}}
{\left(1-e^{\frac{4\pi}{\lambda-1}(Z_c\sqrt {|\xi|}-\eta+\xi)}\right)^2}d\eta+\nonumber\\
&&+\int_{{\cal C}_2}
{\cal O}\left(\frac{\eta-\xi}{|\xi|^2}\right)
\frac{e^{\frac{4 \pi}{\lambda-1}(Z_c\sqrt {|\xi|}-\eta+\xi)}}
{\left(1-e^{\frac{4\pi}{\lambda-1}(Z_c\sqrt {|\xi|}-\eta+\xi)}\right)^2}d\eta+
{\cal O}\left(e^{-a|\xi|}\right)\,\,\hbox{as}\,\,|\xi|\to +\infty.
\eean

The first term in the right hand side can be computed explicitly and gives $-(\lambda-1)/4\pi \xi$. The second one is estimated using the form of the contour ${\cal C}_2$  and is bounded by ${\cal O}\left(|Z_c|/|\xi| \right)$.\\  Therefore, the first  term in the right hand side of (\ref{S5Elajota}) is estimated as ${\cal O}(1/|\xi|)$ as $|\xi|\to +\infty$. The second term is bounded using similar arguments by  ${\cal O}(1/|\xi|^{3/2})$ as $|\xi|\to +\infty$. It then follows that
\bean
J(\xi)={\cal O}\left(\frac{1}{|\xi|}\right)\,\,\hbox{as}\,\,|\xi|\to +\infty.
\eean
Using (\ref{S5Enumeroahi}) we obtain
\bean
\frac{\partial ^2 \Phi}{\partial Z^2}(\xi, Z_c, t)=\frac{2\, i\, \sqrt{|\xi|}}{(\lambda-1)\, Z_c}+{\cal O}\left(\frac{1}{t}\right)\,\,\,
\hbox{as}\,\,|\xi|t^2\to +\infty
\eean
whence Lemma \ref{S5Tsecondder} follows.
\qed
\begin{lem} 
\label{S5Ttercerder}
Suppose that $Z_c$, $B$ and $\delta_0$ are as in Lemma \ref{S5Lcriticalpt}. Then the following asymptotics holds:
\bean
\left|\frac{\partial ^3 \Phi}{\partial Z^3}(\xi, Z, t)\right|={\cal O}\left(\frac{\sqrt{|\xi|}}{t^2}\right)
\,\,\,\hbox{as}\,\,\,|\xi|\, t^2\to +\infty.
\eean
uniformly in $0\le t \le 1$, $Z\in D(\xi, B)$, $|Z|\le B$.
\end{lem}

\noindent
\textbf{Proof of Lemma \ref{S5Ttercerder}. }
\bear
\label{S10Enumeroals}
&&\frac{\partial ^3 \Phi}{\partial Z^3}(\xi, Z, t)=-\frac{8 \pi |\xi|^{3/2} i}{(\lambda-1)^2}
\int_{{\cal C}_2}
\left(\frac{\Phi' (\eta)}{\Phi (\eta)}\right)'
\frac{e^{\frac{4 \pi}{\lambda-1}(Z\sqrt {|\xi|}-\eta+\xi)}}
{\left(1-e^{\frac{4\pi}{\lambda-1}(Z\sqrt {|\xi|}-\eta+\xi)}\right)^2}d\eta-\nonumber \\
&&-\frac{2\, i \sqrt{|\xi|}}{(\lambda-1)\, Z^2}-\frac{1}{ Z^3}
\eear
The last two terms of (\ref{S10Enumeroals}) are bounded as $C(\sqrt{|\xi|}/t^2 +1/t^3)$. and this can be estimated as $C\sqrt{|\xi|}/t^2$ for $|\xi|t^2 >>1$. On the other hand we may bound the first term in the right hand side of (\ref{S10Enumeroals}) using 
\bean
\left|\left(\frac{\Phi' (\eta)}{\Phi (\eta)}\right)'\right|\le \frac{C}{1+|\eta|^2}.
\eean
the form of the contour ${\cal C}_2$. The term under consideration is then bounded by a constant. Combining all these estimates for the terms in the right hand side of (\ref{S10Enumeroals}) the Lemma follows. \qed 

\noindent
Combining Lemma \ref{S5TPhi} and Lemma \ref{S5Lcriticalpt} it follows 

\begin{cor}
\label{S5TPhiCrit}
For all $0<t<1:$
\bean
\Phi(\xi, Z_c, t)=-\sqrt{|\xi|}\sqrt{2 \pi}\,t\,(1+iQ)-\frac{1}{2}\ln \left(|\xi|^{1/2} \right)
-\frac{1}{2}\ln \left( \frac{2\, i\, Z_c}{\lambda-1}\right)+{\cal O}(1)\,\,\,\hbox{as}\,\,|\xi|t^2\to +\infty.
\eean
\end{cor}
\textbf{Proof.}
Using (\ref{S5Ecriticalpt}) and (\ref{S5E2JKLMn25}) we deduce
\bear
\label{S5E2JKLMn23}
&&\Phi(\xi, Z_c, t)=
-\sqrt{|\xi|}\frac{2 i Z_c}{\lambda-1}\left[1+\ln (t) -\ln \left(\frac{2 i Z_c}{\lambda-1} \right)
+\ln\left(2\sqrt{\pi}e^{iQ\frac{\pi}{4}} \right)\right]-\frac{1}{2}\ln \left( \frac{2\, i\, Z_c}{\lambda-1}\right)+{\cal O}(1)\nonumber \\
&&=-\sqrt{|\xi|}\sqrt{2 \pi}\,t\,(1+iQ)-\frac{1}{2}\ln \left(|\xi|^{1/2} \right)-\frac{1}{2}\ln \left( \frac{2\, i\, Z_c}{\lambda-1}\right)+{\cal O}(1)
\eear
as $|\xi|t^2\to +\infty$.
\qed
\begin{rem} We must emphasise that the sign of the real part of the main term in the asymptotic expansion of $\Phi$ as $ t^2\, |\xi|\to \infty$ is fundamental for the construction of our solutions. 
\end{rem}
In the next Lemma we prove that the function $\Phi(\xi, Z, t)$ is ``well behaved'' in a region $|Z-Z_c|\ge \delta t$ and $|Z|\le M\, t$ for $M>0$ large.

\begin{lem}
\label{A10estimderivada} For all $\delta>0$ and $M>0$ large,  there exists $\varepsilon_0>0$ and $L>0$ such that the function $\Phi(\xi, Z, t)$  satifies:
\bean
\Re e\, \Phi(\xi, Z, t)\le \Re e\, \Phi(\xi, Z_c, t)-\varepsilon_0 \sqrt{|\xi|}\, t
\eean
when $Z$ lies in the curve
\bean
&&\gamma(M)=\gamma_1(M)\cup\gamma_2(M)\cup\gamma_3(M)\setminus \left\{Z;\,\,|Z-Z_c|\le \delta\, t\right\}
\eean
where
\bean
&&\gamma_1(M)=\left\{Z;\,\,Z=Z_c+\lambda,\,\,\lambda\in \RR, |\lambda|\le M\, t \right\}\\
&&\gamma_2(M)=\left\{Z;\,\,Z=Z_c+M\, t+\lambda\, i,\,\,\lambda\in \left[0,\,\, |\Im m Z_c|+\gamma_1\right]  \right\}\\
&&\gamma_3(M)=\left\{Z;\,\,Z=Z_c-M\, t+\lambda\, i,\,\,\lambda\in \left[0,\,\, |\Im m Z_c|+\gamma_1\right]  \right\}
\eean
for all $\xi$ and $t$ such that $|\xi|\, t^2>L$.
\end{lem}
\textbf{Proof.}
Let us consider the auxiliary function:
\bear
\label{A10ETheta}
&&\Theta(\xi, \Omega, t)=-2\sqrt{\pi}\sqrt{|\xi|}\,\Omega\, t\left[1-\ln (\Omega)+\ln (\Omega_0) \right] \\
&&\Omega=\frac{i Z}{\lambda-1}\frac{1}{\sqrt{\pi}\, t}\label{A10EOmega}\\
&&\Omega_0=e^{iQ\frac{\pi}{4}}\label{A10EOmegacero}.
\eear
By Lemma \ref{S5TPhi}:
\bean
\Phi(\xi, Z, t)=\Theta(\xi, \Omega, t)-\frac{1}{2}\ln \left(|\xi|^{1/2} \right)+h(\xi, Z, t).
\eean
Moreover, it is easily checked that $\Omega_0$ is the critical point of the function $\Theta(\xi, \Omega, t)$.
By Lemma \ref{S5Lcriticalpt} we already know that $Z_c$, the critical point of the function  $\Phi$, converges to $\Omega_0$ as $|\xi|\, t^2\to +\infty$. We are now going to study the behaviour of the function $\Theta$ along the curve obtained from $\gamma(M)$ using the change of variable (\ref{A10EOmega}). Due to the convergence properties of the function $\Phi$ and its critical point $Z_c$ when $|\xi|\, t^2\to +\infty$ this will be enough in order to prove  the statement in Lemma \ref{A10estimderivada}.\\
We first consider the curve corresponding to $\gamma_1(M)$. It is then enough to consider $\Re e\,  \Theta(\xi, \Omega, t)$ along the points:
$
\Omega=\Omega_0+ \frac{\lambda}{\sqrt 2 \, }\, i,\quad \lambda\in \RR.
$
A straightforward calculation yields:
\bean
&&\Re e\, \Theta(\xi, \Omega, t)=-\sqrt{2\, \pi}\sqrt{|\xi|}\,t\, \psi(\sigma) \\
&&\sigma=1+Q \lambda
\\
&&\psi(\sigma)=1-\frac{1}{2}\ln\left(1+\sigma^2\right) +\frac{1}{2}\ln 2
-\sigma\left( \frac{\pi}{4}-arctg \sigma\right).
\eean
Since
$
\psi'(\sigma)=-\frac{\pi}{4}+arctg (\sigma)
$
the point $\sigma_0=1$ is a strict minimum for the function $\psi$. It is also easily checked that
\bean
&&\psi(\sigma)\sim \frac{\pi}{4}\quad \hbox{as}\,\,\,\sigma \to +\infty\\
&&\psi(\sigma)\sim -\frac{3\,\pi}{4}\quad \hbox{as}\,\,\,\sigma \to -\infty.
\eean
It follows that, for any $\delta>0$ there exists $\varepsilon >0$ such that if $|\sigma-1|>\delta$,
\bean
\psi (\sigma)-\psi(1)\ge \varepsilon .
\eean
Arguing by continuity this yields the statement of the Lemma when $Z$ lies in the curve $\gamma_1(M)$. \\
The evolution of the function $\theta$ along the two curves corresponding to $\gamma_2(M)$ and $\gamma_3(M)$ is studied with very similar arguments. We  then only consider the case of the curve $\gamma_2(M)$. It is then enough to consider $\Re e\,  \Theta(\xi, \Omega, t)$ along the points:
\bean
&&\Omega= r\, e^{i\, \varphi},\\
&&\Im m \left( r\, e^{i\, \varphi}-\Omega_0 \right)= \pm \frac{M}{\pi (\lambda -1)}
\eean
a straightforward calculation gives:
\bean
\Re e\, \Theta(\xi, \Omega, t)=-2\sqrt{\pi}\sqrt{|\xi|}\,\Omega\, t\left[
r\, \cos\varphi (1-\ln r)-r\, \sin \varphi \left(\frac{\pi Q}{4}-\varphi\right) \right].
\eean
For $Z\in \gamma_2(M)$ and the constant $M$ sufficiently large, we have that $\varphi >\pi/4+\delta$, $r\cos\varphi\le 2$ and $r>M/2$.  Similarly, if $Z\in \gamma_3(M)$ and the constant $M$ sufficiently large, we have that $\varphi<-\pi/4-\delta$, $r\cos\varphi\le 2$ and $r>M/2$. Therefore, we have in both cases:
\bean
r\, \cos\varphi (1-\ln r)-r\, \sin \varphi \left(\frac{\pi Q}{4}-\varphi\right)>\varepsilon>0.
\eean
for $M$ large enough.

Using again the convergence properties of the function $\Phi$ and its critical point $Z_c$ when $|\xi|\, t^2\to +\infty$ this yields  the statement in Lemma \ref{A10estimderivada} when $Z$ lies in the curves $\gamma_2(M)$, $\gamma_3(M)$.\qed

In the following Lemma we extend the  behaviour of $\Phi(\xi, Z, t)$ to the region  $|Z-Z_c|\ge \delta t$ and $|Z|\le \varepsilon _1\sqrt{|\xi|}$ for some $\varepsilon_1>0$ sufficiently small.
\begin{lem}
\label{A10estimderivadamas} For all $\delta>0$ and $M>0$ large,  there exists $a>0$, $\varepsilon_1>0$ and $L>0$ such that the function $\Phi(\xi, Z, t)$  satifies:
\bean
\Re e\, \Phi(\xi, Z, t)\le -a\, \sqrt{|\xi|}\, |Z|,
\eean
for all $Z\in {\cal C}_1$ such that $M\, t\le |Z|\le \varepsilon_1 \sqrt{|\xi|}$ and all  $\xi$ and $t$ such that $|\xi|\, t^2>L$.
\end{lem}
\textbf{Proof.} We only need to check that the function $\Theta$ defined in the proof of the previous Lemma behaves linearly when $|\Omega|\to +\infty$ 
and $\Re e (\Omega)$ remains constant. This follows from
\bean
\Re e(\Theta(\xi, \Omega, t))\le  -\frac{\pi^{3/2}}{2}\sqrt{|\xi|}\, t\,|\Omega|
\eean
uniformly for $\Omega=i\, |\Omega|+{\cal O}(1)$. Using (\ref{A10EOmega}) we deduce
\bean
\Re e(\Theta(\xi, \Omega, t))\le -\frac{\pi}{2}\sqrt{|\xi|}\,\frac{|Z|}{\lambda-1}
\eean
for all $Z\in {\cal C}_1$ such that $|Z|\ge M\, t$ assuming that $M>0$ is sufficiently large. Using Lemma \ref{S5TPhi} we have:
\bean
\Re e\, \Phi(\xi, Z, t)\le \Re e(\Theta(\xi, \Omega, t))+C\, \left(Z^2+{\cal O}\left(\frac{1}{|\xi|} \right)\right) \le -\frac{\pi}{2}\sqrt{|\xi|}\,\frac{|Z|}{\lambda-1}+C\, \left(Z^2+{\cal O}\left(\frac{1}{|\xi|} \right)\right)\\
\le -\sqrt{|\xi|}\, |Z|\left(\frac{\pi}{2(\lambda-1)}-\varepsilon_1+{\cal O}\left(\frac{1}{|\xi|^{3/2}\, t} \right)\right)
\eean
for all  $Z\in {\cal C}_1$ such that $M\, t\le |Z|\le \varepsilon_0\sqrt{|\xi|}$. The result follows for $\varepsilon_1$ small enough and $|\xi|^{1/2}\, t\to +\infty$. \qed

\begin{lem}
\label{A11Tderivadapsi} For all $B>0$ there exists $\xi_0$ and $C>0$ such that 
\bear
\label{A11TderivadadospsiE1}
&&\left|\frac{\partial^\ell \Psi}{\partial \xi^\ell}\left(\xi, Y, t\right)\right|\le C\frac{|Y|}{|\xi|^\ell}, \;\;\ell=1, 2
\eear
for $Y=Z\sqrt{|\xi|}$, $|Y|\le \frac{|\xi|}{8}$, $Z\in D(\xi, B)$ and $|\Re e(\xi)|>\xi_0$.
\end{lem}
\textbf{Proof.} Differentiating the function $\Psi$:
\bean
\Psi(\xi, Y, t) & = & \frac{2}{(\lambda-1)\, i}
\int_{Im\, \eta  =\beta_1}
\ln\left(-\Phi (\eta)\right)
\Theta(\eta-\xi, Y)d\eta-\\
&&-\frac{2 i Y}{\lambda-1}\ln (t)-\frac{2 i Y}{\lambda-1}
+\left(\frac{2 i Y}{\lambda-1}-\frac{1}{2} \right)\ln \left(\frac{2 i Y}{\lambda-1} \right)
\eean
with respect to $\xi$ we obtain:
then,
\bean
\frac{\partial \Psi}{\partial \xi}(\xi, Y, t) & = & \frac{2}{(\lambda-1)\, i}
\int_{Im\, \eta  =\beta_1}
\ln\left(-\Phi (\eta)\right)
\frac{\partial \Theta}{\partial \xi}(\eta-\xi, Y)d\eta\\
& = &  -\frac{2}{(\lambda-1)\, i}
\int_{Im\, \eta  =\beta_1}
\ln\left(-\Phi (\eta)\right)
\frac{\partial \Theta}{\partial \eta}(\eta-\xi, Y)d\eta\\
& = & \frac{2}{(\lambda-1)\, i}
\int_{Im\, \eta  =\beta_1}
\frac{\Phi' (\eta)}{\Phi(\eta)}
\Theta(\eta-\xi, Y)d\eta
\eean
By Proposition \ref{S3T1} we have
\bean
\left|\frac{\Phi' (\eta)}{\Phi(\eta)}\right|\le \frac{C}{1+|\eta|}
\eean
in the domain $Y/\sqrt{|\xi|}\in D(\xi, B)$.  We deform the contour of integration to ${\mathcal C}_2$ defined in Figure 4. Then we split the integral in two pieces:
\bean
|\frac{\partial \Psi}{\partial \xi}(\xi, Y, t)|\le 
C\int_{\eta \in {\mathcal C}_2, |\eta-\xi|\le \frac{|\xi|}{4}}
\frac{1}{1+|\eta|} |
\Theta(\eta-\xi, Y)| |d\eta|\\+
C\int_{\eta \in {\mathcal C}_2, |\eta-\xi|\ge \frac{|\xi|}{4}}
\frac{1}{1+|\eta|}
|\Theta(\eta-\xi, Y)||d\eta|=J_1+J_2.
\eean
By the exponential decay of the function $\Theta$:
\bean
J_2\le Ce^{-a|\xi|}
\eean
for some positive constant $a$. On the other hand,
\bean
\Theta(\eta-\xi, Y)=\frac{1}{1-e^{-\frac{4\pi}{\lambda-1} (\eta-\xi)}} 
-\frac{1}{1-e^{\frac{4\pi}{\lambda-1} (-(\eta-\xi)+Y) }}.
\eean
The integral $J_1$ is then divided in two parts. The first, $J_{1, 1}$ is the integral along the ``vertical part of the curve'' ${\mathcal C}_2$. The second, $J_{1, 2}$ is along the horizontal part of that curve, where $\Im m(\eta)=\beta_1$. In the integral $J_{1,1}$, $\Re e(\eta)$ is bounded and therefore
$
|\Theta(\eta-\xi, Y)|\le C.
$

Since the total length of the integration curve of $J_{1,1}$ is of order $|Y|$ we deduce that $J_{1,1}\le C|Y|/(1+|\xi|)$.  We split the integral $J_{1, 2}$ as follows:
\bean
J_{1,2}\le \frac{C}{1+|\xi|}\left(\int_{\Im m(\eta)=\beta_1, |\eta-\xi|\le 2\,|Y|}|\Theta(\eta-\xi, Y)|d\eta+
\int_{\Im m(\eta)=\beta_1, |\eta-\xi|\ge 2\, |Y|}|\Theta(\eta-\xi, Y)|d\eta\right)
\eean
\bean
&&\int_{\Im m(\eta)=\beta_1, |\eta-\xi|\ge 2\, |Y|}|\Theta(\eta-\xi, Y)|d\eta
\le\int_{\Im m(\eta)=\beta_1, |\sigma|\ge 2\, |Y|}
\left|\frac{e^{-\frac{4\pi}{\lambda-1} \sigma}- e^{\frac{4\pi}{\lambda-1} (Y-\sigma) }}
{(1-e^{-\frac{4\pi}{\lambda-1} \sigma})(1-e^{\frac{4\pi}{\lambda-1} (Y-\sigma) })} 
\right|d\eta.
\eean
We use now that, if $\Re e(\sigma)>2\,|Y|$:
\bean
\left|\frac{e^{-\frac{4\pi}{\lambda-1} \sigma}- e^{\frac{4\pi}{\lambda-1} (Y-\sigma) }}
{(1-e^{-\frac{4\pi}{\lambda-1} \sigma})(1-e^{\frac{4\pi}{\lambda-1} (Y-\sigma) })}\right|\le Ce^{-\frac{2\, \pi}{\lambda-1}|\sigma|}
\eean
and if $\Re e(\sigma)< -2 |Y|$
\bean
\left|\frac{e^{-\frac{4\pi}{\lambda-1} \sigma}- e^{\frac{4\pi}{\lambda-1} (Y-\sigma) }}
{(1-e^{-\frac{4\pi}{\lambda-1} \sigma})(1-e^{\frac{4\pi}{\lambda-1} (Y-\sigma) })}\right|\le 
C\left|\frac{e^{\frac{4\pi}{\lambda-1} Y }}
{(e^{\frac{4\pi}{\lambda-1} \sigma}-1)(e^{\frac{4\pi}{\lambda-1} (\sigma-Y) }-1)}\right|\\
\le Ce^{\frac{4\pi}{\lambda-1} Y }\, e^{-\frac{6\, \pi}{\lambda-1}|\sigma|}.
\eean
The last remaining term  is easily estimated by:
\bean
\int_{\Im m(\eta)=\beta_1, |\eta-\xi|\le 2\,|Y|}|\Theta(\eta-\xi, Y)||d\eta|
\le C\int_{\Im m(\eta)=\beta_1, |\eta-\xi|\le 2\,|Y|}|d\eta|\le C|Y|.
\eean
It then follows that $J_{1, 2}\le Ce^{-a|Y|}/(1+|\xi|)$ for positive constant $C$ and $a$. This ends the proof of (\ref{A11TderivadadospsiE1}) for $\ell=1$. Similarly, 
\bean
\frac{\partial^2 \Psi}{\partial \xi^2}(\xi, Y, t)  =  \frac{2}{(\lambda-1)\, i}
\int_{Im\, \eta  =\beta_1}
\left(\frac{\Phi' (\eta)}{\Phi(\eta)}\right)'
\Theta(\eta-\xi, Y)d\eta
\eean
with
\bean
\left|\left(\frac{\Phi' (\eta)}{\Phi(\eta)}\right)' \right|\le \frac{C}{(1+|\eta|)^2}
\eean
for $Y/\sqrt{|\xi|}\in D(\xi, B)$, again by Proposition \ref{S3T1}. The proof of (\ref{A11TderivadadospsiE1}) follows then from the same arguments as those of the proof of (\ref{A11TderivadadospsiE1}) for $\ell=2$. \qed
\\ \\
\textit{Acknowledgements.} Both authors acknowledge the hospitality and support  of the MPI for Mathematics in the Sciences (Leipzig) where this research was began. JJLV is supported by Grant MTM2007-61755. He thanks Universidad Complutense for its hospitality and also thanks the support of the Humboldt Foundation." M.E. is supported by Grants MTM2008-03541 and IT-305-07.


\end{document}